\documentclass[twocolumn,showpacs,showkeys,preprintnumbers,amsmath,amssymb,floatfix,aps,prc,10pt]{revtex4-1}
\usepackage[latin9]{inputenc}

\usepackage{feynmp-auto}
\usepackage{color}
\usepackage{graphicx}
\usepackage{float}
\usepackage{mathtools}
\usepackage{makecell}
\usepackage{multirow}
\usepackage{siunitx}
\usepackage{array,booktabs,tabularx,longtable}
\usepackage{afterpage} 
\usepackage[pagewise]{lineno}
\usepackage[colorlinks,urlcolor=blue]{hyperref}
\usepackage{breakurl}

\usepackage[normalem]{ulem}
\newcounter{univ_counter}
\addtocounter{univ_counter} {1} 
\edef\ANL{$^{\arabic{univ_counter}}$ }

\addtocounter{univ_counter} {1} 
\edef\BRESCIA{$^{\arabic{univ_counter}}$ }

\addtocounter{univ_counter} {1} 
\edef\UCR{$^{\arabic{univ_counter}}$ }

\addtocounter{univ_counter} {1} 
\edef\CSUDH{$^{\arabic{univ_counter}}$ }

\addtocounter{univ_counter} {1} 
\edef\CANISIUS{$^{\arabic{univ_counter}}$ }

\addtocounter{univ_counter} {1} 
\edef\CNU{$^{\arabic{univ_counter}}$ }

\addtocounter{univ_counter} {1} 
\edef\UCONN{$^{\arabic{univ_counter}}$ }

\addtocounter{univ_counter} {1} 
\edef\DUKE{$^{\arabic{univ_counter}}$ }

\addtocounter{univ_counter} {1} 
\edef\DUQUESNE{$^{\arabic{univ_counter}}$ }

\addtocounter{univ_counter} {1} 
\edef\FU{$^{\arabic{univ_counter}}$ }

\addtocounter{univ_counter} {1}
\edef\FERRARAU{$^{\arabic{univ_counter}}$ }

\addtocounter{univ_counter} {1} 
\edef\FIU{$^{\arabic{univ_counter}}$ }

\addtocounter{univ_counter} {1}
\edef\GSIFFN{$^{\arabic{univ_counter}}$ }

\addtocounter{univ_counter} {1} 
\edef\GWUI{$^{\arabic{univ_counter}}$ }

\addtocounter{univ_counter} {1} 
\edef\GLASGOW{$^{\arabic{univ_counter}}$ }

\addtocounter{univ_counter} {1} 
\edef\INFNFE{$^{\arabic{univ_counter}}$ }

\addtocounter{univ_counter} {1} 
\edef\INFNFR{$^{\arabic{univ_counter}}$ }

\addtocounter{univ_counter} {1} 
\edef\INFNGE{$^{\arabic{univ_counter}}$ }

\addtocounter{univ_counter} {1} 
\edef\INFNPAV{$^{\arabic{univ_counter}}$ }

\addtocounter{univ_counter} {1} 
\edef\INFNRO{$^{\arabic{univ_counter}}$ }

\addtocounter{univ_counter} {1} 
\edef\INFNTUR{$^{\arabic{univ_counter}}$ }

\addtocounter{univ_counter} {1} 
\edef\JLUGiessen{$^{\arabic{univ_counter}}$ }

\addtocounter{univ_counter} {1} 
\edef\JMU{$^{\arabic{univ_counter}}$ }

\addtocounter{univ_counter} {1} 
\edef\KSU{$^{\arabic{univ_counter}}$ }

\addtocounter{univ_counter} {1} 
\edef\KNU{$^{\arabic{univ_counter}}$ }

\addtocounter{univ_counter} {1} 
\edef\MIT{$^{\arabic{univ_counter}}$ }

\addtocounter{univ_counter} {1} 
\edef\MISS{$^{\arabic{univ_counter}}$ }

\addtocounter{univ_counter} {1} 
\edef\UNH{$^{\arabic{univ_counter}}$ }

\addtocounter{univ_counter} {1} 
\edef\NMSU{$^{\arabic{univ_counter}}$ }

\addtocounter{univ_counter} {1} 
\edef\NSU{$^{\arabic{univ_counter}}$ }

\addtocounter{univ_counter} {1} 
\edef\OHIOU{$^{\arabic{univ_counter}}$ }

\addtocounter{univ_counter} {1} 
\edef\ODU{$^{\arabic{univ_counter}}$ }

\addtocounter{univ_counter} {1} 
\edef\ORSAY{$^{\arabic{univ_counter}}$ }

\addtocounter{univ_counter} {1} 
\edef\URICH{$^{\arabic{univ_counter}}$ }

\addtocounter{univ_counter} {1} 
\edef\ROMAII{$^{\arabic{univ_counter}}$ }

\addtocounter{univ_counter} {1} 
\edef\SACLAY{$^{\arabic{univ_counter}}$ }

\addtocounter{univ_counter} {1} 
\edef\MSU{$^{\arabic{univ_counter}}$ }

\addtocounter{univ_counter} {1} 
\edef\SCAROLINA{$^{\arabic{univ_counter}}$ }

\addtocounter{univ_counter} {1} 
\edef\TEMPLE{$^{\arabic{univ_counter}}$ }

\addtocounter{univ_counter} {1} 
\edef\UTFSM{$^{\arabic{univ_counter}}$ }

\addtocounter{univ_counter} {1}
\edef\JLAB{$^{\arabic{univ_counter}}$ }

\addtocounter{univ_counter} {1} 
\edef\VIRGINIA{$^{\arabic{univ_counter}}$ }

\addtocounter{univ_counter} {1} 
\edef\WM{$^{\arabic{univ_counter}}$ }

\addtocounter{univ_counter} {1} 
\edef\YEREVAN{$^{\arabic{univ_counter}}$ }

\addtocounter{univ_counter} {1} 
\edef\YORK{$^{\arabic{univ_counter}}$ }

\begin{document}

\preprint{Phys. Rev. C}

\title{Inclusive Electron Scattering in the Resonance Region off a Hydrogen Target with CLAS12}

\newcommand{\orcidauthorA}{0000-0002-1280-0983} 
\newcommand{\orcidauthorB}{0000-0001-9833-3695} 
\newcommand{\orcidauthorC}{0000-0002-4557-1320} 

\author{
V.~Klimenko,\UCONN$\!\!^,$\ANL\
D.S.~Carman,\JLAB\
R.W.~Gothe,\SCAROLINA\
K.~Joo,\UCONN\
N.~Markov,\UCONN$\!\!^,$\JLAB\
V.I.~Mokeev,\JLAB\
G.~Niculescu,\JMU\
P.~Achenbach,\JLAB\
J.S.~Alvarado,\ORSAY\
W.~Armstrong,\ANL
% ...
\\
(CLAS Collaboration)}

\affiliation{\ANL Argonne National Laboratory, Argonne, Illinois 60439}
\affiliation{\BRESCIA Universit\`{a} degli Studi di Brescia, 25123 Brescia, Italy}
\affiliation{\UCR University of California Riverside, 900 University Avenue, Riverside, California 92521}
\affiliation{\CSUDH California State University, Dominguez Hills, Carson, California 90747}
\affiliation{\CANISIUS Canisius College, Buffalo, New York 14208}
\affiliation{\CNU Christopher Newport University, Newport News, Virginia 23606}
\affiliation{\UCONN University of Connecticut, Storrs, Connecticut 06269}
\affiliation{\DUKE Duke University, Durham, North Carolina 27708}
\affiliation{\DUQUESNE Duquesne University, 600 Forbes Avenue, Pittsburgh, Pennsylvania 15282}
\affiliation{\FU Fairfield University, Fairfield, Connecticut 06824}
\affiliation{\FERRARAU Universit\`{a} di Ferrara, 44121 Ferrara, Italy}
\affiliation{\FIU Florida International University, Miami, Florida 33199}
\affiliation{\GSIFFN GSI Helmholtzzentrum fur Schwerionenforschung GmbH, D-64291 Darmstadt, Germany}
\affiliation{\GWUI The George Washington University, Washington, D.C. 20052}
\affiliation{\GLASGOW University of Glasgow, Glasgow G12 8QQ, United Kingdom}
\affiliation{\INFNFE INFN, Sezione di Ferrara, 44100 Ferrara, Italy}
\affiliation{\INFNFR INFN, Laboratori Nazionali di Frascati, 00044 Frascati, Italy}
\affiliation{\INFNGE INFN, Sezione di Genova, 16146 Genova, Italy}
\affiliation{\INFNPAV INFN, Sezione di Pavia, 27100 Pavia, Italy}
\affiliation{\INFNRO INFN, Sezione di Roma Tor Vergata, 00133 Rome, Italy}
\affiliation{\INFNTUR INFN, Sezione di Torino, 10125 Torino, Italy}
\affiliation{\JLUGiessen II Physikalisches Institut der Universitaet Giessen, 35392 Giessen, Germany}
\affiliation{\JMU James Madison University, Harrisonburg, Virginia 22807}
\affiliation{\KSU King Saud University, Riyadh, 11362 Kingdom of Saudi Arabia}
\affiliation{\KNU Kyungpook National University, Daegu 702-701, Republic of Korea}
\affiliation{\MIT Massachusetts Institute of Technology, Cambridge, Massachusetts 02139}
\affiliation{\MISS Mississippi State University, Mississippi State, Mississippi 39762}
\affiliation{\UNH University of New Hampshire, Durham, New Hampshire 03824}
\affiliation{\NMSU New Mexico State University, Las Cruces, New Mexico 88003}
\affiliation{\NSU Norfolk State University, Norfolk, Virginia 23504}
\affiliation{\OHIOU Ohio University, Athens, Ohio 45701}
\affiliation{\ODU Old Dominion University, Norfolk, Virginia 23529}
\affiliation{\ORSAY Universit\'{e} Paris-Saclay, CNRS/IN2P3, IJCLab, 91405 Orsay, France}
\affiliation{\URICH University of Richmond, Richmond, Virginia 23173}
\affiliation{\ROMAII Universit\`{a} di Roma Tor Vergata, 00133 Rome, Italy}
\affiliation{\SACLAY IRFU, CEA, Universit\'{e} Paris-Saclay, F-91191 Gif-sur-Yvette, France}
\affiliation{\MSU Skobeltsyn Nuclear Physics Institute and Physics Department at Lomonosov Moscow State University, 119899 Moscow, Russia}
\affiliation{\SCAROLINA University of South Carolina, Columbia, South Carolina 29208}
\affiliation{\TEMPLE Temple University, Philadelphia, Pennsylvania 19122}
\affiliation{\UTFSM Universidad T\'{e}cnica Federico Santa Mar\'{i}a, Casilla 110-V Valpara\'{i}so, Chile}
\affiliation{\JLAB Thomas Jefferson National Accelerator Facility, Newport News, Virginia 23606}
\affiliation{\VIRGINIA University of Virginia, Charlottesville, Virginia 22901}
\affiliation{\WM College of William and Mary, Williamsburg, Virginia 23187}
\affiliation{\YEREVAN Yerevan Physics Institute, 375036 Yerevan, Armenia}
\affiliation{\YORK University of York, York YO10 5DD, United Kingdom}

\date{\today}

\begin{abstract}
  Inclusive electron scattering cross sections off a hydrogen target at a beam energy of 10.6~GeV have been measured with data 
  collected from the CLAS12 spectrometer at Jefferson Laboratory. These first absolute cross sections from CLAS12 cover a wide 
  kinematic area in invariant mass $W$ of the final state hadrons from the pion threshold up to 2.5~GeV for each bin in virtual 
  photon four-momentum transfer squared $Q^2$ from 2.55 to 10.4~GeV$^2$ owing to the large scattering angle acceptance of the CLAS12
  detector. Comparison of the cross sections with the resonant contributions computed from the CLAS results on the nucleon resonance
  electroexcitation amplitudes has demonstrated a promising opportunity to extend the information on their $Q^2$ evolution up to
  10~GeV$^2$. Together these results from CLAS and CLAS12 offer good prospects for probing the nucleon parton distributions at large
  fractional parton momenta $x$ for $W < 2.5$~GeV, while covering the range of distances where the transition from the strongly coupled
  to the perturbative regimes is expected.
\end{abstract}

\maketitle

\section{Introduction}
\label{sec:intro}

Studies of inclusive electron scattering off protons represent an effective tool for the exploration of the structure of the proton 
ground state in terms of parton distribution functions (PDFs). The global Quantum Chromodynamics (QCD)-driven analyses of the 
experimental results on inclusive electron scattering off nucleons, $pp$ and $pd$ Drell-Yan cross sections, lepton and $W$ boson charge
asymmetries, and jet and $\gamma$+jet production - with the analyses dominated by the $p(e,e')X$ inclusive data - have provided detailed
information on the quark and gluon PDFs in a wide range of fractional parton momenta $x$ from $10^{-4}$ to above 0.9 and at photon 
virtualities $Q^2$ from $\approx$1 -- 10$^4$~GeV$^2$~\cite{accardi,alekh,ball,revpdf}. At large $x$, corresponding to the nucleon
resonance excitation region of invariant mass $W \lesssim 2$~GeV, the PDFs have been less explored compared with those at smaller 
$x$. The extraction of PDFs from inclusive electron scattering data at large $x$ is faced by constraints imposed by the applicability 
of the factorization of the perturbative and non-perturbative processes. Isolation of the factorizable contributions requires, 
beyond accounting for higher-twist effects and target-mass corrections~\cite{accardi,alekh,tmc1,tmc2}, the evaluation of the 
contributions from nucleon resonance electroexcitations that are clearly seen as peak structures in the $W$-dependence of the inclusive
electron scattering cross sections~\cite{bodek}.

Dedicated studies of inclusive electron scattering cross sections off nucleons in the resonance region have been carried out at the
Thomas Jefferson National Accelerator Facility (Jefferson Laboratory - JLab)~\cite{osipenko,malace09,tvaskis,liang}. Measurements of 
inclusive electron scattering with the CLAS detector in Hall~B~\cite{mecking} have been provided for $x$ up to 0.9 (or $W$ from the 
pion threshold up to 2.5~GeV) and for $Q^2$ from 0.25 -- 4.5~GeV$^2$~\cite{osipenko}. Because of the large electron scattering angle
acceptance of CLAS, the data span the entire kinematically allowed range of $W$ in each given bin of $Q^2$ with a bin width 
$\Delta Q^2$ = 0.05~GeV$^2$. The inclusive electron scattering experiments in Hall~C at JLab~\cite{malace09} extended the $Q^2$ coverage 
of the data from 3.6 -- 7.5~GeV$^2$ with almost the same range over $x$ (or $W$) as for the CLAS experiments, while also providing
information on the longitudinal and transverse components of the inclusive cross section~\cite{tvaskis,liang}. Because of the small
acceptance of the detector in Hall~C ($\approx$35~mrad for the scattered electron), these $p(e,e')X$ cross sections are available 
within highly correlated $(W,Q^2)$ values. Therefore, the data from CLAS and Hall~C offer complementary information on inclusive 
electron scattering off protons. The available published data in the deep inelastic scattering regime, $W > 2$~GeV, were taken at 
SLAC in the 1980s and span $Q^2$ up to 9.5~GeV$^2$. See Ref.~\cite{christy10} for an overview.

In the analyses of inclusive electron scattering to date, the contribution of excited nucleon ($N^*$) states has been treated
\cite{meln05,malace09} within the framework of quark-hadron duality~\cite{bh1,bh2}. Studies of exclusive $\pi N$, $\eta N$, and 
$\pi^+\pi^-p$ electroproduction off protons with CLAS have provided information on the electroexcitation amplitudes (also known as
the $\gamma_vpN^*$ electrocouplings) of most $N^*$ states in the mass range up to 1.8~GeV for $Q^2 < 5$~GeV$^2$
\cite{Bu12,Mokeev:2018zxt,Mo19,Car20,mokeev23,Car24}. Consistent results for the $\gamma_vpN^*$ electrocouplings of the $N^*$ states have 
been achieved within independent analyses of the major $\pi N$ and $\pi^+\pi^-p$ electroproduction channels, demonstrating their 
reliable extraction~\cite{mokeev23,Car24,Mo12,Mo16,Car23}. This allows for the evaluation of the resonant contributions to the inclusive
electron scattering cross sections~\cite{blin19,blin21} and spin structure functions~\cite{blin23} from the experimental results on the
full set of the $\gamma_vpN^*$ electrocouplings and their total hadronic decay widths. This paves the way for the first time to gain
insight into the ground state nucleon PDFs in the resonance region with $N^*$ parameters taken from the studies of exclusive meson 
photo-, electro-, and hadroproduction data.

The data with the new CLAS12 detector in Hall~B~\cite{clas12nim} allow for a significant extension of the studies of inclusive electron
scattering in the resonance region. In this paper data are presented from the first measurements of inclusive electron scattering off
protons using CLAS12 for $W < 2.5$~GeV and $Q^2$ from 2.55--10.4~GeV$^2$. Owing to the large acceptance of CLAS12, these data further 
the understanding of the ground state nucleon PDFs as a function of $x$ (or $W$) over the entire resonance region for $W$ from the pion
threshold up to 2.5~GeV in each bin of $Q^2$ for the first time at $Q^2 > 4$~GeV$^2$. The advances in the developments of the novel
pseudo- and quasi-PDF concepts~\cite{Qiu2019,Radyushkin2019} in connection with lattice QCD~\cite{Lin21,Ale23} and advances in continuum 
QCD approaches~\cite{Cr22,Cr23}, allow for the computation of the ground state nucleon PDFs in a wide range of $x$ from the QCD 
Lagrangian. In order to confront these theory expectations with experimental data in the resonance region it is important to have 
a) a reliable evaluation of the resonant contributions from exclusive meson electroproduction data within the framework described in
Refs.~\cite{blin19,blin21,blin23} and b) the measured observables for inclusive electron scattering $p(e,e')X$ with broad coverage over
$W$. The studies at large $x$ in the resonance region are of particular importance for the exploration of strong QCD, which refers to 
the study of phenomena in the regime where the QCD running coupling is large ({\it i.e.} comparable with unity), which is responsible 
for the generation of the ground and excited states of the nucleon from quarks and gluons.

Future analyses of our results at $Q^2 > 4$~GeV$^2$ within theoretical approaches with a connection to the QCD Lagrangian
\cite{Radyushkin2019,Qiu2019,Lin21,Ale23,Cr22,Cr23,croberts1,croberts2} will allow the exploration of the evolution of the ground 
state nucleon PDFs in the resonance region for the first time at distances where the transition from the strongly coupled to the 
perturbative QCD regimes is expected. Furthermore, the data from CLAS12 will provide new information relevant for studies of 
quark-hadron duality~\cite{malace09,meln05} and will shed light on the prospects for the exploration of $N^*$ structure in exclusive
electroproduction experiments with CLAS12~\cite{Bu18,Bu20,Mo19,Car23}. 

The organization of the remainder of this paper is as follows. Section~\ref{sec:detector} provides details on the CLAS12 spectrometer
employed for these measurements. Section~\ref{sec:analysis} provides an overview on the analysis cuts employed to isolate the electron
scattering events of interest from the data and Section~\ref{sec:monte} discusses the Monte Carlo and event generator.
Section~\ref{sec:cs-extraction} discusses the extraction of the inclusive cross sections and all data correction procedures.
Section~\ref{sec:iterations} reviews the iterative procedure to improve the event generator model by matching to the data. 
Section~\ref{statsyserr} describes the statistical uncertainties and details the different sources of bin-by-bin and scale-type 
systematic uncertainties in this measurement. Section~\ref{sec:results} presents the final cross section results and discusses their 
role to offer insight into the evolution of the partonic structure of ground state nucleons at distances where the transition from the
strongly coupled to perturbative QCD regimes is expected. Finally, a summary of this work and our conclusions are presented in
Section~\ref{sec:summary}.

\section{CLAS12 Detector}
\label{sec:detector}

\begin{figure}[!htbp]
\includegraphics[width=1.0\columnwidth]{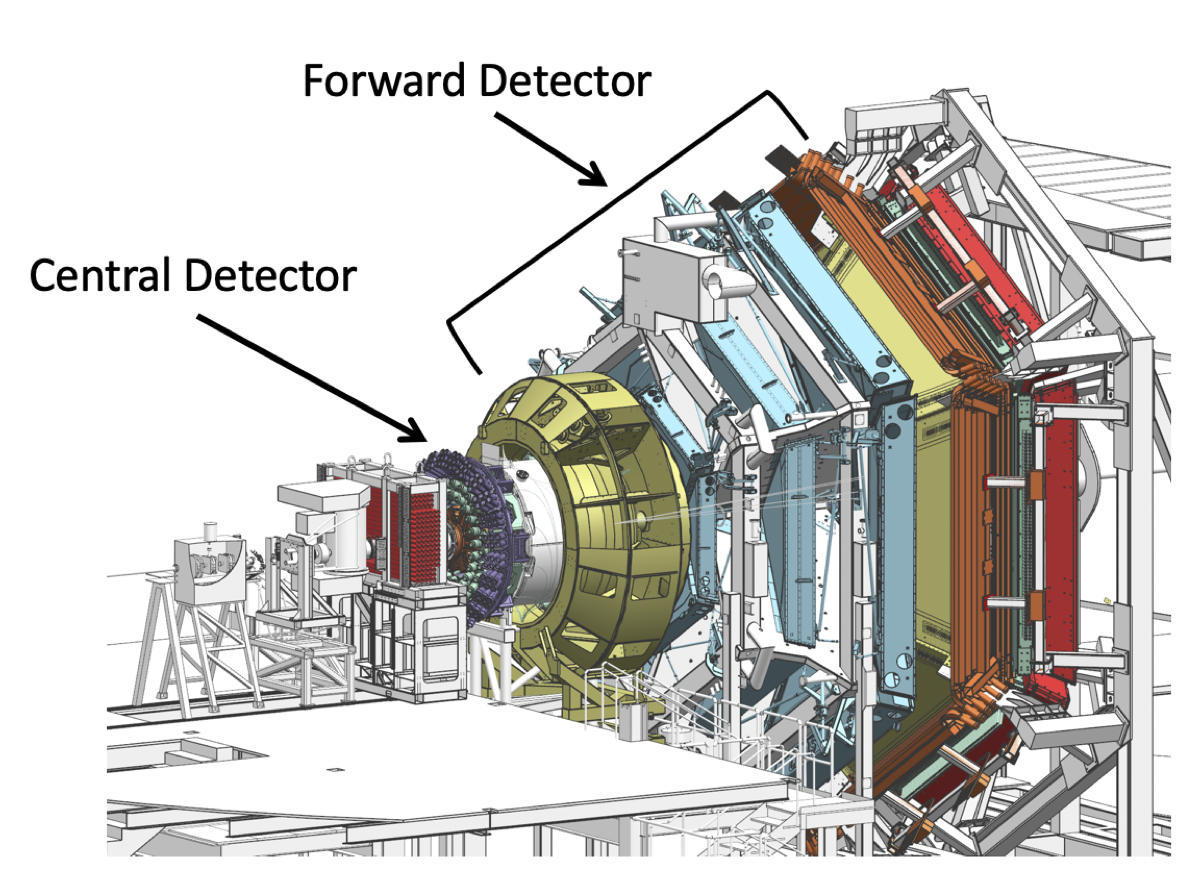}
\caption{Model of the CLAS12 detector highlighting the Forward Detector used for detection of the scattered electron and the Central
Detector about the target. The beam is incident from the left. The CLAS12 detector extends for 13~m along the beamline. See
Ref.~\cite{clas12nim} for details.} 
\label{clas12}
\end{figure}

The CEBAF Large Acceptance Spectrometer for operation at 12~GeV beam energy (CLAS12)~\cite{clas12nim} in Hall~B at JLab is used to study 
electro-induced nuclear and hadronic reactions. CLAS12 was developed as part of the energy-doubling project of JLab's Continuous Electron
Beam Accelerator Facility (CEBAF). This spectrometer provides efficient detection of charged and neutral particles over a large fraction
of the full solid angle. It is based on a dual-magnet system with a superconducting torus magnet in the Forward Detector region spanning
polar angles from 5$^\circ$ to 35$^\circ$ that provides a largely azimuthal field distribution, and a superconducting solenoid magnet in
the Central Detector region covering polar angles from $35^\circ$ to $125^\circ$ with full azimuthal coverage. See Fig.~\ref{clas12} for 
a model representation of CLAS12.

In the forward direction, the trajectories of charged particles are reconstructed using a multi-layer drift chamber (DC) system
consisting of 3 sets of detectors called Region~1 (R1), Region~2 (R2), and Region~3 (R3)~\cite{dc-nim}. Electron identification requires 
a matching of this track with a signal in the High Threshold Cherenkov counter (HTCC)~\cite{htcc-nim} close to the target and a cluster 
reconstructed in the forward electromagnetic sampling calorimeters (ECAL)~\cite{ecal-nim}. Charged particle timing is provided by the 
Forward Time-of-Flight (FTOF) system~\cite{ftof-nim}. The momentum resolution of the scattered electron is presently at the level of 
$\Delta p/p\!\approx$0.5-1\%. The data acquisition system~\cite{daq-nim} and detector design allow for operations at a beam-target 
luminosity of $1 \times 10^{35}$~cm$^{-2}$s$^{-1}$. The data included in this work were acquired in fall 2018 with a beam energy of 
10.604~GeV and a 5-cm-long liquid-hydrogen target. This data collection period was part of CLAS12 Run Group A (RG-A). Typical event 
rates were 15~kHz and 500~MB/s with a data acquisition livetime above 90\%.

The RG-A data acquired in fall 2018 were taken with two detector settings. The first part had an inbending torus field (that bent 
negatively charged particles toward the beamline) at maximum field strength ($\int B d\ell$ at nominal full field ranges from 2.8~Tm 
at 5$^\circ$ to 0.54~Tm at 40$^\circ$) and the second part had an outbending torus field (that bent negatively charged particles away 
from the beamline) at maximum field strength. The beam current varied between 45~nA and 55~nA. The cross section data presented in this
work used only the inbending torus polarity runs in order to provide the maximal coverage over $Q^2$.

Several hardware triggers were defined for the collection of the RG-A fall 2018 inbending data. The specific trigger configuration
selected for this data analysis was defined as

\begin{eqnarray}
{\rm DC_{roads}} ~\cdot~ {\rm HTCC_{nphe}} > 2 ~\cdot~ {\rm ECAL > 300~MeV} ~\cdot~ \nonumber \\ [0.2ex]
{\rm PCAL > 60~MeV} ~\cdot~ {\rm (ECin+ECout) > 10~MeV}.
\end{eqnarray}

\noindent
This electron trigger required pre-defined DC trajectories (called roads) that matched to a cluster in the HTCC with a threshold of 2
photoelectrons, a minimum deposited energy in the ECAL of 300~MeV, and a minimum energy deposited in specific ECAL sub-layers (called
PCAL, ECin, ECout) in the same CLAS12 sector~\cite{trigger-nim}.

\begin{figure}[!htbp]
\centering
\includegraphics[width=0.85\columnwidth]{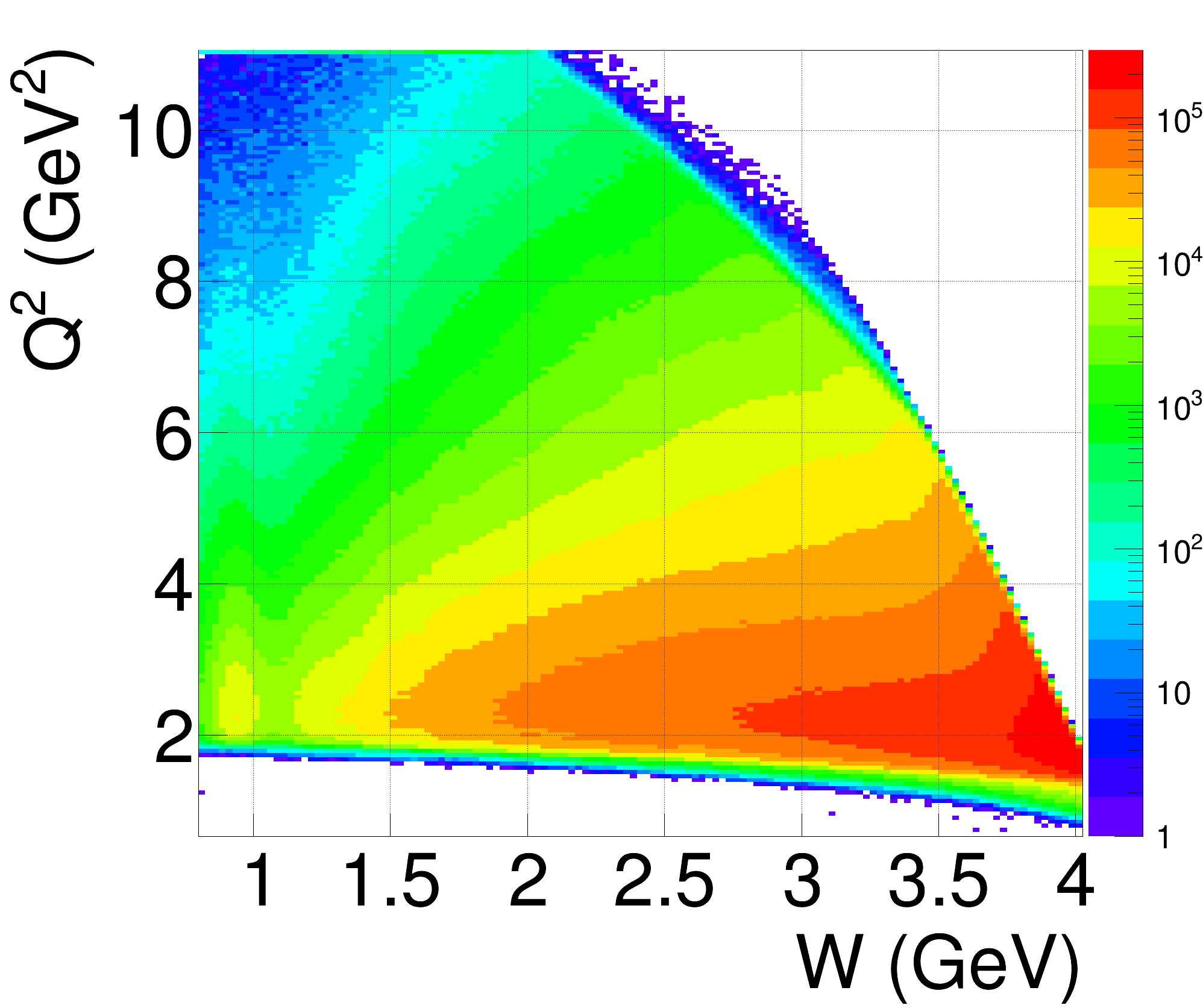}
\caption{Data from CLAS12 with a 10.6~GeV electron beam incident on a hydrogen target showing the coverage in $Q^2$ vs.~$W$. For these 
data the polarity of the torus magnet was set to bend negatively charged particles toward the electron beamline.}
\label{kin}
\end{figure}

The large forward acceptance of CLAS12 enabled detection of the final-state electron in the inclusive scattering reaction $p(e,e')X$ 
over a range of $W$ up to 4~GeV and $Q^2$ from 0.5 to 12~GeV$^2$. Figure~\ref{kin} shows the full kinematic phase space of the scattered
electron for these data. With the six coils of the torus magnet, the forward acceptance is divided into six active sectors. Due to the 
width of the torus coils, the azimuthal acceptance varies with polar angle $\theta$ from 50\% at 5$^\circ$ to 90\% at 35$^\circ$, which 
is the maximum polar angle for the final state particles in the laboratory frame covered by the Forward Detector. The design of the 
system allows for six independent measurements that are ultimately combined to present the final measured cross sections. However, 
each separate measurement allows for cross checks of the assigned systematic uncertainties. 

\section{Analysis Cut Overview}
\label{sec:analysis}

A series of criteria, or cuts, was applied to the multiple detector responses to identify negatively charged tracks that are electron
candidates. The cuts were designed to discriminate against minimum-ionizing particles (MIPs), such as negative pion ($\pi^-$) tracks. 
The CLAS12 Event Builder (EB)~\cite{eb-nim} protocol first assigns electron identification to tracks with responses in the HTCC and ECAL
that satisfy the criteria in Table~\ref{tab:eb_cuts}, with an associated geometrically matched hit in FTOF. After the EB identifies 
electron candidates, additional cuts were applied to the data to select a refined sample of candidate electrons prior to proceeding with 
the remainder of the analysis.

\begin{table}[htp]
\begin{center}
\begin{tabular}{|c|c|} \hline
 Cut              &  Limits \\ [0.5ex] \hline
 Electric Charge                 & negative \\ \hline
 Number of Photoelectrons        & $N_{phe}>2$ \\ \hline
 Min. PCAL Energy                & PCAL$_{dep} >$ 60 MeV \\ \hline
 Sampling Fraction vs.~$E_{dep}$  & $\pm5\sigma$ \\ \hline
\end{tabular}
\caption{EB electron assignment requirements. The ECAL sampling fraction used by the EB is parameterized as a function of the total 
energy deposited in the ECAL. Note that the ECAL is subdivided into three module stacks, PCAL, ECin, and ECout~\cite{ecal-nim}.}
\label{tab:eb_cuts}
\end{center}
\end{table}

The electric charge of a candidate particle will dictate its curvature as it traverses through the toroidal magnetic field. The field
will deflect particles along the polar angle, either making their curvature ``inbending" or ``outbending" based on the field polarity.
Reconstruction algorithms take the track curvature into account to assign a charge. Since electrons need to be identified, only negatively
charged tracks were selected. Electrons in CLAS12 can only be identified in the Forward Detector. The HTCC aids in reducing negative pion
contamination in the electron sample for candidate tracks up to 4.9~GeV (the threshold for $\pi^-$ to begin generating an HTCC signal 
in the CO$_2$ radiator gas), based on reading out photomultiplier tubes (PMTs) to determine the number of photons emitted via
Cherenkov radiation by a charged track traversing the detector volume. Up to the $\pi^-$ momentum threshold it is sufficient to cut on 
the number of photoelectrons ($N_{phe}$) produced in the detector. An electron candidate track will typically produce more than 2
photoelectrons (the CLAS12 trigger threshold), which is the minimum threshold for this cut. This cut is automatically enforced when 
using the EB particle identification assignment to select electrons in the Forward Detector. Beyond the HTCC efficiency corrections
detailed in Section~\ref{htccMap}, no HTCC fiducial cuts were applied, since the HTCC provides only a crude and discrete hit 
distribution from its PMT information.

\begin{figure}
\centering
\includegraphics[width=0.9\columnwidth]{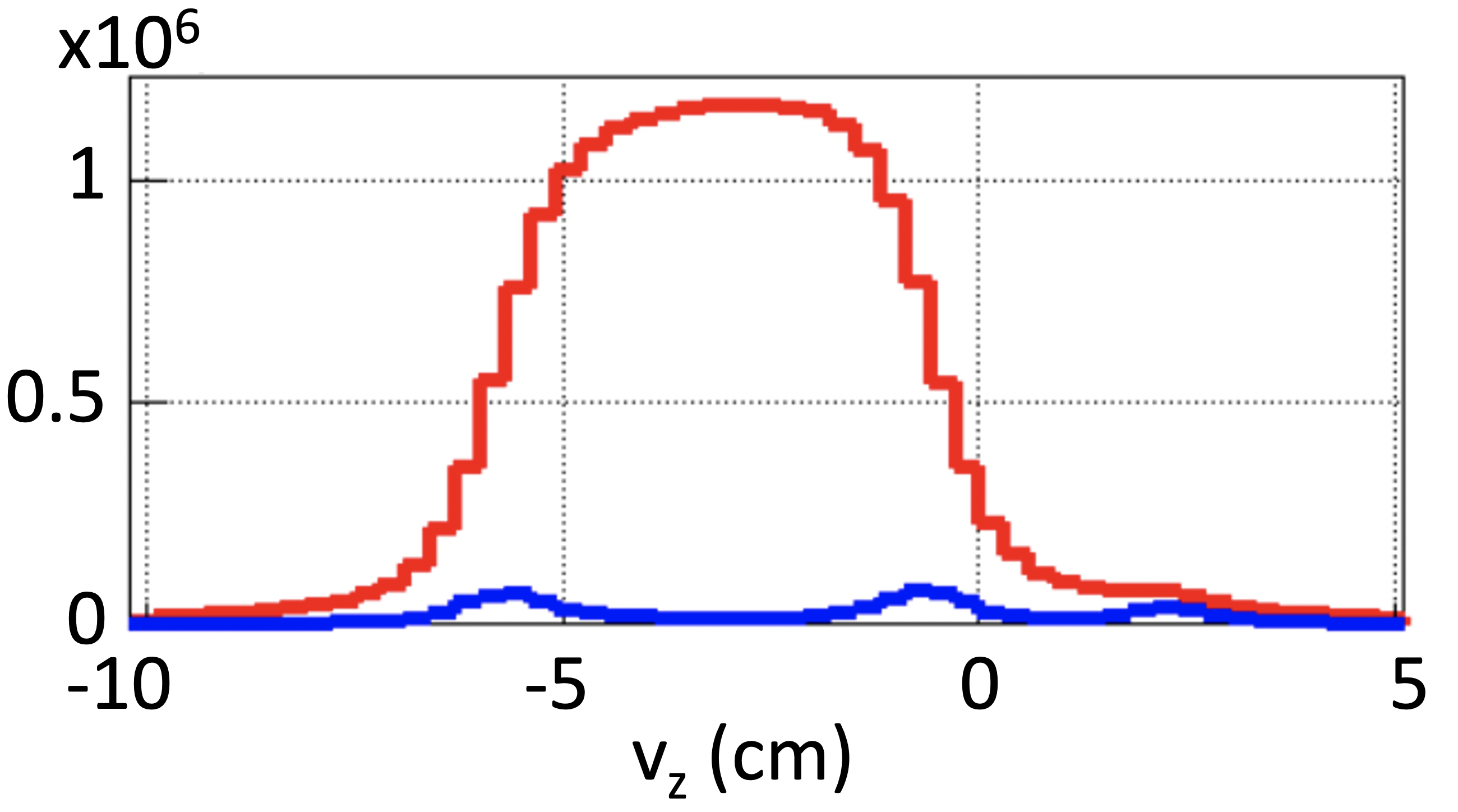}
\caption{Vertex coordinate distributions ($v_z$) for full target runs (red) and for empty target runs (blue) normalized by charge for 
a single CLAS12 sector.}
\label{fig:VzCutH}
\end{figure}

The target in the experiment was a 5-cm-long cell filled with liquid-hydrogen. Using tracking information from the drift chambers
projecting back to the $e+p$ interaction point, a cut on the reconstructed vertex coordinate along the beamline (labeled as the
$z$-axis of CLAS12) was applied to ensure the event originated from the target location. A cut on the vertex $z$-coordinate from
$(-8.0,2.0)$~cm was applied to account for the $\approx$1~cm drift chamber trace-back resolution. Figure~\ref{fig:VzCutH} illustrates 
the $v_z$ distributions for a single CLAS12 sector for the full (red) and empty (blue) target data.

Drift chamber fiducial boundary cuts were applied based on the reconstructed hit coordinates to select charged tracks that were 
detected in regions where the detector efficiency is high and uniform. The cuts were placed within a small distance from the detector 
borders to make sure that only tracks in the fiducial volume were selected for analysis. The cuts are sector independent and depend only 
on the layer ({\it i.e.} the R1, R2, and R3 drift chambers). A similar fiducial cut in the PCAL plane was also applied to remove events 
outside of the fiducial volume between the DC and PCAL. The cut was based on the PCAL scintillation strip coordinate information along 
the three sides of the triangular module geometries.

\begin{figure}
\centering
\includegraphics[width=0.95\columnwidth]{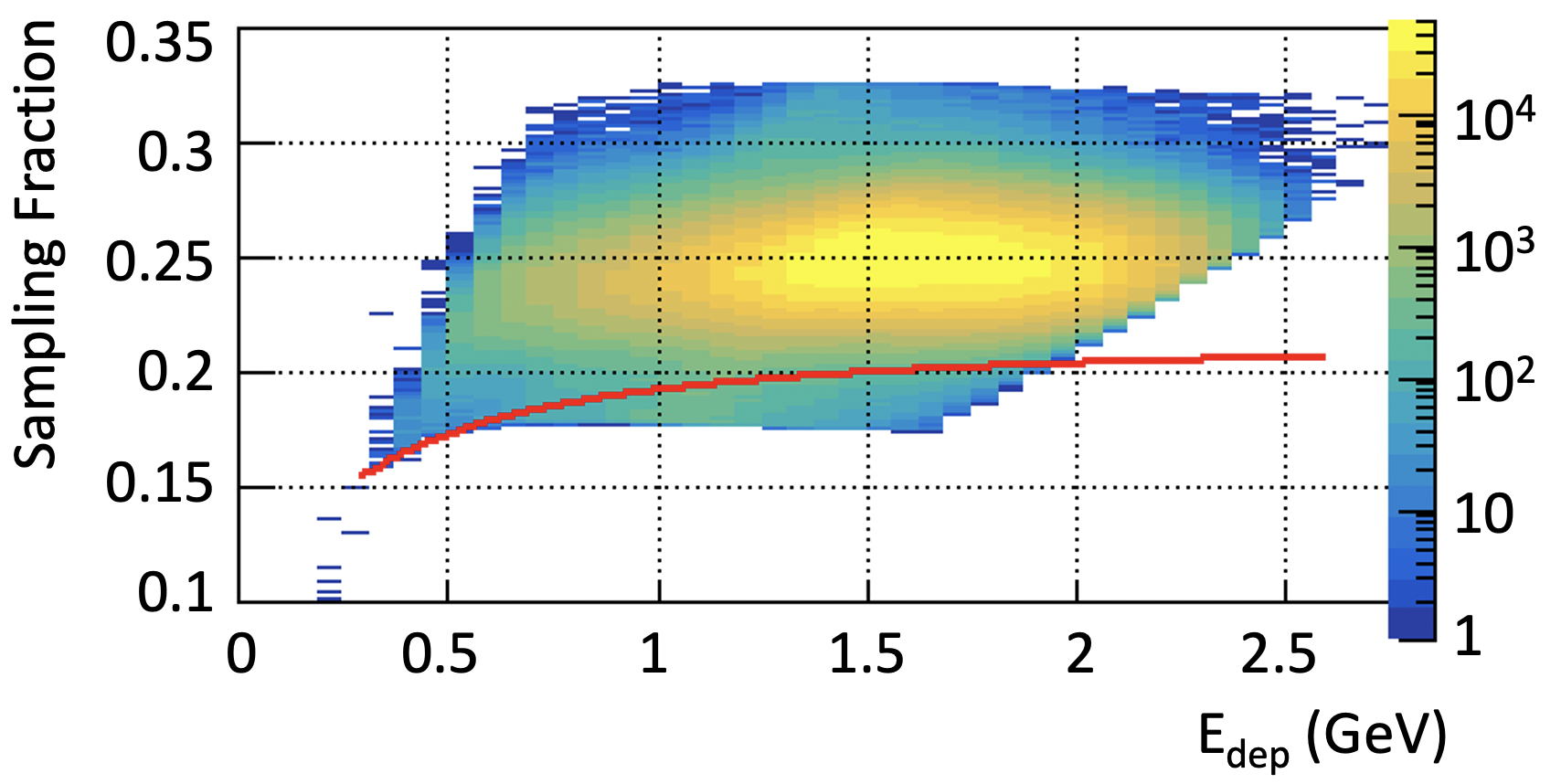}
\caption{ECAL sampling fraction as a function of deposited energy in all layers for a single sector. The red line is the $3.5 \sigma$ 
cut. The kinematic cuts from $0.8 < W < 2.7$~GeV and $2.18 < Q^2 < 10.6$~GeV$^2$ were applied.}
\label{fig:SFcut}
\end{figure}

An ECAL sampling fraction cut was applied to remove non-minimum-ionizing pion tracks. The sampling fraction is defined as the ratio of 
the total energy deposited in all layers of the ECAL to the reconstructed momentum, $P$, of the track determined from the drift chambers.
The sampling fraction signature for electrons is nearly constant at a value of $\sim$0.25 across all momenta. This implies the deposited
energy scales with the momentum or deposited energy of the electron. The electron tracks are selected with a $3.5\sigma$ cut below the
parameterized sampling fraction (no upper cut is necessary) as a function of the deposited energy $E_{dep}$ as seen in 
Fig.~\ref{fig:SFcut}. The cut, which was designed to be tighter than that employed by the EB, was developed by fitting the sampling
fraction for successive bins in deposited energy. A Gaussian function was fit to determine the mean $\mu_b$ and sigma $\sigma_b$ for 
each bin $b$. A polynomial was fit to the mean and sigma to create an energy-dependent sampling fraction cut for each CLAS12 sector. 
The parameters are the result of fits to the sampling fraction distribution in bins of deposited energy, separately, for each sector 
in data and a sector-independent fit in simulation.

\begin{figure}
\centering
\includegraphics[width=0.9\columnwidth]{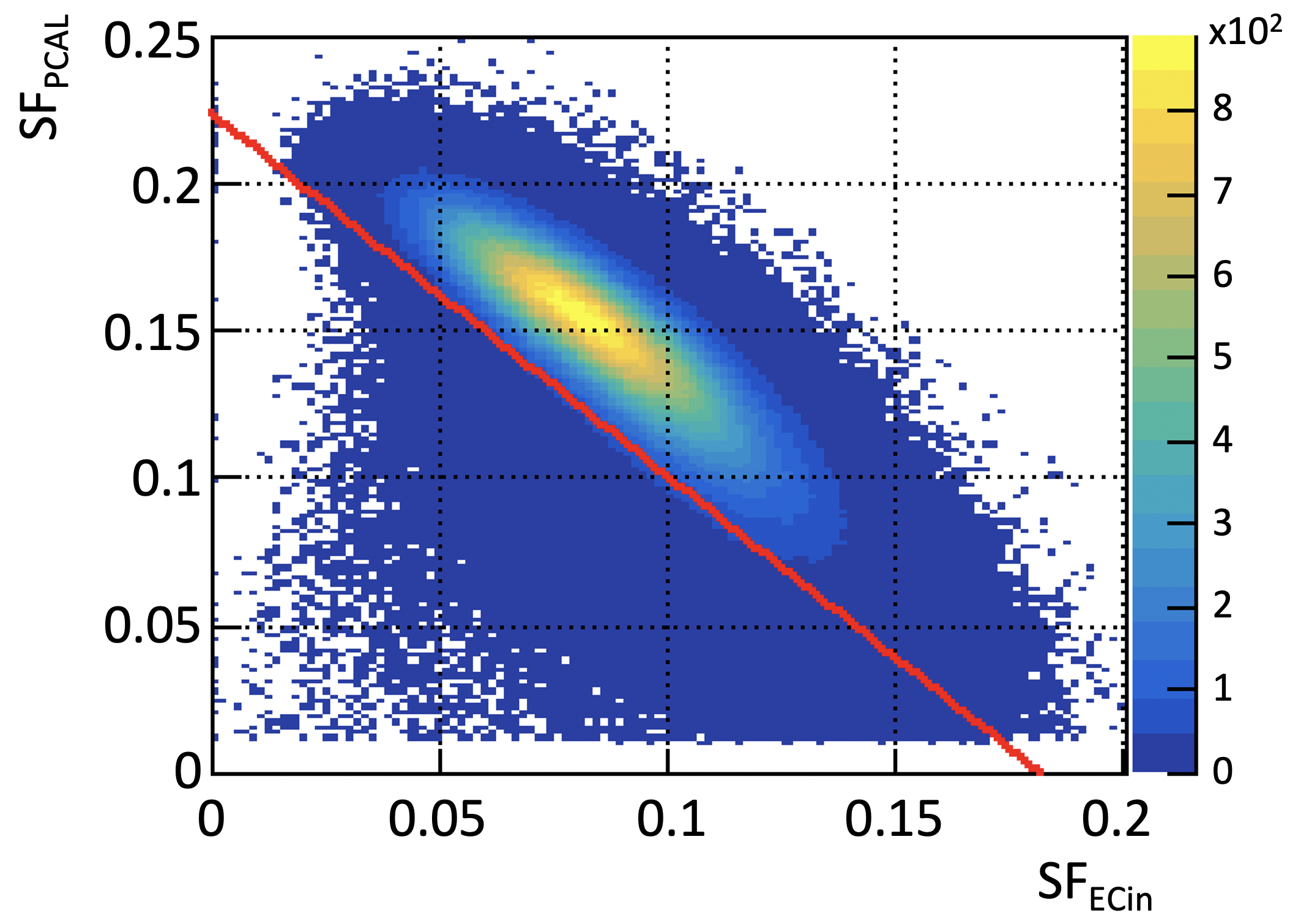}
\caption{Sampling fractions from PCAL vs.~ECin for a single sector for a representative momentum bin ($6 < P < 7$~GeV). The partial
sampling fraction cut employed is shown by the red line.}
\label{fig:trigCutPic_data_123}
\end{figure}

Sometimes pions can be identified as electrons, especially for higher momentum tracks. To further reduce the pion contamination, the
correlation of the sampling fractions calculated separately from the PCAL and ECin was studied. A linear cut in the sampling fraction 
plane was introduced based on simulation studies to reduce the leakage of pions into the electron sample (see
Fig.~\ref{fig:trigCutPic_data_123}). This cut is referred to as the partial sampling fraction cut.

Inaccuracies in the magnetic field map and unaccounted-for detector misalignments resulted in systematic misreconstruction of the 
momentum for charged particles. These effects in turn result in misreconstruction of $Q^2$ and $W$ for the events. A correction was
developed as a function of $P$, $\theta$, and $\phi$ for the electron detected in each of the six sectors of the CLAS12 Forward Detector
to minimize such deviations. These corrections were below the level of 0.5\%. A detailed discussion of the applied momentum correction
method can be found in Ref.~\cite{cn-2023-001}.

\section{Monte Carlo}
\label{sec:monte}

\subsection{GEMC - Geant4 Monte Carlo}

The CLAS12 detector simulation has been implemented within the GEMC software framework~\cite{simulation-nim}. GEMC is used to accurately
calculate the CLAS12 acceptance, including the detector response, geometrical acceptance, and charged particle tracking efficiency. GEMC
is a C++ framework that uses Geant4~\cite{geant4} to simulate the passage of particles through matter. Particles are transported through 
the detector materials and produce radiation, hits, and secondaries. GEMC then collects the Geant4 results and produces the digitized 
output in the same data format as that from the detector. All of the elements within the particle trajectory paths to any of the CLAS12 
detectors are included in the simulation. In addition, the simulation has been tuned in order to reproduce beam-related background
hits in the detectors to a good level of accuracy in order to properly model charged particle tracking efficiencies. The simulations
performed for this analysis employed the same version of the reconstruction code as was used for data processing. 

\subsection{Event Generator}

The event generator (EG) developed by Sargsyan used for the inclusive electron scattering Monte Carlo (MC) was based on the Bodek
parameterization of the world data~\cite{misakEG}. It describes both the elastic and inelastic parts of inclusive electron scattering and 
allows for these parts to be switched on or off for the purpose of event generation. Radiative effects are accounted for using the Mo 
and Tsai approach~\cite{moTsaiRC} for both the elastic and inelastic parts of the cross section. Note that the code also has the option
to turn these effects on or off, which was used to account for internal radiative corrections in the $p(e,e')X$ cross section extraction
from the measured data. The EG code was designed to account for external radiation in the target cell and target material before the 
$e+p$ interaction point. While the momentum loss associated with radiative effects for the incoming and scattered electrons was taken 
into account, the radiative photon itself was not generated, and only the scattered electron was available in the final state. The
external radiation after the $e+p$ interaction point was accounted for by GEMC.

The simulations were carried out accounting for all non-functioning detector elements that were seen in the data, including holes in the
drift chambers due to broken wires or bad readout boards and missing channels in the ECAL and FTOF. In this way the geometrical 
acceptance of the MC was matched closely to the data. The simulations used the same code as was used for analysis of the data. To 
ensure that the ultimate EG model was realistic and did not introduce significant bias in the measurements, the original EG was tuned 
to match the extracted cross sections as discussed in Section~\ref{sec:iterations}.

\subsection{Background Merging}
\label{bck-merge}

The simulations for this analysis employed a background merging option in order to best match the backgrounds in the various CLAS12
detector subsystems~\cite{cn-2020-005}. Of relevance for this analysis is the hit merging for the DC, FTOF, and ECAL systems. This 
background was developed based on CLAS12 data collected with a random trigger at the production beam current that was merged with the 
MC event-by-event. As the background was determined based on actual data from CLAS12 taken in the exact same conditions as the data used 
for analysis, it matches well the overall background distribution in the detector. This includes the local DC hit occupancy necessary to 
properly model the tracking efficiency across the full Forward Detector acceptance at the different beam currents used for production data 
collection in this experiment. Merging of background events with simulation was performed at the level of raw ADC and TDC hits.

\begin{figure}[htp]
\centering
\includegraphics[width=0.75\columnwidth]{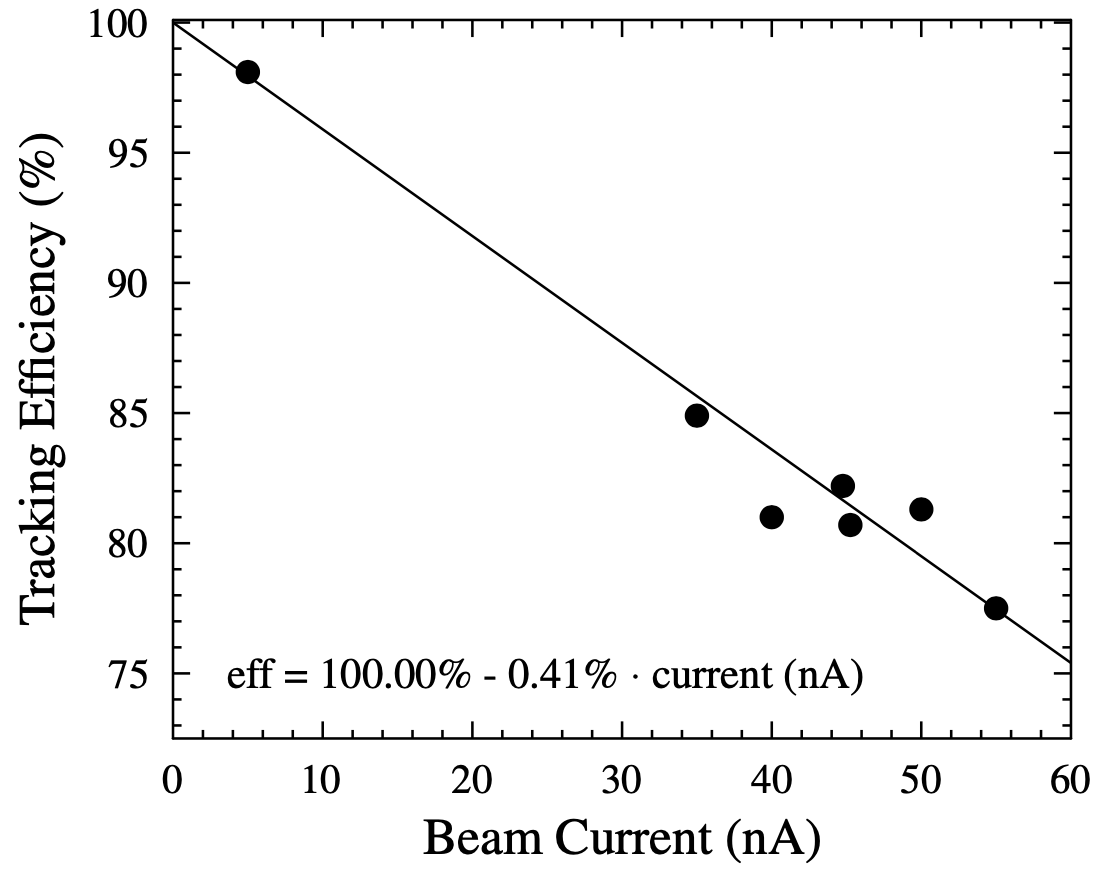}
\caption{Efficiency vs.~beam current for negatively charged particles in the CLAS12 Forward Detector from luminosity scan data (black
data points) using the reconstruction software version employed for this analysis. The black line shows a linear fit with a slope of
-0.41\%/nA.}
\label{eff-current}
\end{figure}

The background merging procedure is essential to match the charged track reconstruction efficiency from simulation to the efficiency
measured with data. The analysis of the tracking efficiency from data was completed using data taken from a luminosity scan analysis
as shown in Fig.~\ref{eff-current}. The inefficiency increases linearly with increasing beam current with a slope of -0.41\%/nA. This
inefficiency factor is accounted for in combining the data in this analysis taken at different beam currents from 45--55~nA.

\subsection{Monte Carlo Smearing}
\label{smearing}

Studies of the missing mass distributions from exclusive meson electroproduction show that the CLAS12 MC used for this analysis gives 
a better momentum resolution for charged tracks than is seen in the data by a factor of $1.5-2$. The GEMC reconstruction includes two 
primary deviations from the generated (true) momentum values

\begin{itemize}
\item $\Delta P^{res}$ - deviation based on momentum resolution, which is centered around zero,
\item $\Delta P^{loss}$ - deviation based on physical energy losses ({\it e.g.} radiative effects) or detector imperfections, which is
typically not centered around zero.
\end{itemize}
 
Our goal was to smear the GEMC momentum resolution ($\Delta P^{res}$) without amplifying or altering any legitimate physical effects
($\Delta P^{loss}$). To parameterize the GEMC resolution model as a function of the kinematic variables, a simulation sample of 
$ep \to e'\pi^+n$ events was generated with the AAO\_GEN EG~\cite{aao_gen}. The simulation was performed in the same $W$ and $Q^2$ range
used in our inclusive measurement. The simulation sample was split into a kinematic grid in polar angle $\theta$ and the distributions 
of $(P^{gen}-P^{rec})/P^{rec}$ were fit for the $e'$ and $\pi^+$ for each cell of the grid. The extracted widths were given by
$\sigma^{GEMC}(\theta_g)$, where $g$ is the grid cell identifier. Considering that only a narrow range of momentum was associated with 
each $\theta$ bin, the resolution did not show significant momentum dependence across a given cell. With this procedure the resolution 
for both the $e'$ and $\pi^+$ was obtained as a function of $\theta$. In order to compare the resolution in the data with that determined
from the MC, we calculated $\delta P/P$ for both particles from data and simulation, where $\delta P$ is the difference between the
calculated or generated momentum and the reconstructed momentum. The calculated momentum for the electron from data was obtained from 
the electron polar angle and the pion four-momentum and similarly for the pion. The electron momentum smearing for the inclusive 
analysis was based on using

\begin{equation}
\label{eq:smearingAPL}
	 P^{new} = P^{rec} + P^{rec} \cdot \sigma^{GEMC}(\theta) \cdot gaus(0,1) \cdot F(\theta),
\end{equation}

\noindent
where $P^{rec}$ and $P^{new}$ are the electron momentum before and after smearing, respectively, $\sigma^{GEMC}(\theta)$ is the 
resolution function, $F$ is the determined smearing factor, and $gaus(mean,sigma)$ represents a random number sampled from a Gaussian
distribution with the given mean and width.

\begin{figure}
\centering
\includegraphics[width=0.7\columnwidth]{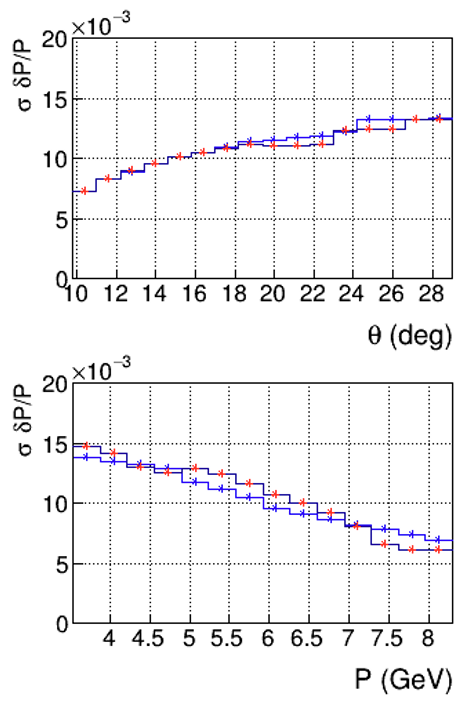}
\caption{$\delta P/P$ vs.~$\theta$ and $P$, where $\delta P/P$ represents $(P^{calc/gen}-P^{rec})/P^{rec}$, for electrons from data 
(red) and simulation (blue) with smearing.}
\label{fig:eres_pth}
\end{figure}

Starting with $F = 1.0$ (an arbitrary value), $F$ was increased separately for all kinematic bins until good agreement (matching 
$\delta P/P$ for data and simulation) was achieved. The comparisons with the final functional $F(\theta)$ with resolution
nearly twice worse than the original MC gave reasonable agreement (at the level of 10\%) of the reconstructed momenta between data 
and simulation as shown in Fig.~\ref{fig:eres_pth}.

\begin{figure}
\centering
\includegraphics[width=0.98\columnwidth]{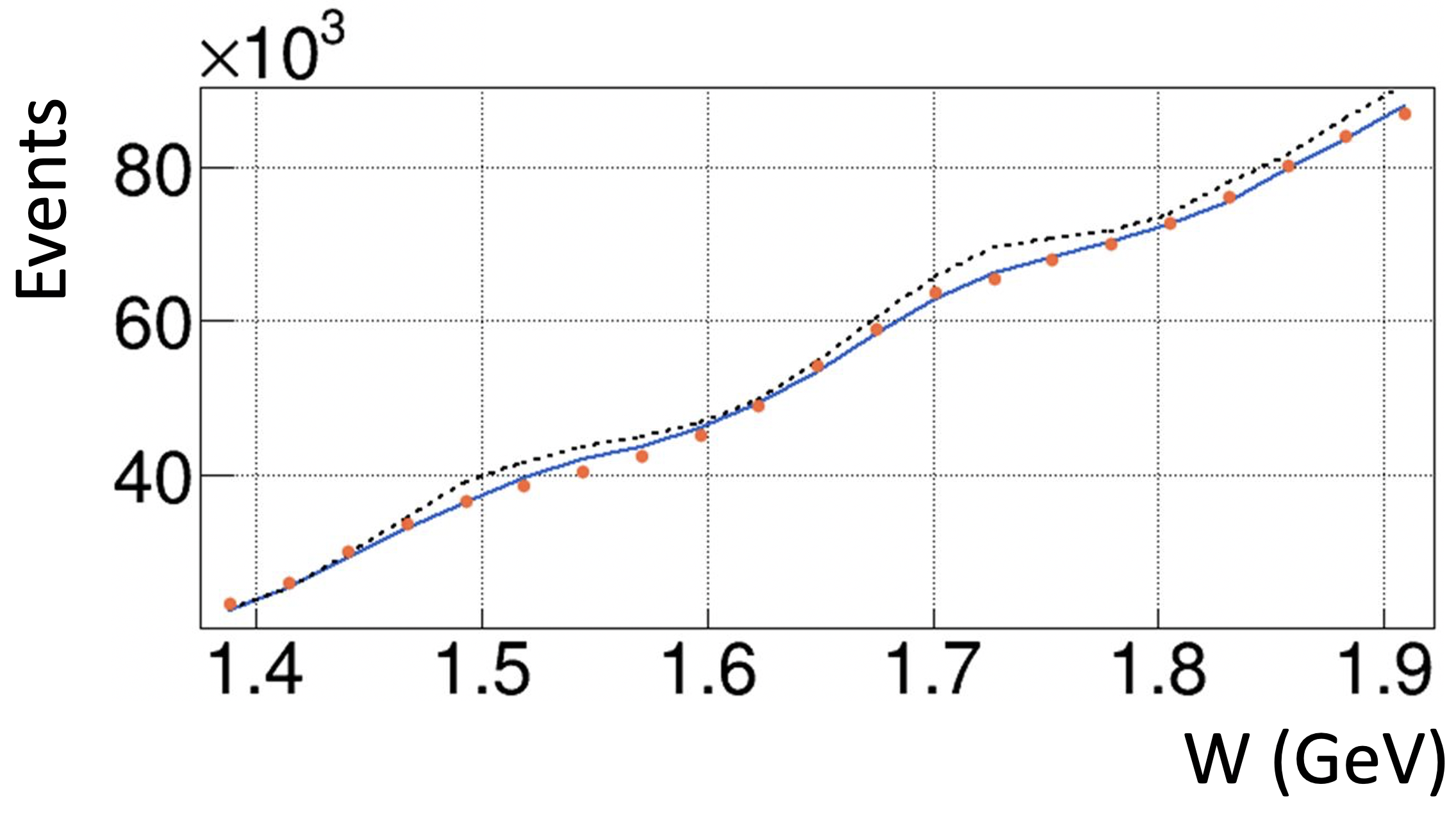}
\caption{$W$ spectra for a representative $Q^2$ bin ($4.08 < Q^2 < 4.78$~GeV$^2$) for data (orange points), MC without smearing (black
dashed line), and MC with the applied smearing function (blue line). The resolution smearing procedure matches the MC distributions to 
the data.}
\label{fig:smearingW}
\end{figure}

\begin{figure}
\centering
\includegraphics[width=0.98\columnwidth]{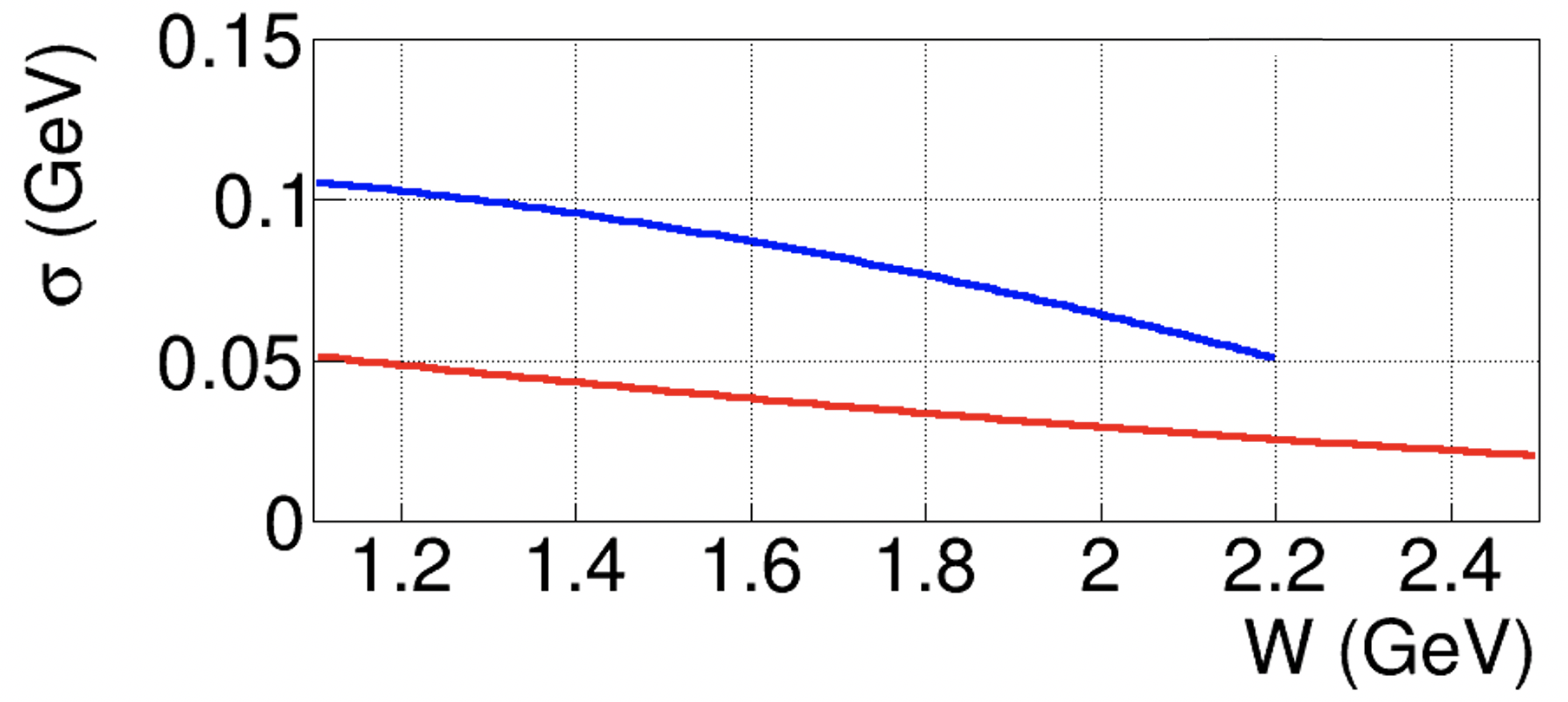}
\caption{Resolution vs.~$W$ from MC applying smearing for the reconstructed $p(e,e')X$ events within the bin of minimal $Q^2$ from 
2.55--2.99~GeV$^2$ (red) and maximal $Q^2$ from 8.94--10.4~GeV$^2$ (blue).}
\label{fig:resolutionW}
\end{figure}

For the inclusive scattering case, we applied the smearing factor ($F$) obtained from the exclusive reaction study. The MC momentum
smearing aligns the $W$ spectra quite well as shown in Fig.~\ref{fig:smearingW}. Finally, a realistic average resolution function was
estimated for the CLAS12 data as a function of $W$ using our smearing procedure. Figure~\ref{fig:resolutionW} shows the width $\sigma$ 
of $W_{gen} - W_{rec}$ as a function of $W$ for our lowest and highest $Q^2$ bin values.

\subsection{Comparison of MC to Data}

\begin{figure*}
\centering
\includegraphics[width=0.8\textwidth]{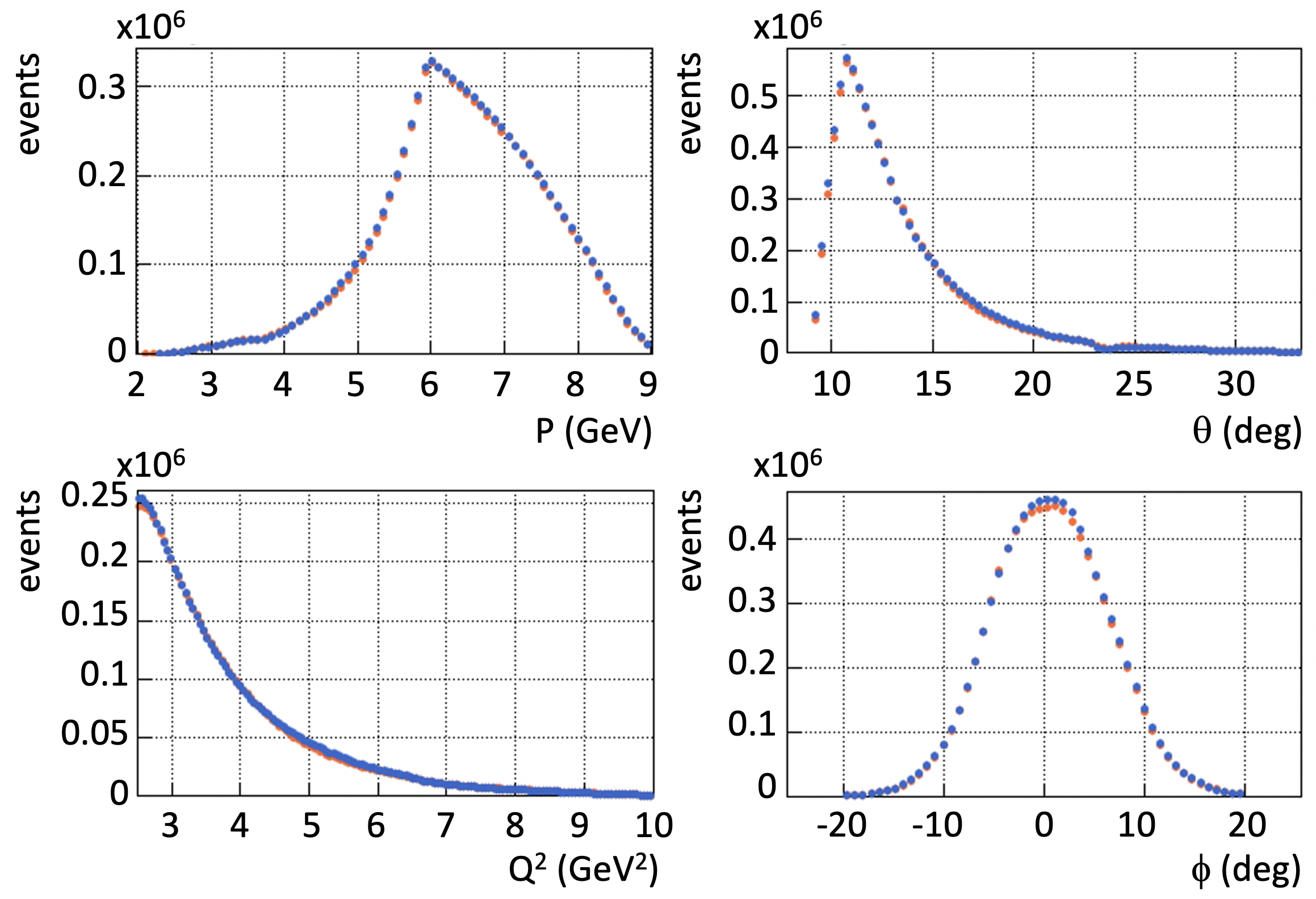}
\caption{Comparison of the measured (blue) to reconstructed MC event distributions (orange) for a single representative CLAS12 sector for
(UL) momentum, (UR) polar angle $\theta$, (LL) $Q^2$, and (LR) azimuthal angle $\phi$. The MC distributions were normalized to the data 
by the luminosity factor of Eq.(\ref{mc_factor}).}
\label{kin-mc-data}
\end{figure*}

For the purpose of a direct comparison of our MC and the data, both the elastic and inelastic parts were invoked in the MC, as well as 
full radiative effects in the EG. Here the final version of the EG was used that was obtained after the adjustment procedure described 
in Section~\ref{egAdj}. The simulation was normalized to the data luminosity by the factor
\begin{equation}
\label{mc_factor}
\frac{\int\!\!{\cal L}dt \cdot \sigma_{EG}}{N_{total}},
\end{equation}
\noindent
where
\begin{itemize}
\itemsep 0em
\item $\int\!\!{\cal L}dt$ is the integrated luminosity given by the product of the number of incident beam electrons and the number 
of target particles per area,
\item $\sigma_{EG}$ is the integrated cross section obtained from the EG for the generated kinematic region, and
\item $N_{total}$ is total number of generated events.
\end{itemize}
Figure~\ref{kin-mc-data} shows comparisons between the measured data and reconstructed simulation for a single representative CLAS12
sector for the momentum, polar angle, $Q^2$, and azimuthal angle distributions. Dips in the polar angle distributions correspond to 
knocked out elements in the ECAL or other CLAS12 Forward Detector subsystem elements. Overall the distributions show very close 
agreement between the data and the simulation model.

\section{Cross Section Extraction}
\label{sec:cs-extraction}

The following expression was used to compute the differential cross section in each kinematic bin $i$

\begin{equation}
\label{dcs}
\frac{d\sigma_i}{dQ^2 dW} =  \frac{1}{\left( \Delta Q^2 \Delta W \right)}
\cdot \frac{N_i}{\eta_i \cdot N_0 \cdot R_i \cdot BC_i} \cdot \frac{CMB}{\left(N_A \rho t / A_w \right)}. 
\end{equation}

The terms in this expression are defined as follows:

\begin{itemize}
\item $\Delta Q^2 \Delta W$ : kinematic bin volume
\item $N_i$ : number of reconstructed electrons in each bin that pass our cuts
\item $\eta_i$ : acceptance and efficiency correction
\item $N_0$ : number of incident beam electrons
\item $R_i$ : radiative correction
\item $BC_i$ : bin-centering correction
\item $N_A \rho t / A_w$ : target number density with $N_A$ Avogadro's number, $\rho$ the liquid-hydrogen density 
(0.07151~g/cm$^3$), $t$ the target length, and $A_w$ the atomic weight of hydrogen (1.00794~g/mol)
\item $CMB$ : cm$^2$ to $\mu$b conversion factor of $1 \times 10^{30}$.
\end{itemize}

\subsection{Electron Events}

\subsubsection{Empty Target Subtraction}

In the event distribution, there is a contribution that comes from the 30-$\mu$m-thick aluminum end-caps of the cryotarget cell. This
contribution is not related with inclusive electron scattering off hydrogen, so it must be removed. Our dataset contains special 
runs that were collected with an almost empty target ({\it i.e.} only residual cold hydrogen gas in the target cell). The events
obtained from these runs were used for the subtraction of the empty target contribution. We extracted events from the empty 
target runs with exactly the same electron requirements and cuts as were used for the full target runs. However, we put an additional 
cut on $v_z$, namely, we removed the middle part of the target cell from $-4.7 < v_z < -1.65$~cm in order to minimize the 
over-correction for the residual hydrogen gas in the target cell. The empty and full target runs have different Faraday cup charges 
(the full target runs had $\sim$12 times more charge), so the empty target events were properly normalized. The events after the
empty target contribution subtraction are given by

\begin{equation}
N_{hydrogen} = N_{full} - N_{empty} \cdot \frac{Q_{full}}{Q_{empty}},
\label{ETcorrectedYield}
\end{equation}

\noindent
where $N_{hydrogen}$, $N_{full}$, and $N_{empty}$ are the events for inclusive events from hydrogen, the full target cell, and
the empty target cell, respectively, and $Q_{full}$ and $Q_{empty}$ are the corresponding Faraday cup charges for the full target and 
empty target runs. Note that this approach does not take into account the small tracking efficiency difference between the full and 
empty target data.

\subsubsection{Cherenkov Counter Efficiency Correction}
\label{htccMap}

We looked at the HTCC photoelectron response in bins of $x$ and $y$, where $x$ and $y$ are the coordinates of intersection of the track
with the mirror in the lab frame. For every bin the photoelectron spectrum was fit with a Poisson distribution. The ratio of the Poisson
distribution integrals obtained for the number of photoelectrons in the range from $[2,50]$ to $[0,50]$ was used to determine the HTCC 
efficiency map for each $(x,y)$ bin that accounted for the efficiency loss due to the two photoelectron requirement in the trigger and 
the EB. The procedure was applied to both the data and MC.

\begin{figure}[htp]
\centering
\includegraphics[width=0.9\columnwidth]{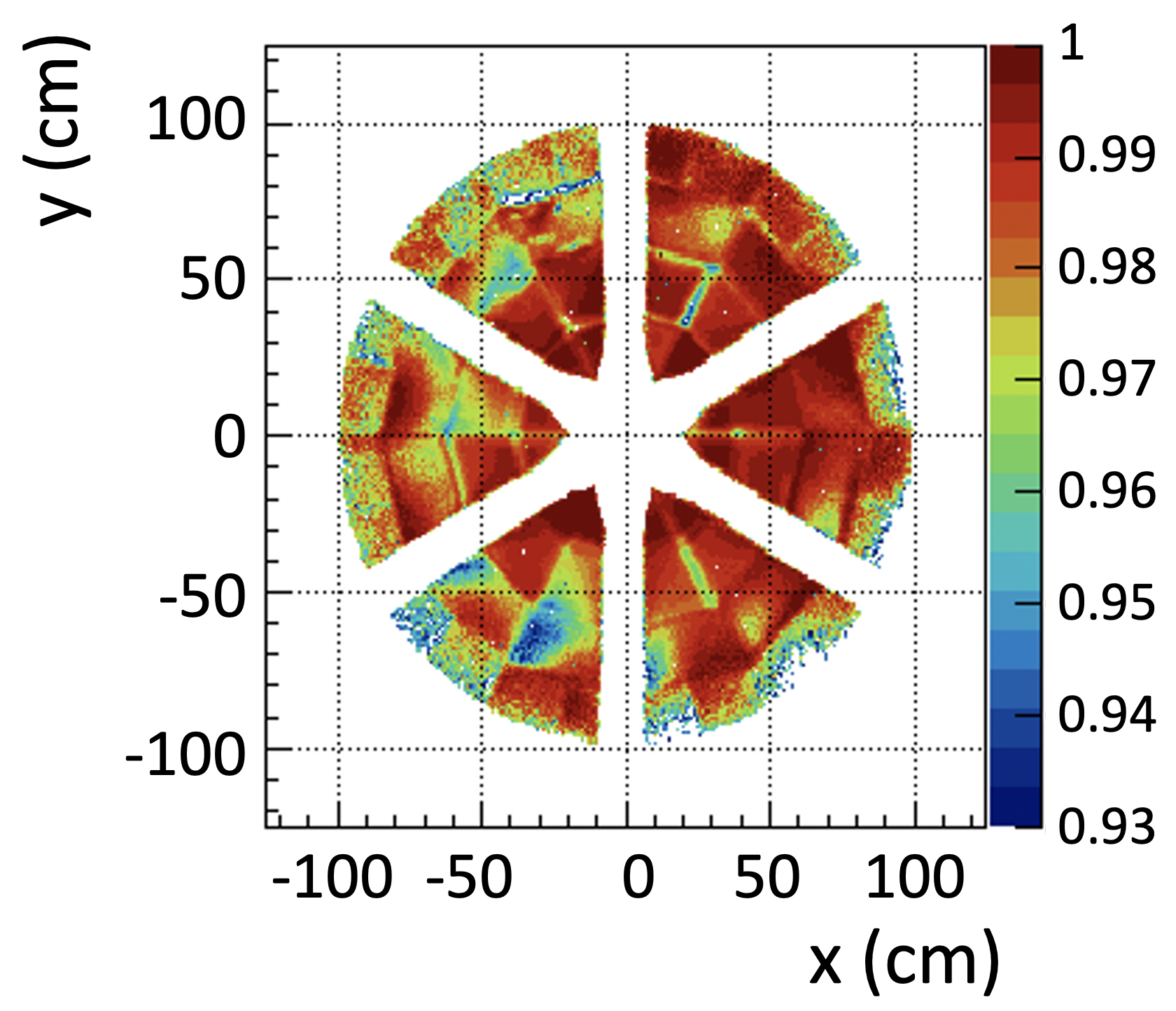}
\caption{Efficiency correction map for the HTCC over the $y$ vs.~$x$ coordinate acceptance of the detector. Sector 1 is located at
$x > 0$ and centered around $y = 0$. The sector number increases counterclockwise. Note the truncated $z$ scale for the efficiency.}
\label{fig:HTCC_map}
\end{figure}

The difference between the HTCC efficiency from data and simulation is attributed to non-uniformities in the mirror surface (both in 
terms of reflectance and geometry), which results in signal dispersion and non-optimal overlap with the acceptance of the HTCC PMTs
(which themselves could be slightly misaligned). To obtain the correction factor for every $(x,y)$ bin, we took the ratio of 
efficiencies for data to MC. The final HTCC efficiency map that was used in the analysis is shown in Fig.~\ref{fig:HTCC_map}. The map
contains efficiency corrections for all $x$ and $y$ bins in the range $(-125,125)$~cm, with a bin size of 1~cm. To correct for the HTCC
inefficiency, a weight of $1/\epsilon$, where $\epsilon$ is the efficiency from the map, was applied to the reconstructed events.

\subsubsection{Forward Tracking Efficiency}
\label{CurrentCorrection}

As the beam current increases, the forward tracking efficiency decreases. A linear dependence of the tracking efficiency as a function
of beam current was determined. In our dataset, we have runs collected with beam currents from 45 to 55~nA. After applying the efficiency
correction factor from the study associated with negatively charged particles (see Section~\ref{bck-merge}) to our dataset, we see good 
agreement between the cross sections determined using the 55~nA and 45~nA data runs using only a 45~nA simulation sample.

This correction was applied directly to the Faraday cup charge using

\begin{equation}
\begin{aligned}
  C_{eff} = 1 - 0.0041/{\rm nA} \cdot (I_{run} - 45~{\rm nA}), \\ 
  ~~Q_{corr} = Q_{run} \cdot C_{eff}, 
\label{fluxeq2}
\end{aligned}
\end{equation}
\noindent
where $I_{run}$ is the nominal beam current in nA for a given run, $Q_{corr}$ the Faraday cup charge in nC after correction for the
detector efficiency loss, $Q_{run}$ the Faraday cup charge in nC, and $C_{eff}$ is the correction factor that depends on beam current.
$Q_{corr}$ was used for the integrated experiment electron current. Strictly speaking, this correction accounts for the decrease of
the forward tracking efficiency at the higher beam current ({\it i.e.} higher luminosity) that affects the reconstructed inclusive 
events. Our approach to account for this by applying the correction to the Faraday cup charge amounts to the same thing but
was done in this way as a convenience in the data analysis.

\subsection{Data Binning}

The CEBAF 10.6~GeV electron beam, combined with the broad polar and azimuthal angle acceptance of CLAS12, provides a wide kinematic 
range over both $W$ and $Q^2$, with the ability to cover the full resonance range in $W$ for any given $Q^2$. This feature is especially 
important for studies of inclusive electron scattering in the resonance region of $W \lesssim 2$~GeV. The peaks seen in the first, second, 
and third resonance regions (at $W\!\approx$1.2~GeV, $\approx$1.5~GeV, and $\approx$1.7~GeV) make application of interpolation procedures 
over 
$W$ and $Q^2$ rather questionable. The validation by the $p(e,e')X$ cross sections measured within a broad $W$-range in any given 
$Q^2$ bin is definitely needed. 

Accounting for the nucleon resonance contributions in inclusive electron scattering also dictates the desired binning over $W$, as we 
want to map out the structures in the $W$ spectrum with as fine detail as possible. On the other hand, the finite resolution of the 
CLAS12 detector sets a lower bound on the practical size of the $W$ bins. A reasonable compromise is achieved with a $W$ bin width of 
50~MeV for the covered $W$ range from 1.125~GeV to 2.525~GeV. As detailed in Fig.~\ref{fig:resolutionW} where the CLAS12 $W$ resolution 
is shown, the chosen $W$ bin width is consistent with the detector resolution function.

The prominent feature of inclusive electron scattering, namely its fast drop-off with increasing $Q^2$, prevents us from using uniform
binning. Hence, we chose to employ a logarithmic binning over $Q^2$. To define the $Q^2$ bins we selected the $Q^2$ range from 
$Q^2_{min} = 2.557$~GeV$^2$ to $Q^2_{max} = 10.456$~GeV$^2$ in $N_Q = 9$ bins as described in Eq.(\ref{eq:q2Binning}). The last line 
in Eq.(\ref{eq:q2Binning}) gives the bin number as a function of $Q^2$. 

\begin{equation}
 \begin{aligned}
Q^2_{min} &=& 2.557~{\rm GeV}^2\\
Q^2_{max} &=& 10.456~{\rm GeV}^2\\
N_Q &=& 9\\
\Delta Q^2 &=& \log(Q^2_{max}/Q^2_{min})/N_Q \\
{\rm bin}(Q^2) &=& \log(Q^2/Q^2_{min})/\Delta Q^2
\label{eq:q2Binning}
\end{aligned}
\end{equation}

\begin{figure}[htp]
\centering
\includegraphics[width=0.95\columnwidth]{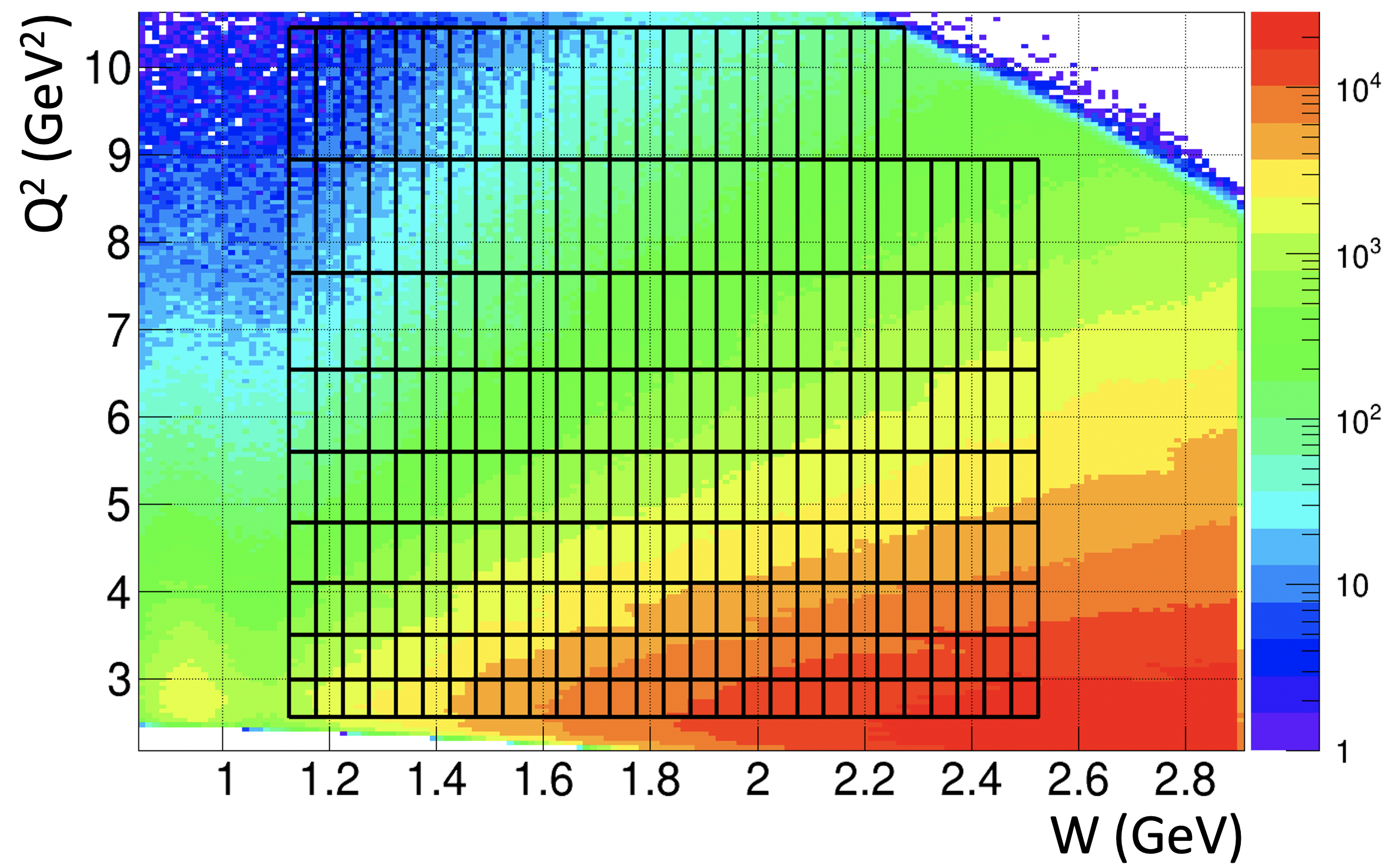}
\caption{$Q^2$ vs.~$W$ distribution for the inclusive electrons for the RG-A fall 2018 inbending dataset. The black lines show the 
kinematic binning used for this cross section analysis in the nucleon resonance region.}
\label{fig:binning}
\end{figure}

The $Q^2$-dependence of the reconstructed electrons shows a strong acceptance-related fall-off at $Q^2 < 2.5$~GeV$^2$. Therefore, we started 
our analysis at $Q^2 = 2.55$~GeV$^2$ to stay in the area where we have well-defined control over the electron detection efficiency. We have 
9 bins over $Q^2$ in the range from 2.55--10.4~GeV$^2$. For $W$, we started at 1.15~GeV (which is the center of the lowest bin, so the
minimum $W$ is 1.125~GeV), which is close to the inelastic threshold value (1.08~GeV). The cross section is small at low $W$, so we were
limited by statistics and did not go lower. The focus of this analysis is the resonance region, so the last $W$ bin center was set to 
$W = 2.5$~GeV for all but the last $Q^2$ bin. In the last $Q^2$ bin we had to stop at $W = 2.25$~GeV because of the detector acceptance.
The $Q^2$ vs.~$W$ phase space of the RG-A fall 2018 inbending dataset used for this analysis is shown in Fig.~\ref{fig:binning} overlaid
with the chosen binning grid.

\subsection{Data Unfolding Procedure}
\label{sec:migration}

\begin{figure*}
\centering
\includegraphics[width=0.95\textwidth]{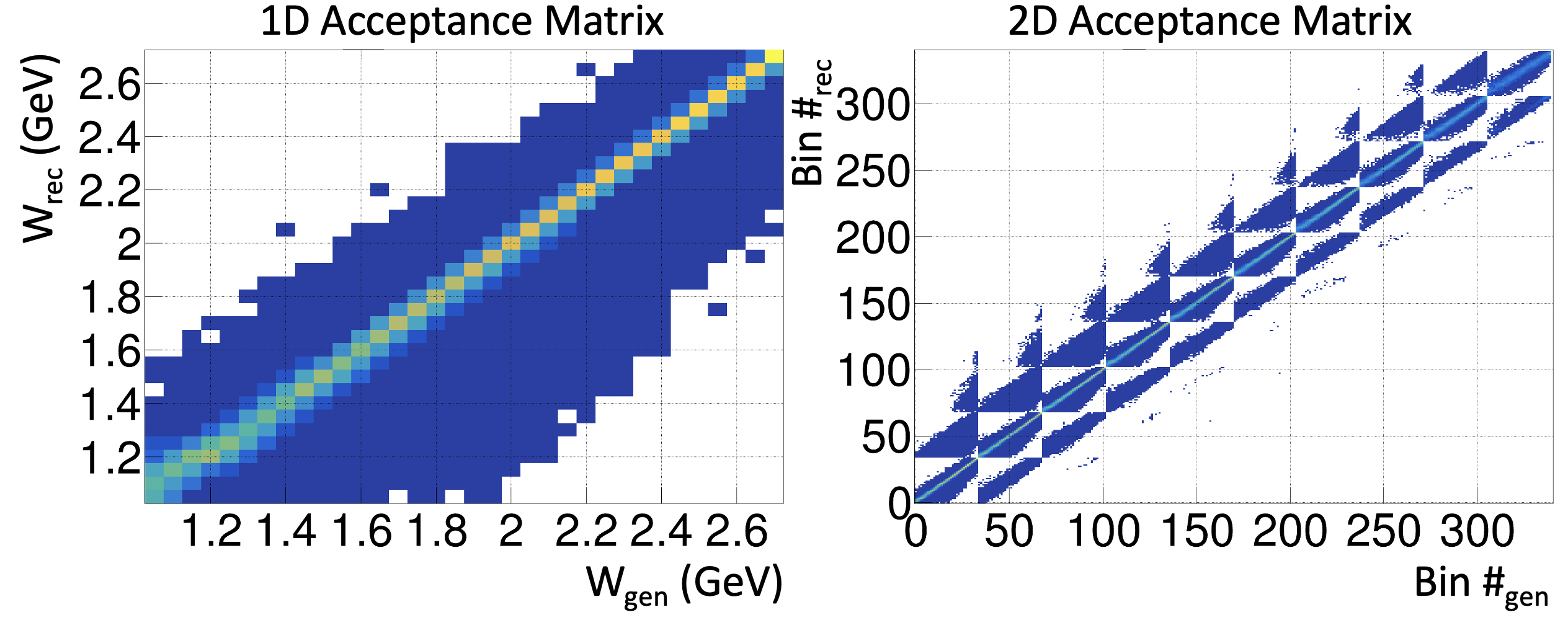}
\caption{Acceptance matrix of bin migration for the 1D case (left) in terms of the reconstructed $W_{rec}$ distributions for the generated
events in the $W_{gen}$ bins summed over all $Q^2$ bins and the 2D case (right) where each rectangular panel shows similar distributions 
as for the 1D case but in the different $(W,Q^2)$ intervals. Each primary group is binned in $W$. The $x$- and $y$-axes show the 2D-group 
bin numbers.}
\label{fig:accMatrix}
\end{figure*}

To mitigate the problem of bin migration (the misidentification of a kinematic bin due to effects of finite resolution, acceptance, and
distortions, among others), we employed our MC with its response function matching the data. The bin migration can be accommodated by a 
matrix $R_{i,j}$ transformation

\begin{equation}
\label{problemUnf}
x_i = \sum\limits_{j=0}^n R_{i,j}y_j,
\end{equation}

\noindent
where $y$ represents the ``true" distribution and $x$ represents what is reconstructed. The response function $R_{i,j}$ is the probability
for an inclusive $p(e,e')X$ event generated in bin $j$ to be reconstructed in bin $i$. It can be obtained from simulation as the number 
of events generated in bin $j$ but reconstructed in bin $i$ divided by the total number of generated events in bin $j$. In our analysis 
the bins $i$ and $j$ represent $(W,Q^2)$ bins. The acceptance matrix is shown for the one-dimensional (1D) and two-dimensional (2D) cases 
in Fig.~\ref{fig:accMatrix}. In our nominal analysis we account for 2D $(W,Q^2)$ bin migration effects.

The deconvoluted distributions $y_j$ can be obtained by

\begin{equation}
\label{deCONV}
y_j = R_{i,j}^{-1} x_i.
\end{equation}

\noindent
Even if $R_{i,j}^{-1}$ is well defined, it can cause large variations in the deconvoluted distributions, so it is crucial to find a way 
to come up with a reasonable and applicable $R_{i,j}^{-1}$. A detailed explanation of the theory and formalism behind practical unfolding
methods can be found in Ref.~\cite{RooUnfoldArticle}. In the remainder of this subsection, the two different deconvolution approaches
considered for this work are discussed along with a comparison of the methods.

\subsubsection{Bin--by--Bin Method}
\label{bbbMethSec}

The most common method to estimate $R^{-1}$ is the bin-by-bin method. In the 1D case, the relevant matrix is given by $R_i$, 
the ratio of reconstructed events in bin $i$ relative to the number of generated events in bin $i$
\begin{equation}
\label{bin_by_bin}
y_i = \frac{N_{rec_i}}{N_{gen_i}} \cdot x_i.
\end{equation}

\noindent
This method does not track the probability for an event generated in bin $i$ to be reconstructed in a different bin $j$, and its ability
to account for event migration between different bins of $(W,Q^2)$ depends directly on the quality of the event generator and the
detector simulation. The simulation provides the matrix $R^{-1}$, so consequently this method is free from uncertainties
related to the matrix inversion mentioned above.

\subsubsection{Richardson-Lucy (Iterative Bayes) Deconvolution}

Deconvolution methods employ the matrix $R$ determined from simulation evaluated according to Eq.(\ref{problemUnf}). The deconvoluted
distributions can be evaluated from Eq.(\ref{deCONV}) using the inverse matrix $R^{-1}$. In our studies, we used the Richardson-Lucy or
Iterative Bayes deconvolution method. A detailed description of this method is available in Ref.~\cite{bayes_article}. Using the same
notation as above, with the true distribution $y = y(y_1,...,y_n)$ and the measured distribution $x = x(x_1,...,x_n)$, then
\begin{equation}
\label{Bays_1}
y_i  = \frac{1}{\epsilon_i}\sum\limits_{j=0}^n x_jP(y_i|x_j),
\end{equation}
\noindent
where $P(y_i|x_j)$ is the probability of an event reconstructed in bin $j$ to be generated in bin $i$
\begin{equation}
\label{prob_1}
P(y_i|x_j) = \frac{P(x_j|y_i) \cdot P(y_i)}{\sum\limits_{l=0}^n P(x_j|y_l)\cdot P(y_l)}.
\end{equation}
\noindent
There is a chance that some events are not reconstructed at all, so an efficiency $\epsilon_i$ was introduced, which is 
\begin{equation}
\label{prob_2}
\epsilon_i = \sum\limits_{j=0}^n P(x_j|y_i).
\end{equation}
\noindent
All of the probabilities can be determined from MC. After deconvolution $P(y_i)$ can be re-estimated, which is our new best estimate 
(better than the generated distribution estimate)
\noindent
\begin{equation}
\label{prob_3}
P_1(y_i) = \frac{y_i}{\sum\limits_{i=0}^n y_i}.
\end{equation}
\noindent
Using this new $P_1(y_i)$, we can start over and perform a second deconvolution iteration. 

\begin{figure*}
\centering
\includegraphics[width=0.98\textwidth]{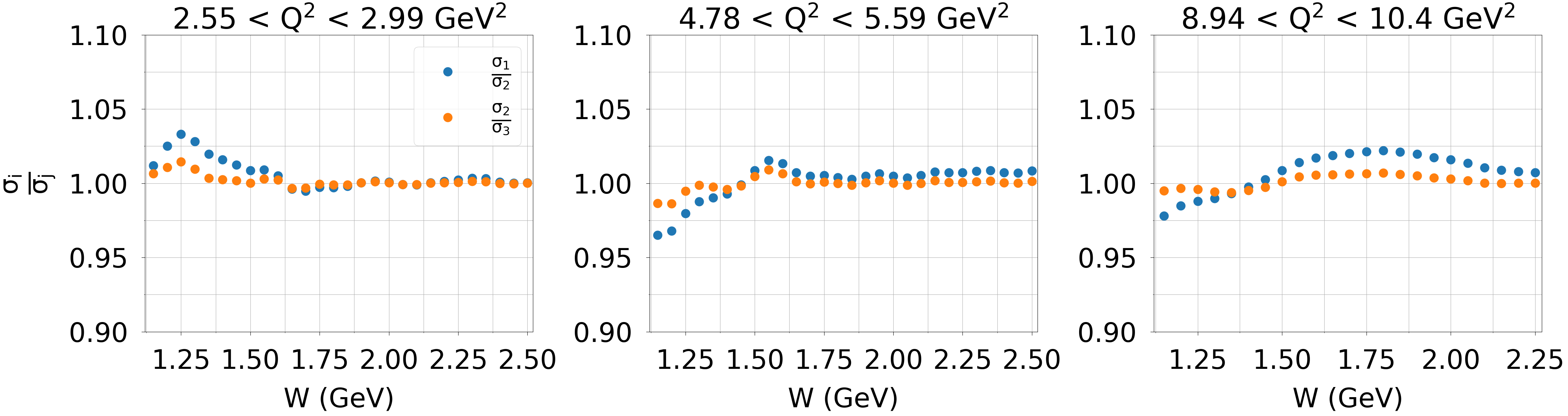}
\vspace{-6mm}
\caption{Ratio of deconvoluted cross sections with the 2D Iterative Bayes algorithm for iteration $i$ = 1 to $j$ = 2 (blue) and for
$i$ = 2 to $j$ = 3 (orange) for 3 representative $Q^2$ bins.}
\label{fig:baysIter_ratio}
\end{figure*}

This procedure was applied separately to each sector. We constructed response matrices for each sector, estimated events, and 
performed the deconvolution procedure. Figure~\ref{fig:baysIter_ratio} shows comparisons for three iterations. Even after two
iterations, the events for all $Q^2$ bins converged and did not change with iteration number. Therefore we used two iterations in our
deconvolution procedure.

\subsubsection{Method Comparison}
\label{UnfCompSect}

As discussed above, we considered two unfolding different methods in the cross section extraction (bin-by-bin and Bayesian) to account 
for bin migration effects. The deconvolution for the Bayesian method was done using the RooUnfold package~\cite{RooUnfoldCode}. The 
bin-by-bin method was implemented using the definition from Section~\ref{bbbMethSec}. For the Bayesian deconvolution, we filled response 
matrices for each sector.

To properly account for event migration, the simulation should be performed within the same or an even broader area over $W$ and $Q^2$ 
than for the region populated by the measured events. We deconvoluted the extracted events in the $W$ range from 0.825 to 2.825~GeV 
and $Q^2$ from 2.18 to 12.0~GeV$^2$. The bin-by-bin and Bayesian deconvolution methods were seen to give consistent results for all $W$ 
and $Q^2$ bins, with variations of only a few percent. Our final choice of unfolding method for this analysis was the Bayesian method.

\subsection{Radiative Corrections}
\label{rc}

\begin{figure*}[htp]
\centering
\includegraphics[width=0.9\textwidth]{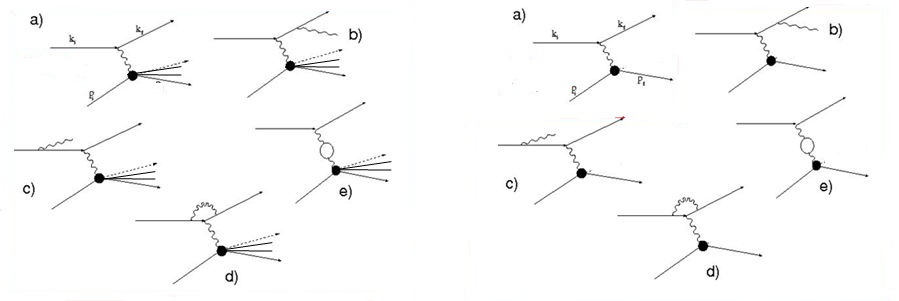}
\caption{Feynman diagrams contributing to the Born amplitude and the radiative effects for inclusive electron scattering (left) and 
elastic scattering (right): (a) Born amplitude, (b) and (c) internal bremsstrahlung, (d) vertex correction, and (e) vacuum polarization.}
\label{fig:rc_diag}
\end{figure*}

Radiative effects are present in both inclusive and elastic electron scattering as shown in Fig.~\ref{fig:rc_diag}. Our inclusive EG 
incorporated internal radiative effects based on the Mo and Tsai~\cite{moTsaiRC} approach with the option to run with them ``on" or 
``off". This feature allowed us to conveniently evaluate radiative corrections by calculating the cross section with and without 
radiative effects for each $W$ and $Q^2$ bin for which we report our results. As noted in Section~\ref{sec:monte}, our radiative model
described both the elastic and inelastic parts of inclusive electron scattering. The final version of the EG that we obtained after the
adjustment procedure described in Section~\ref{egAdj} was used. We divided every $W$ and $Q^2$ bin into 11 sub-bins in $W$ and 21 
sub-bins in $Q^2$ (these sub-bins were uniform in both $W$ and $Q^2$). There are 231 bins in total for each $W$ and $Q^2$ bin 
($11 \times 21$). The reason for smaller binning in $Q^2$ is due to the fact that we used wide bins in $Q^2$ and narrow bins in $W$. 
We calculated the cross section with and without radiative effects in the center of each sub-bin. Using the cross section values in 
each smaller bin $i$, we calculated the average cross section over the entire $W$ and $Q^2$ bin. The radiative correction factor $R$ 
was computed as the ratio of the radiative to non-radiative cross sections as

\begin{equation}
 R(W, Q^2) = \frac{\sum\limits_{i=1}^{231} \sigma_{RC_i}(W_i, Q_i^2)}{\sum\limits_{i=1}^{231}  \sigma_{noRC_i}(W_i, Q_i^2)}.
\end{equation}

\begin{figure*}[htp]
\centering
\includegraphics[width=0.98\textwidth]{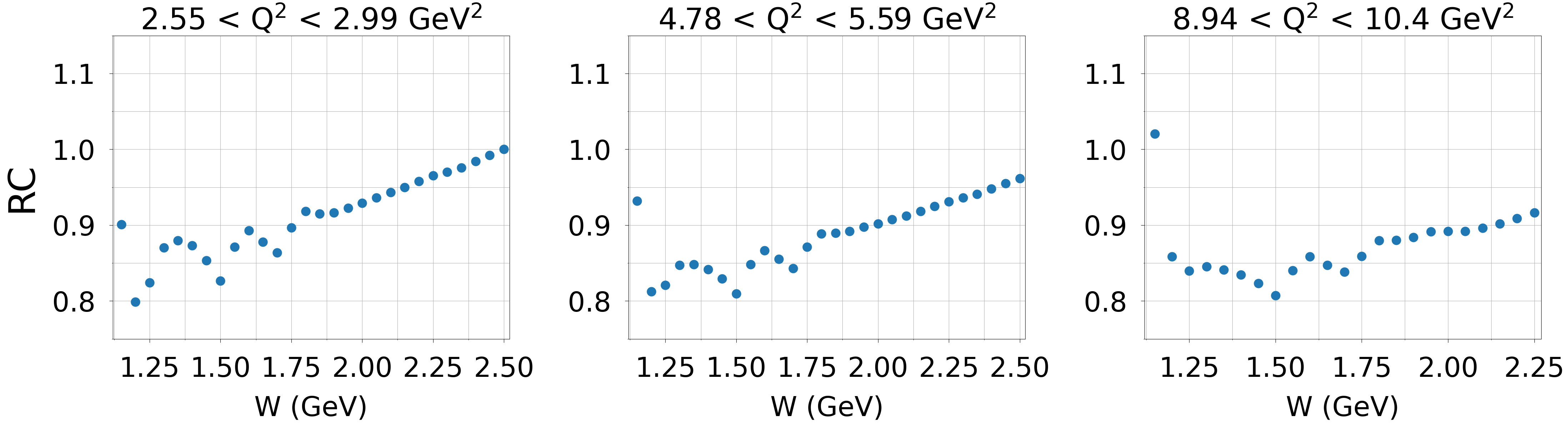}
\caption{Radiative corrections vs.~$W$ for 3 representative $Q^2$ bins calculated as the ratio of the radiative to non-radiative cross 
sections.}
\label{fig:rcIllustration1}
\end{figure*}

\noindent
Figure~\ref{fig:rcIllustration1} illustrates the results of the radiative correction calculation procedure. Note that all external 
radiation effects after the $e+p$ interaction point are modeled within GEMC and accounted for by our analysis cuts and unfolding 
procedures.

\subsection{Bin-Centering Corrections}
\label{bc}

When calculating the cross section, one actually obtains an average value within the 2D $(W, Q^2)$ bin. Assigning this value to the 
center point of the bin is justified in the case of a linear dependence across the bin, although in more realistic situations, 
this is not the case.

A procedure was developed to calculate the correction that takes this effect into account. By dividing each of our $28 \times 9$
2-fold $(W, Q^2)$ bins into 11 by 21 sub-bins (for a total of $\approx$60000 bins as mentioned in Section~\ref{rc}), the cross section 
was calculated in each sub-bin using the EG. The computed cross sections were then averaged to obtain the bin-centering correction 
defined by
 
\begin{equation}
\label{Eq:bccEq}
 BC(W, Q^2) = \frac{\sum\limits_{i=1}^{231} \sigma_{i}(W_i, Q_i^2)}{\sigma_{center}(W, Q^2)\cdot 231}.
\end{equation}

\begin{figure*}[htp]
\centering
\includegraphics[width=0.98\textwidth]{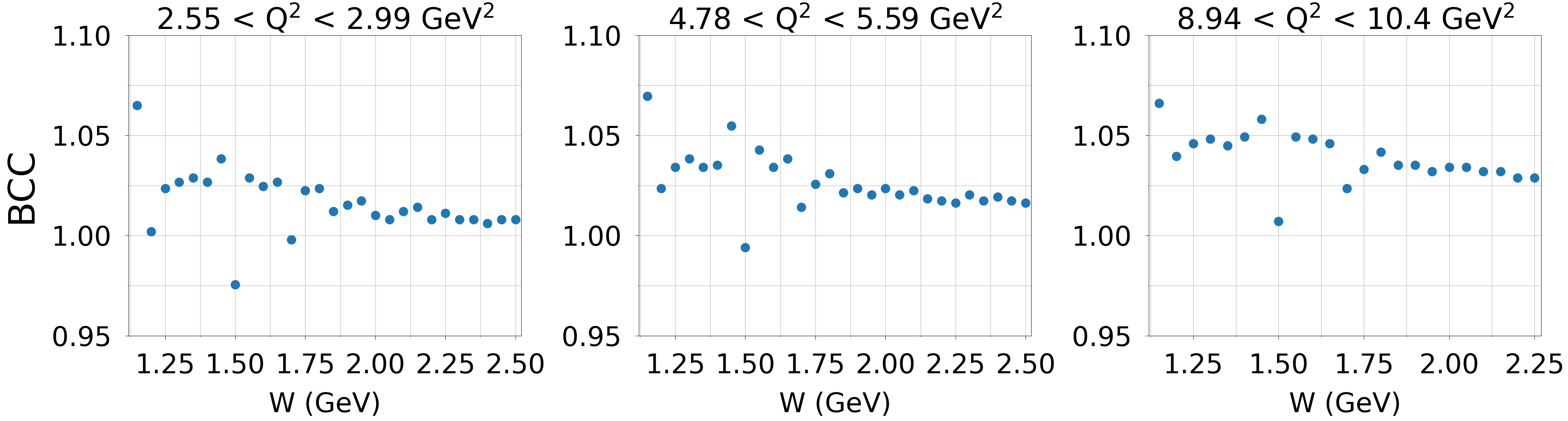}
\caption{Bin-centering corrections vs.~$W$ for 3 representative $Q^2$ bins calculated as the ratio of the mean cross section over the bin 
to the cross section in the center of the bin.}
\label{fig:bcIllustration2}
\end{figure*}

\noindent
Figure~\ref{fig:bcIllustration2} illustrates the results of the bin-centering correction calculation procedure. The structures in the 
corrections as a function of $W$ correspond to the resonance peaks in the second and third resonance regions where the central point of 
the bin is located at the maxima and then the ratio in Eq.(\ref{Eq:bccEq}) decreases due to the maxima in the denominator. There are no
structures in the $Q^2$ evolution of the cross sections so it does not introduce any other structures in our correction factor. 
Table~\ref{bin-values} details the $Q^2$ bin limits for the 9 bins used in this analysis and provides the bin-centered $Q^2$ value
for each bin. Note that the sub-plots in the following figures are labeled with the $Q^2$ bin limits instead of the geometric center
bin values.

\begin{table}[htp]
\begin{center}
\begin{tabular}{|c| c|} \hline
  $Q^2$ Bin Limits & $Q^2$ Bin Center \\
  (GeV$^2$)        & (GeV$^2$)        \\ \hline
  2.557 - 2.990  & 2.774 \\ \hline
  2.990 - 3.497  & 3.244 \\ \hline
  3.497 - 4.089  & 3.793 \\ \hline
  4.089 - 4.782  & 4.436 \\ \hline
  4.782 - 5.592  & 5.187 \\ \hline
  5.592 - 6.539  & 6.066 \\ \hline
  6.539 - 7.646  & 7.093 \\ \hline
  7.646 - 8.942  & 8.294 \\ \hline
  8.942 - 10.456 & 9.699 \\ \hline
\end{tabular}
\caption{Listing of the bin limits in $Q^2$ for each of the 9 bins used in this cross section analysis and the associated 
bin-centered values at the geometric center of the bin.}
\label{bin-values}
\end{center}
\end{table}

\section{Iterations}
\label{sec:iterations}

Having discussed all of the terms in Eq.(\ref{dcs}), we can obtain the preliminary inclusive cross sections $\sigma_0$. However, to make
sure that our EG model is realistic and that we do not introduce significant bias in our measurements, we estimated the cross sections
iteratively. Our approach was to use the preliminary extracted cross sections $\sigma_0$ to modify the EG model, and then to perform 
new MC simulations to re-determine the acceptance/efficiency, deconvolution, radiative, and bin-centering corrections. We performed three 
iterations, so we ultimately ran four sets of simulations and correction estimations.

\subsection{Event Generator Adjustments}
\label{egAdj}

\begin{figure*}[htp]
\centering
\includegraphics[width=0.98\textwidth]{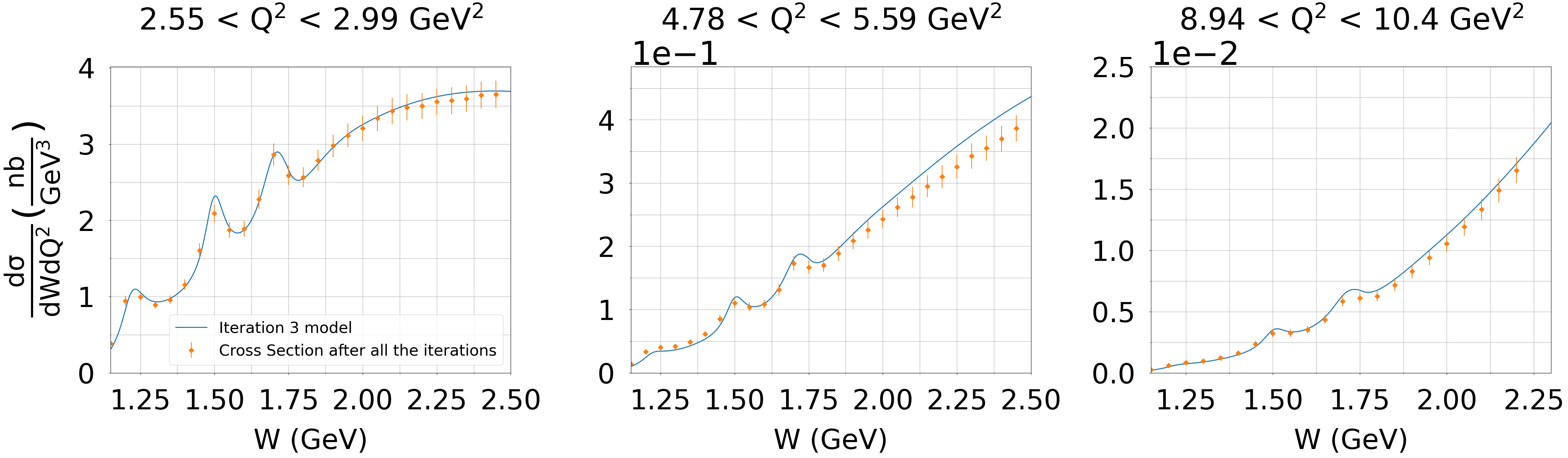}
\caption{Inclusive cross sections $\sigma_3$ (orange points with error bars) obtained with the iteration 3 model $\sigma_{sim3}$ (blue 
curves) for 3 representative $Q^2$ bins.}
\label{fig:modelIter0}
\end{figure*}

The inclusive EG has a built-in background parameterization that is given by a seventh-order polynomial along with the Bodek 
parameterization~\cite{bodek} for the resonance contributions. To come up with a new, more realistic EG, we calculated the $\chi^2$ 
for all preliminary cross section data points relative to the EG values (over the full $W$ and $Q^2$ coverage) and tried to minimize 
it varying the parameters in the EG model to adjust the initial cross section model $\sigma_{sim0}$. This was done in two steps. In the
first step, we varied the background function to come up with the correct normalization. After that, we varied all resonance parameters 
to align the peak positions, widths, and the evolution of these parameters with $Q^2$. Figure~\ref{fig:modelIter0} shows the inclusive
cross sections $\sigma_3$ from the third EG iteration model compared to the $\sigma_{sim3}$ EG model.

\begin{figure*}[htp]
\centering
\includegraphics[width=0.98\textwidth]{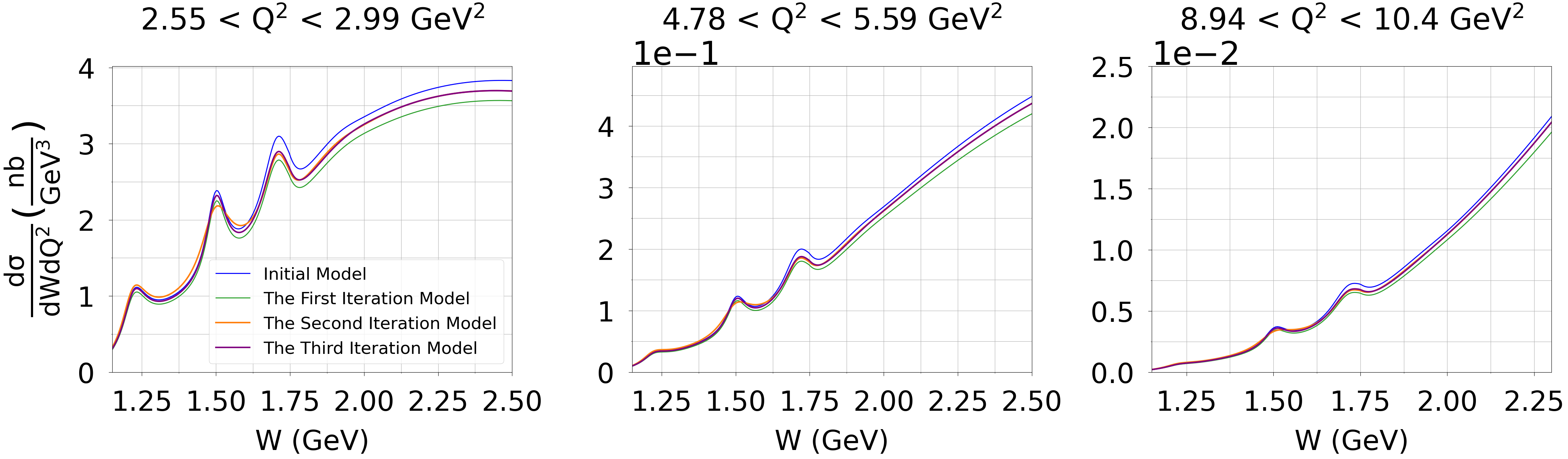}
\caption{Initial inelastic model cross section from EG (blue), iteration $N=1$ model (green), and iteration $N=2$ model (orange). The
purple curve is the model cross section for iteration 3, which is discussed separately in Section~\ref{thirdIter} for 3 representative
$Q^2$ bins.}
\label{fig:modelIter2}
\end{figure*}

The $N=1$ iteration model $\sigma_{sim1}$ was used for the new MC simulation to re-estimate the radiative (see Section~\ref{RC_tweak}) 
and bin-centering corrections (see Section~\ref{BCC_tweak}). Using this information and Eq.(\ref{dcs}), we estimated the cross sections 
$\sigma_1$ after iteration $N=1$. This $\sigma_1$ cross section after the first iteration was then used as a grid for a new, third even 
more realistic EG. Three models: initial EG model $\sigma_{sim0}$ (blue), iteration $N=1$ $\sigma_{sim1}$ (green), and iteration $N=2$ 
$\sigma_{sim2}$ (orange) are shown in Fig.~\ref{fig:modelIter2}. It is seen that the resonance peaks are shifted to slightly lower $W$ 
values after two iterations. This occurred as the peak position in the initial model and our measurement did not fully coincide.

\subsection{Acceptance Iterations}
\label{acc_tweak}

As a result of the EG adjustments based on the data measurements, we developed a new, more realistic model $\sigma_{sim1}$ and a new 
EG based on the $\sigma_{sim1}$ model that was used for the acceptance correction estimation following the procedure described in 
Section~\ref{sec:migration}. Using the new $R_1$, $BC_1$, and $\eta_1$ obtained with model $\sigma_{sim1}$, we estimated preliminary 
cross sections after one iteration $\sigma_1$ that made it possible to start iteration 2. The second iteration procedure is identical 
to the first iteration. 

\subsection{Radiative Correction Iterations}
\label{RC_tweak}

\begin{figure*}[htp]
\centering
\includegraphics[width=0.98\textwidth]{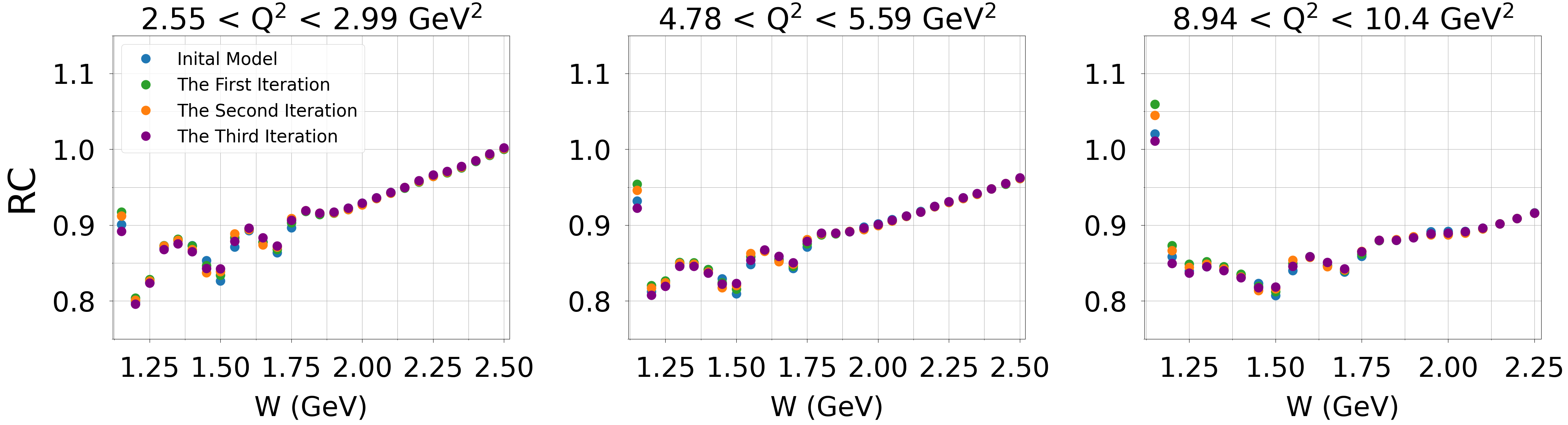}
\caption{Radiative corrections $R_0$ obtained with the initial model $\sigma_{sim0}$ (blue), the $R_1$ values obtained with the first
iteration model $\sigma_{sim1}$ (green), and the $R_2$ values obtained with the second iteration model $\sigma_{sim2}$ (orange) for 3
representative bins in $Q^2$. The purple points are from iteration 3, which is discussed Section~\ref{thirdIter}.}
\label{fig:rc_iterfig}
\end{figure*}

Similar to the acceptance corrections, the radiative corrections for all three models can be compared. The radiative corrections 
$R_0$ obtained with the initial model $\sigma_{sim0}$, the $R_1$ values obtained with the first iteration model $\sigma_{sim1}$,
and the $R_2$ values obtained with the second iteration model $\sigma_{sim2}$ are shown in Fig.~\ref{fig:rc_iterfig}.  

A slight shift in the peak position to lower $W$ with increasing iteration number can be seen. This is a result of different peak
positions in the measured $W$ distribution and the initial model $W$ distribution. There was no need for more iterations since the
different iterations are were consistent with each other.

\subsection{Bin-Centering Correction Iterations}
\label{BCC_tweak}

\begin{figure*}[htp]
\centering
\includegraphics[width=0.98\textwidth]{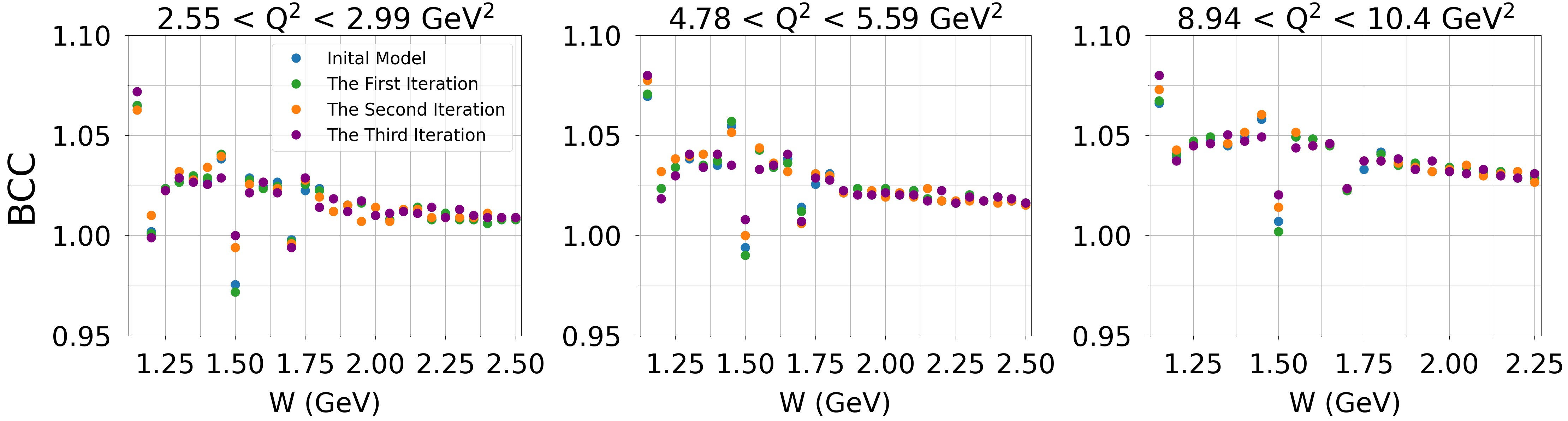}
\caption{Bin-centering corrections $BC_0$ obtained with the initial model $\sigma_{sim0}$ (blue), the $BC_1$ values obtained with the
first iteration model $\sigma_{sim1}$ (green), and the $BC_2$ values obtained with the second iteration model $\sigma_{sim2}$ (orange)
for 3 representative $Q^2$ bins. The purple points are from iteration 3, which is discussed in Section~\ref{thirdIter}.}
\label{fig:bcc_iterfig}
\end{figure*}

Similar to the acceptance corrections, the bin-centering corrections for all three models can be compared. The bin-centering corrections
$BC_0$ obtained with the initial model $\sigma_{sim0}$, the $BC_1$ values obtained with the first iteration model $\sigma_{sim1}$, and 
the $BC_2$ values obtained with the second iteration model $\sigma_{sim2}$ are shown in Fig.~\ref{fig:bcc_iterfig}.  

It is seen that the bin-centering corrections do not change much because there is no significant change in the cross section shape. 
There is no need for more iterations since the iterations were consistent with each other.

\subsection{Cross Section After Iterations}
\label{XSEC_iter}

\begin{figure*}[htp]
\centering
\includegraphics[width=0.98\textwidth]{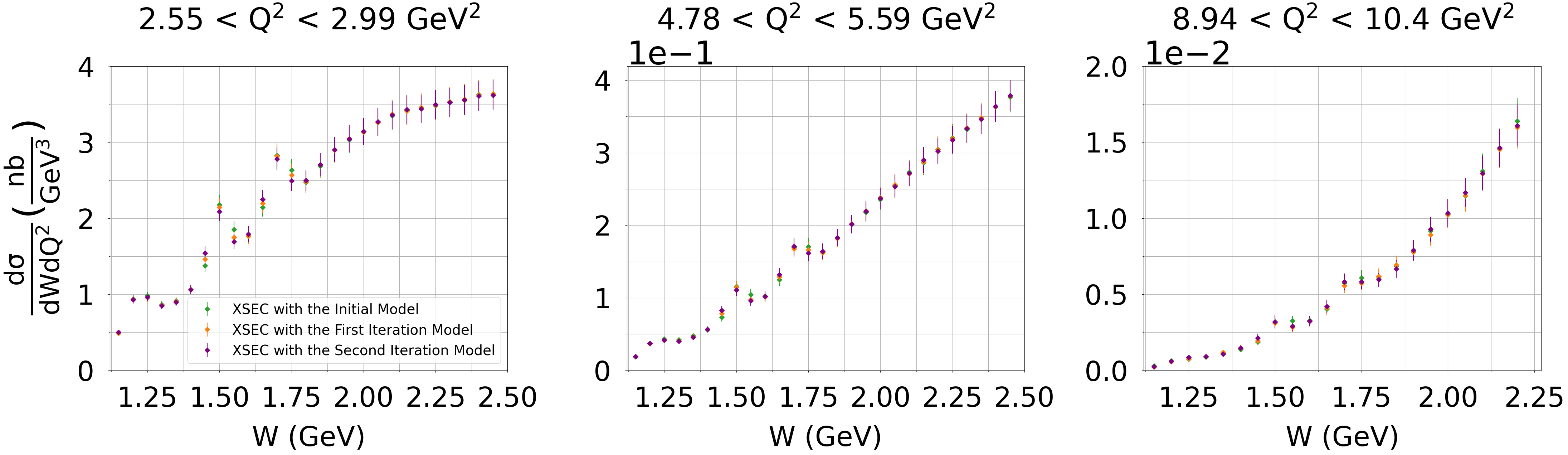}
\caption{Cross sections $\sigma_0$ obtained with the initial model $\sigma_{sim0}$ (green), the $\sigma_1$ values obtained with the 
first iteration model $\sigma_{sim1}$ (orange), and the $\sigma_2$ values obtained with the second iteration model $\sigma_{sim2}$
(purple) for 3 representative bins in $Q^2$.}
\label{fig:xsec_iterfig}
\end{figure*}

\begin{figure*}[htp]
\centering
\includegraphics[width=0.98\textwidth]{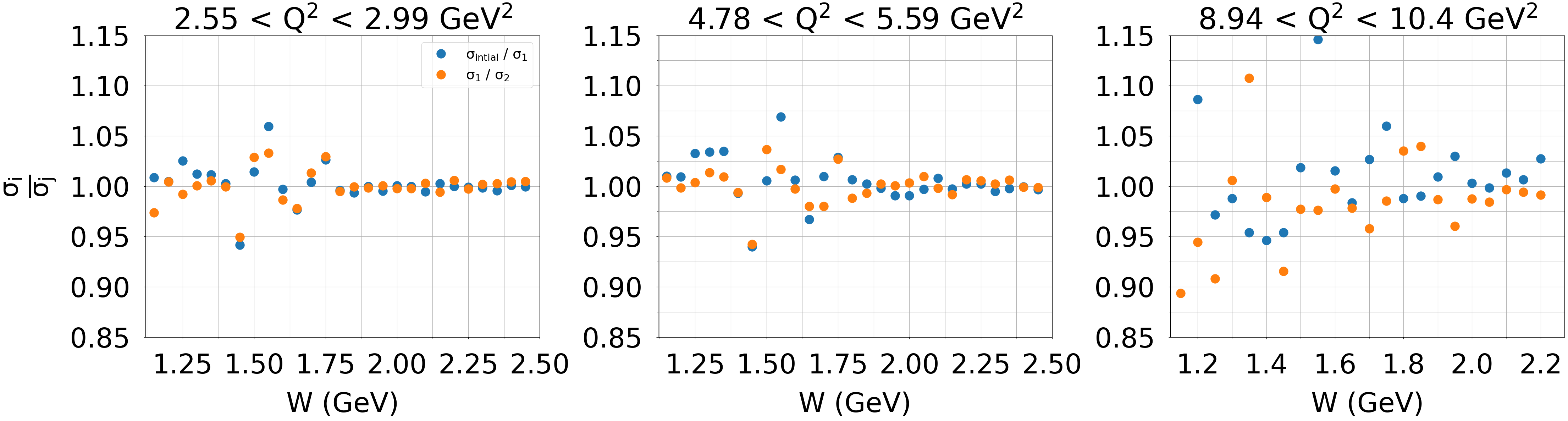}
\caption{Ratio of cross sections obtained with the initial model $\sigma_0$ to those obtained with the first iteration model
$\sigma_{sim1}$ (blue) and to those obtained with the second iteration model $\sigma_{sim2}$ (orange) for 3 representative $Q^2$ bins.}
\label{fig:xsec_iterfig1}
\end{figure*}

Combining $R$, $BC$, and $\eta$ obtained with the different models, the cross sections for every EG model can be calculated. There is a 
minimal visible effect on the peaks in the second and third resonance regions comparing the initial cross sections extracted with the
original EG model relative to those extracted from the second iteration model. There is no change seen with iteration version where 
the cross section is smoothly varying. Figure~\ref{fig:xsec_iterfig} shows the extracted cross sections for each of our iteration 
models and Fig.~\ref{fig:xsec_iterfig1} shows the ratio of the cross sections after each iteration relative to the original EG model.

\subsection{Third Iteration}
\label{thirdIter}

Several corrections and procedures in the cross section extraction were introduced or updated after performing the iteration studies.
These updates called for an additional iteration that would make our MC simulation up to date. The additional iteration was performed
exactly the same way as for the two previous iterations. The EG parameterization was fit to reproduce the extracted cross sections 
after all of the aforementioned improvements. Using the updated EG, updated acceptance, radiative, and bin-centering corrections were
found. The radiative correction (see Section~\ref{rc}), bin-centering corrections (see Section~\ref{bc}), and bin-migration studies 
(see Section~\ref{UnfCompSect}) were done with this third iteration version EG. This is the most realistic EG possible for this
measurement. We would like to point out that since the updates to the cross section extraction procedure were made, the difference 
between iteration 2 and 3 comes not only from the iteration procedure itself but also from those updates. The most significant effect 
of iteration 3 can be seen in the very first $W$ bin, which is the result of an update to the EG elastic cross section. 
Figures~\ref{fig:rc_iterfig} and~\ref{fig:bcc_iterfig} show the radiative and bin-centering correction  evolution with iteration number. 

\section{Statistical and Systematic Uncertainties}
\label{statsyserr}

\subsection{Statistical Uncertainties}
\label{sec:statistics}

The CLAS12 Forward Detector consists of six separate sectors. We extracted cross sections for the scattered electron separately in each 
sector. As a result we have six separate measurements. For the final cross sections quoted in this work we took the $p(e,e')X$ cross
section as the average over all six sectors.

Statistical uncertainties were estimated assuming normal distributions for inclusive electron scattering events with the dispersion
$\sqrt{N(W,Q^2)}$ for the number of reconstructed events in the $(W,Q^2)$ bins corrected for detection efficiencies and accounting for 
acceptance after performing the deconvolution. In the evaluation of the statistical uncertainty of the $p(e,e')X$ cross section we 
combined the terms in Eq.(\ref{dcs}) into sector-dependent $B_i(W,Q^2)$ and sector-independent $A(W,Q^2)$ factors,

\begin{equation}
\label{eq:statEq1}
\frac{d\sigma}{d\Omega} =  A(W,Q^2) \cdot \sum_{i=1}^6\frac{N_i(W,Q^2)}{B_i(W,Q^2)},
\end{equation}

\noindent
where $N_i(W,Q^2)$ is the number of reconstructed events in sector $i$, $A(W,Q^2)$ is the product of all factors that do not depend on 
sector number such as radiative corrections, luminosity, bin-centering corrections, etc., and $B_i(W,Q^2)$ is the product of all factors 
in sector $i$ that depend on sector number ({\it e.g.} efficiency correction).

By employing the error propagation formula, after computing the derivative over $N_i$ in Eq.(\ref{eq:statEq1}), the statistical 
uncertainty of $\delta \sigma$ of the cross section can be computed as

\begin{equation}
\label{eq:statEq2}
\delta \sigma(W,Q^2) =  A(W,Q^2) \cdot \sqrt{\sum_{i=1}^6\frac{N_i(W,Q^2)}{B_i(W,Q^2)^2}}.
\end{equation}

If we define $N_{i\ corr}(W,Q^2) = N_i(W,Q^2)/B_i(W,Q^2)$, the relative statistical uncertainties in \% for the averaged cross sections 
in each bin of $(W,Q^2)$ can be estimated as
\begin{equation}
\label{eq:statEq3}
100 \cdot \frac{\sqrt{\sum_{i=1}^6\frac{N_{i\ corr}(W,Q^2)}{B_i(W,Q^2)}}}{\sum_{i=1}^6 N_{i\ corr}(W,Q^2)}.
\end{equation}

\noindent
In this approach, we do not have to consider factors that do not depend on sector number because they will be the same in the numerator
and denominator. The statistical uncertainty in each $Q^2$ bin is on average $\lesssim$0.5\% for $Q^2 < 5.59$~GeV$^2$. Overall the
average statistical uncertainty is below 1\%. It increases at high $Q^2$ and $W<1.3$~GeV because of the cross section shape.

Our reconstructed MC event sample has at least a factor of 5 times more statistics than the data sample for all $(W,Q^2)$ bins, so the
statistical uncertainty is dominated by the data events. Given the MC acceptance is computed as 
$\eta(W,Q^2) = N_{rec}(W,Q^2)/N_{gen}(W,Q^2)$, the statistical uncertainty of the acceptance can be computed as

\begin{equation}
\label{eq:stat_MC_Eq1}
\delta \eta(W,Q^2) =  \frac{\sqrt{N_{rec}(W,Q^2) + \frac{N_{rec}^2(W,Q^2)}{N_{gen}(W,Q^2)}}}{N_{gen}(W,Q^2)} .
\end{equation}

If we propagate this statistical uncertainty to the cross sections $\sigma(W,Q^2) = N_i(W,Q^2)/\eta(W,Q^2)$, assuming that the number
of events includes all of the corrections for notation simplification, we compute the relative statistical uncertainty as 

\begin{widetext}
\begin{equation}
\label{eq:stat_MC_Eq2}
\delta \sigma(W,Q^2) = \sqrt{ \Bigl( \frac{\delta N_i(W,Q^2)}{\eta(W,Q^2)} \Bigr)^2 + \Bigl( \frac{\delta \eta(W,Q^2)\cdot N_{i}
(W,Q^2)} {\eta(W,Q^2)^2} \Bigr)^2} \cdot \frac{100}{\sigma(W, Q^2)} .
\end{equation}
\end{widetext}

\begin{figure*}[htp]
\centering
\includegraphics[width=0.98\textwidth]{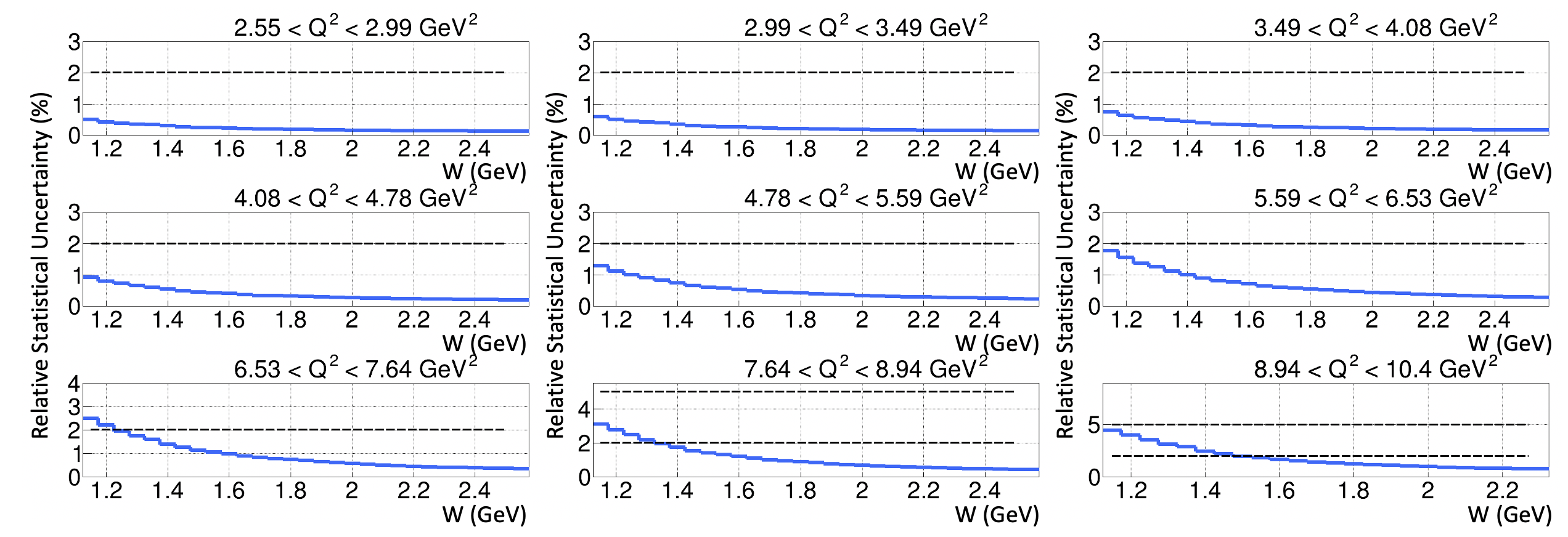}
\caption{Relative statistical uncertainty for each of our $(W,Q^2)$ bins calculated using Eq.(\ref{eq:statEq3}) taking into account the MC
statistical uncertainties. The black dashed lines are at the 2\% and 5\% levels.}
\label{fig:statInt_total}
\end{figure*}

\noindent
This statistical uncertainty takes into account both the data and MC. We then used exactly the same approach as in Eq.(\ref{eq:statEq2}) 
and Eq.(\ref{eq:statEq3}). The total statistical uncertainty for all of our analysis bins is shown in Fig.~\ref{fig:statInt_total}.

\subsection{Systematic Uncertainties}
\label{sec:systematics}

To estimate the systematic uncertainty for every $(W, Q^2)$ bin, the mean cross sections over all six sectors with different values
for each cut or correction were considered, while fixing all other cuts and corrections. In this way we obtained cross section
measurements for loose, nominal, and tight cut versions so we could estimate the effect of a single cut or correction on the final 
cross sections. For this estimation we used

\begin{widetext}
\begin{equation}
\label{eq:sysEq}
\delta \sigma_{cut}(W, Q^2) = \frac{|\sigma(W, Q^2) - \sigma_{loose}(W, Q^2)| + |\sigma(W, Q^2) - \sigma_{tight}(W, Q^2)|}{2} 
\cdot \frac{100}{\sigma(W, Q^2)},
\end{equation}
\end{widetext}

\noindent
where $\sigma(W, Q^2)$, $\sigma_{loose}(W, Q^2)$, and $\sigma_{tight}(W, Q^2)$ represent the mean cross sections for the nominal, loose, 
and tight versions of a cut, respectively. The sources of systematic uncertainty in this analysis are separated into those that vary 
bin-by-bin and those that affect the overall cross section scale. In the remainder of this subsection the different contributions 
are reviewed.

\subsubsection{Bin-by-Bin Systematics}

\noindent
\underline{Momentum corrections}: Momentum corrections were calculated as a multiplicative factor $f_{corr} = \delta P/P + 1$
applied to the reconstructed momentum. For our estimations, $\delta P/P$ was varied by $\pm 10$\%. The systematic uncertainty 
was found to be below 1\% for every $Q^2$ bin but the last one, where it is about 1.5\%.

\vskip 0.2cm
\noindent
\underline{$z$-vertex cuts}: Loose and tight vertex-$z$ cuts were applied using $-8.5 < v_z < 2.5$~cm and $-7.5 < v_z < 1.5$~cm,
respectively. The systematic uncertainty is below 1\% for every $W$ and $Q^2$ bin.

\vskip 0.2cm
\noindent
\underline{Fiducial cuts}: The DC geometric and ECAL shower containment fiducial cuts were studied using both looser and tighter
variations and resulted in systematic uncertainty below 1\% over all analysis bins. Bad detector element selection was accomplished 
with ad hoc coordinate cuts based on the different ECAL layers (PCAL, ECin, and ECout) and on angular regions in $\theta$ vs.~$\phi$ 
for the DC. The width of these cuts was varied and the systematic difference was found to be below 1\% and below 1.5\%, respectively, 
for all analysis bins.

\vskip 0.2cm
\noindent
\underline{Sampling fraction cuts}: The energy-dependent ECAL sampling fraction cut nominally at $3.5 \sigma$ was varied by 
$\pm 0.5\sigma$. The systematic uncertainty was found to be negligible for all analysis bins. A secondary cut on the PCAL and ECin 
partial sampling fractions is also included to reduce $\pi^-$ contamination. This cut was varied and the systematic uncertainty was 
found to be below 1\% for the first 7 $Q^2$ bins but increases to $\sim$3\% at high $Q^2$.

\vskip 0.2cm
\noindent
\underline{Monte Carlo}: As the CLAS12 MC gives better resolutions than the data, the reconstructed momentum in the MC was smeared to 
match the data resolution. In order to estimate the systematic uncertainty that comes from this procedure, the cross sections were
extracted varying the smearing factor by $\pm$5\%. The systematic uncertainty for smearing is less than $1\%$ on average over all bins.

\vskip 0.2cm
\noindent
\underline{Deconvolution method}: In the analysis two different deconvolution methods, bin-by-bin and Bayesian unfolding, were
considered. Our nominal analysis used the Bayesian method because it accounts for multi-dimensional bin migration effects. The cross
sections were calculated with the bin-by-bin and Bayesian deconvolutions methods while keeping all the other corrections exactly the 
same. The difference between the two methods resulted in an assigned systematic of 1\% on average.

\vskip 0.2cm
\noindent
\underline{Empty target subtraction}: In order to estimate the systematic uncertainty associated with the contribution of the target 
end-caps, the contribution for the full $v_z$ spectrum and the empty target spectrum without the $v_z$ region containing just the
residual cold gas was estimated. The first approach over-corrects the number of events because it includes the residual cold hydrogen gas 
in the target cell. It was decided to use 50\% of the difference between the two approaches as the systematic uncertainty for this 
source. Averaging the systematic uncertainty for the empty target contribution subtraction over all kinematic bins gives 0.4\%.

\vskip 0.2cm
\noindent
\underline{Radiative corrections}: The radiative corrections were obtained as a result of an iterative procedure. In our analysis, we
refit the EG cross section parameterization using our measurement to give a new EG. The procedure converges and the difference between 
the last two radiative correction iterations was taken as an estimation of the radiative correction systematic uncertainty. It was found
to be 1\% for all bins.

\vskip 0.2cm
\noindent
\underline{Bin-centering corrections}: The same approach as in the radiative correction case was used for the bin-centering correction
systematic uncertainty estimation. The difference between the last two bin-centering correction iterations was used for estimation of 
the systematic uncertainty. The systematic was found to be less than 1\% for all bins.

\noindent
\underline{$\pi^-$ Contamination}: In order to study possible contamination of our electron sample from non-minimum-ionizing $\pi^-$ 
that have momenta above threshold to give a signal in the HTCC, $p \gtrsim 5$~GeV, using MC we estimated the ratio of electrons to 
$\pi^-$s that pass our electron identification requirements in our kinematic range of interest. The generator used for our MC studies 
was CLASDIS~\cite{clasdis}, which is based on the well-known LEPTO EG~\cite{lepto} that simulates complete events in deep inelastic 
lepton-nucleon scattering (DIS). LEPTO includes a parton-level interaction based on electroweak cross sections implemented in leading 
order for any lepton of arbitrary polarization. First-order QCD matrix elements for boson-gluon fusion and gluon radiation are included,
and higher-order QCD radiation is treated using partonic showers. The hadronization process follows from the LUND string breaking model
implemented in Pythia (Jetset)~\cite{pythia}. Several parameters enter in the tuning of the hadronization process, and those have been
selected to reproduce the electron, photon, and pion distributions obtained in the CLAS12 RG-A data for the semi-inclusive regime 
({\it i.e.} $Q^2 > 1$~GeV$^2$ and $W > 2$~GeV).

In our studies we have made several assumptions:

\begin{itemize}
\item The DIS process is responsible for the creation of the dominant fraction of high-momentum pions in the RG-A dataset. These
pions only appear for $W \gtrsim 2$~GeV.
\item The physics model in the CLASDIS EG contains accurate ratios of final state $\pi^-$ and electrons in its designed deep
inelastic scattering kinematic region to quantitatively estimate the $\pi^-$ contamination for $Q^2 > 1$~GeV$^2$ and $W > 2$~GeV.
\item The contamination of $\pi^-$ in our electron sample for $W > 2$~GeV can be used to set an upper limit on the $\pi^-$ contamination
in the entire kinematic range of this analysis for $W$ from 1.125~GeV to 2.5~GeV.
\end{itemize}

The ratio of $\pi^-$s to electrons is in the range from 0.2\% to 0.5\% with the ratio increasing as the particle momentum decreases.

\begin{figure*}[htp]
\centering
\includegraphics[width=0.98\textwidth]{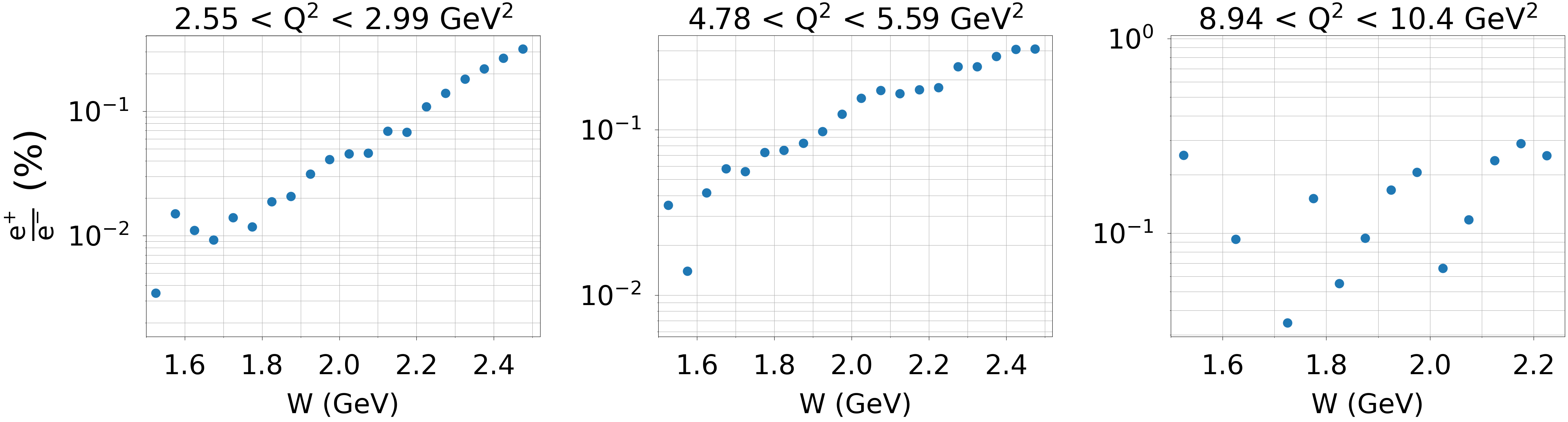}
\caption{Ratio of positron events from the RG-A outbending dataset to the electron events from the RG-A inbending dataset 
(in percent) vs.~$W$ for 3 representative $Q^2$ bins after applying all cuts from this analysis.}
\label{fig:CSB_data}
\end{figure*}

\vskip 0.2cm
\noindent
\underline{Charge Symmetric Background}: For $e+p$ scattering, processes that produce charge-symmetric $e^+e^-$ decays can contaminate 
the inclusive electron sample. The most important source of $e^+e^-$ pairs in CLAS12 is due to the production of $\pi^0$ mesons, which
either decay to $\gamma e^+e^-$ (Dalitz decay mode) or to $\gamma \gamma$, with the latter branch accounting for 98.8\% of the decay
fraction. For these processes $\gamma \to e^+ e^-$ conversions can occur. These secondary electrons can contaminate the events for the 
determination of the inclusive electron scattering cross section. We estimated the contribution of this charge symmetric background 
using two different procedures. One based directly on the RG-A data and one based on available data parameterizations that served as 
a cross check.

The contamination of electrons from the charge symmetric background can be estimated directly from the data by examining positrons 
in the opposite torus polarity dataset collected as part of the same RG-A data run. The positrons in outbending data should behave 
like the charge symmetric background electrons in inbending data. To study this, the inclusive $e^-X$ signal was investigated in the 
inbending runs and the inclusive $e^+X$ signal was investigated in the outbending runs.

After applying our complete set of electron identification requirements to both electrons and positrons, the estimation of the charge
symmetric background in our inclusive kinematic bins was determined by normalizing the $e^+X$ to $e^-X$ ratio by the ratio of 
the inbending to outbending dataset Faraday cup charge ratio (see Fig.~\ref{fig:CSB_data}). As a result of this study, the charge
symmetric background contamination of our inclusive electron sample was shown to be below 0.5\%.

We also estimated the charge symmetric background using the model of Bosted~\cite{bosted} as a cross check. This model is based on a fit 
of the inclusive pion photoproduction reaction from SLAC data and has been carefully checked and shown to be in reasonably good agreement 
with the measured positron cross section. This model was used in the inclusive analysis of CLAS data by Osipenko~\cite{osipenko} and in 
the inclusive analysis of Hall~C data by Malace~\cite{malace09}.

The code generates $\pi^0$s over the full kinematic range of the CLAS12 RG-A data and then decays $\pi^0 \to \gamma \gamma$ over all
possible polar angles and energies. Finally, it accounts for $e^+e^-$ photoproduction. To estimate the contribution of the charge
symmetric background electrons relative to the inclusive electrons as a function of kinematics, an estimate of the ratio of the cross
section for the charge symmetric electrons to the Born cross section is required. To do this the code selects the charge symmetric
background electrons in our $(W,Q^2)$ bins at the input beam energy. The computation shows that we do not have any significant 
contamination (less than 0.1\%) because $W$ is less than 2.525~GeV for all but the last $Q^2$ bin, where it is limited to 2.275~GeV. 
The minimum momentum of electrons in our analysis is more than 2.77~GeV, while the charge symmetric background electrons have lower 
momenta in general. This finding is consistent with the code output and is also consistent with the results based on the RG-A data.

\vskip 0.2cm
\noindent
\underline{Torus field map}: Different finite-element analysis models of the CLAS12 torus magnetic field were available for study based 
on an ideal model of the torus coil geometry and on a model based on survey measurements of the actual coils during their manufacture. 
For most of our analysis bins the difference in the cross sections between these two models was below 3\%, so this difference was 
assigned as a systematic uncertainty.

\vskip 0.2cm
\noindent
\underline{Sector-dependent studies}: In inclusive electron scattering, there is no $\phi$ dependence in the production amplitude, which
should result in uniform cross sections over the azimuthal angle $\phi$. The CLAS12 Forward Detector consists of six sectors that divide
the azimuthal acceptance such that each sector has azimuthal coverage varying from 50\% of 2$\pi$/6 at 5$^{\circ}$ to 80\% of 2$\pi$/6 at 
40$^{\circ}$. However, not all of the sectors shared the same configuration in this dataset. Sectors 1, 2, and 6 had no additional 
Forward Carriage detectors in front of FTOF, while sectors 3, 4, and 5 had Cherenkov detectors installed. The different detector
configurations, non-ideal knowledge of the magnetic field, imperfect understanding of the detector alignment, various inefficiencies, 
and the presence of hot or weak detector channels, can lead to sector-to-sector variations. Some cuts and corrections are sector 
dependent to account for known inefficiencies or sector properties. Ultimately, there are six independent measurements of the inclusive
cross sections. The differences between them are reported as a systematic uncertainty. The sector dependence is estimated as the 
corrected sample standard deviation over all six sectors, {\it i.e.}

\begin{widetext}
\begin{equation}
\label{eq:sysSecEq}
\delta \sigma_{sec}(W, Q^2) = \sqrt{\frac{1}{5}\sum_{i=1}^6(\sigma_{mean}(W, Q^2)-\sigma_{sec_i}(W, Q^2))^2} \cdot 
\frac{100}{\sigma_{mean}(W, Q^2)},
\end{equation}
\end{widetext}

\noindent
where $\sigma_{mean}(W, Q^2)$ is the average cross section over the six sectors and $\sigma_{sec_i}(W, Q^2)$ is the cross section 
determined for sector $i$. The sector dependence is $\approx$5\% for $Q^2 < 6.53$~GeV$^2$. It starts to increase with $Q^2$ and reaches
up to  15\% for $W < 1.4$~GeV in the last $Q^2$ bin, but for $W>1.7$~GeV, it is below 5\% for all $Q^2$ bins. This sector dependence is
the dominant source of bin-by-bin systematic uncertainty in our measurement. 

\subsubsection{Scale-Type Systematics}

\noindent
\underline{Beam charge}: The integrated beam charge has two sources of systematic uncertainty. The first is a 1\% uncertainty associated
with the calibration of the charge integrator on the Faraday cup. The second is a 0.6\% uncertainty associated with the finite number of
charge readings within the data file and with how these charge readings are unpacked in the reconstruction code. Combining these two 
sources gives an assigned scale systematic uncertainty of 1.2\% on the measured Faraday cup charge.

\vskip 0.2cm
\noindent
\underline{Background merging}: The simulations for this analysis employed a background merging approach in order to best match the
backgrounds in the various CLAS12 detector subsystems. The studies of this procedure comparing track reconstruction efficiency in data
and MC as a function of beam current result in a 3\% scale uncertainty in the tracking efficiency in the CLAS12 Forward Detector.

\vskip 0.2cm
\noindent
\underline{Target thickness}: Finite element analysis of the RG-A cryotarget both in its room temperature configuration and at its operating
temperature were performed. The 5.0~cm long target has 30~$\mu$m-thick aluminum entrance and exit windows. The cell base, base tube, and
Kapton are epoxied to a fixture, and their location is within 0.5~mm of the model dimensions in $z$. The cell is manufactured by 
pre-bowing the end windows to minimize stresses at its operating temperature and pressure. The mechanical tolerance is quoted as 
5.00$\pm$0.05~cm for a systematic uncertainty of 1\% on the overall target length. The consideration of the thermal expansion of the
different materials in the target assembly has been shown to have a much smaller effect on the target cell length uncertainty compared 
to the mechanical tolerance and can be neglected. Also included in this contribution is the uncertainty on the average liquid-hydrogen target 
thickness from the fluctuations of the pressure and temperature variations during the experiment. This gives rise to a scale uncertainty of 1.5\%.

\subsubsection{Total Systematic Uncertainty}

The total systematic uncertainty is the quadrature sum of the aforementioned sources,

\begin{equation}
\label{eq:sysTotalEq}
\delta \sigma_{total}(W, Q^2) = \sqrt{\sum_{i=1}^{N_{sources}}\delta \sigma_{i}(W, Q^2)^2},
\end{equation}

\noindent
where the sum runs over the $N_{sources}$ of systematic uncertainty and $\delta \sigma_i(W, Q^2)$ is the systematic uncertainty associated 
with source $i$ in \%.

\begin{figure*}[htp]
\centering
\includegraphics[width=0.98\textwidth]{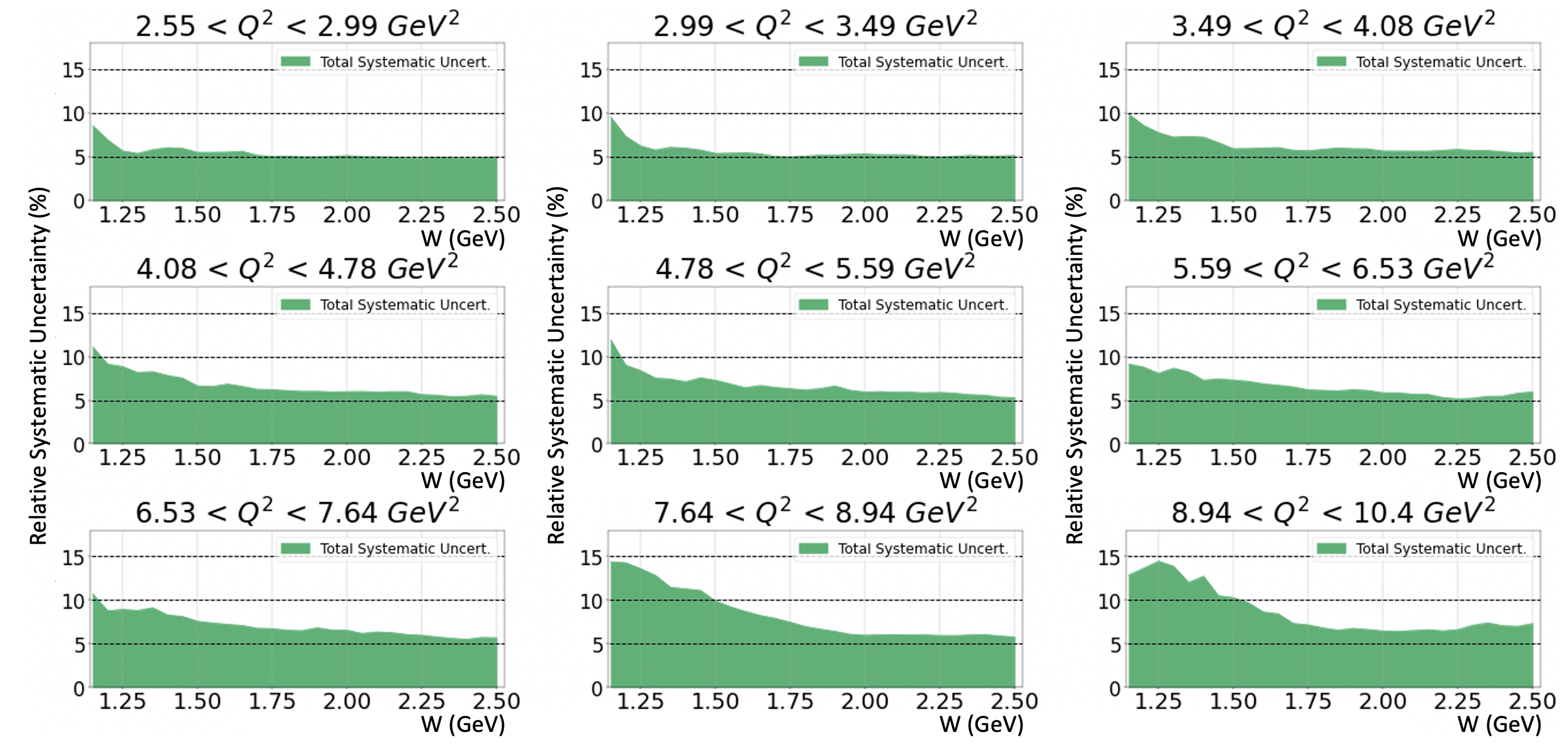}
\caption{Total relative systematic uncertainty for all of our $(W,Q^2)$ bins. The black lines show the 5\%, 10\%, and 15\% levels.}
\label{fig:sys_Total}
\end{figure*}

The relative total systematic uncertainty as a function of $W$ for all $Q^2$ bins is shown in Fig.~\ref{fig:sys_Total}. The systematic
uncertainties averaged over all $Q^2$ and $W$ are given in Table~\ref{tab:totalSysT}. The average bin-by-bin systematic uncertainty for 
this inclusive cross section measurement is 5.8\%.

\begin{table}[htp]
\begin{center}
\begin{tabular}{|c| c|} \hline
  \multicolumn{2}{|c|}{\textbf{Average Systematic Uncertainty}} \\ \hline
  \hline
  \multicolumn{1}{|c|}{Bin-by-Bin Sources} & \multicolumn{1}{c|}{ Uncertainty [\%]}\\ \hline
  Sampling fraction cut & 0.02  \\
  Pion contamination & 0.1 \\
  PCAL fiducial cut & 0.12  \\
  Momentum smearing & 0.28  \\
  Bin-centering corrections & 0.32  \\
  Empty target subtraction & 0.33  \\
  Radiative corrections & 0.36  \\
  Momentum corrections & 0.46  \\
  Charge symmetric background & 0.5 \\
  Deconvolution method & 0.55  \\
  Vertex-$z$ cut & 0.57  \\
  DC fiducial cut & 0.72  \\
  Electron pion separation cut & 0.79  \\
  Torus field map & 3.0 \\
  Sector dependence & 4.41  \\ \hline
  Total Bin-by-Bin & 5.8  \\ \hline \hline
\multicolumn{1}{|c|}{Scale Type Sources} & \multicolumn{1}{c|}{Uncertainty [\%]}\\ \hline
  Beam charge uncertainty & 1.2 \\
  Target thickness uncertainty & 1.8 \\
  Background merging & 3.0   \\ \hline
  Total Scale & 3.7  \\ \hline \hline
  Total Bin-by-Bin and Scale & 6.9  \\ \hline
\end{tabular}
\caption{Systematic uncertainty for the inclusive $p(e,e')X$ cross sections averaged over all $(W, Q^2)$ kinematic bins.}
\label{tab:totalSysT}
\end{center}
\end{table}

\section{Results and Discussion}
\label{sec:results}

\begin{figure}[htp]
\centering
\includegraphics[width=0.9\columnwidth]{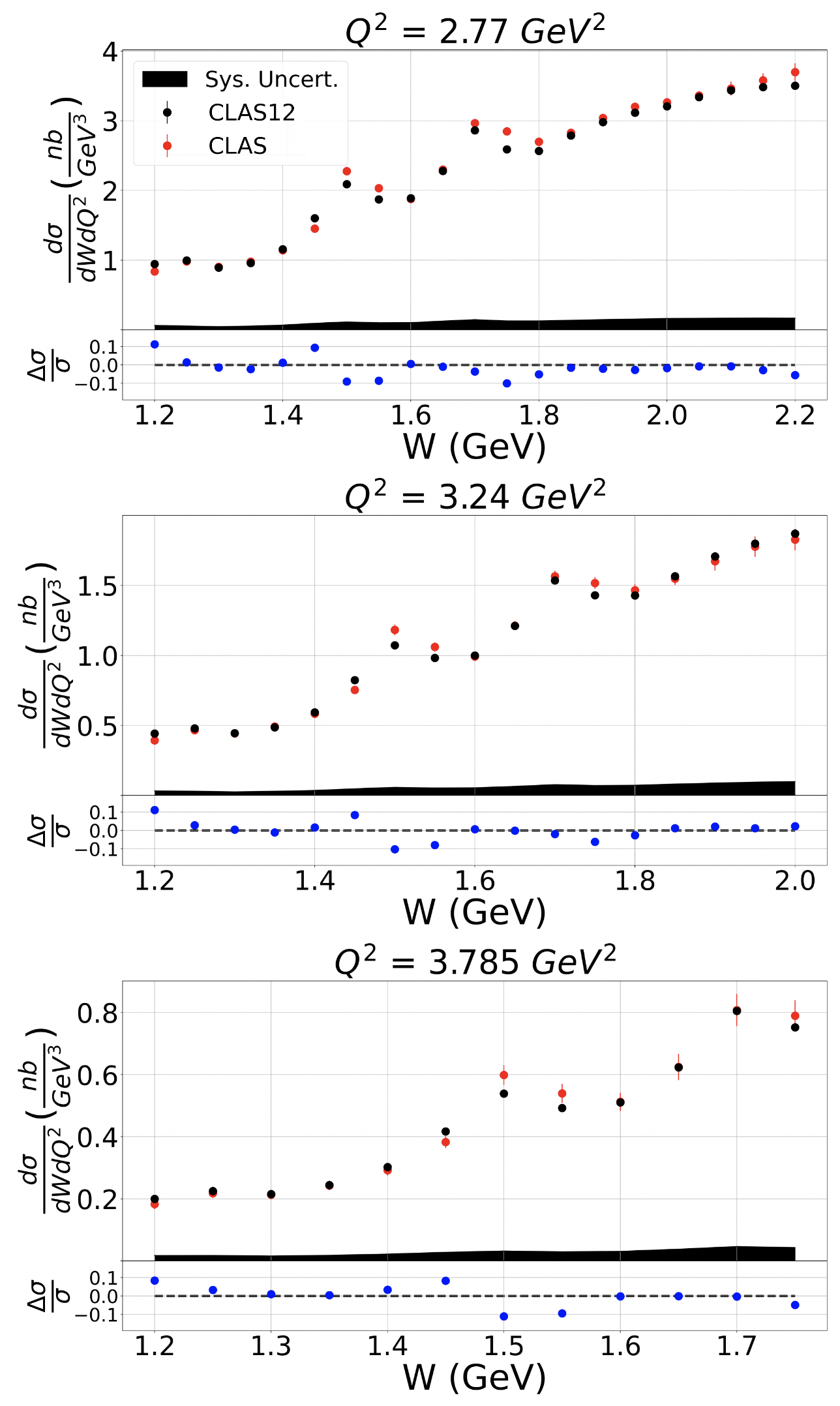}
\caption{Comparison of the CLAS12 inclusive electron scattering cross sections from this work (black points) to those from CLAS (red
points)~\cite{osipenko} after interpolation into the kinematic grid of our experiment. The statistical uncertainties on the CLAS12 data 
are shown but they are smaller than the data point size for the majority of the points. The bin-by-bin systematic uncertainties of the 
CLAS12 data are shown by the filled area at the bottom of each plot. The error bars shown on the CLAS data are the quadrature sum of the 
statistical and systematic uncertainties. Each sub-plot shows the relative cross section difference, 
$(\sigma_{\rm CLAS12}-\sigma_{\rm CLAS})/\sigma_{\rm CLAS12}$, between the two measurements.}
\label{fig:XSEC_res_3bins}
\end{figure}

\begin{figure}
\centering
\includegraphics[width=0.98\columnwidth]{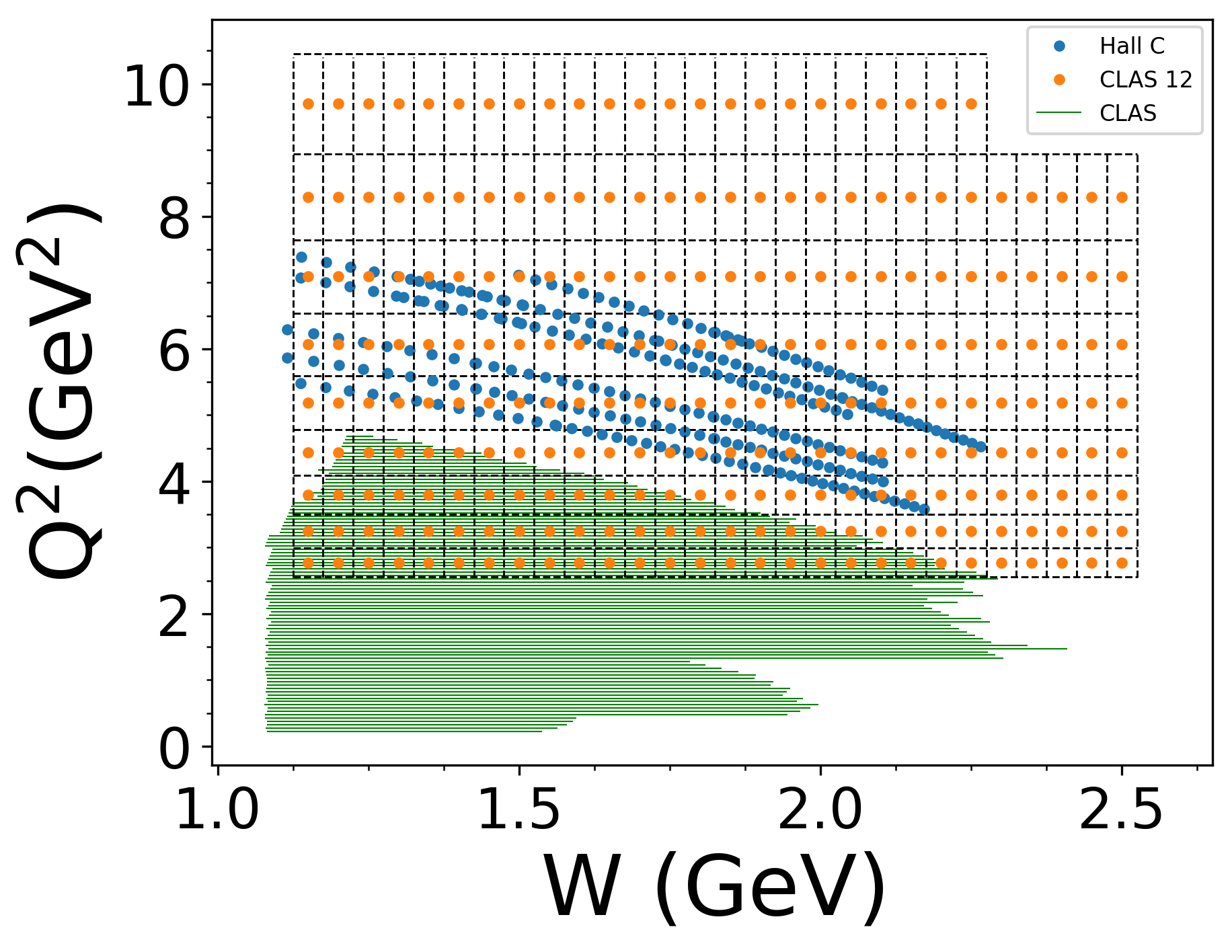}
\caption{Comparison of the kinematic coverage of the CLAS (shown by the green lines) and Hall~C $p(e,e')X$ cross sections (shown by
the blue points) from Ref.~\cite{osipenko} and Ref.~\cite{malace09}, respectively, to that from this analysis of CLAS12 RG-A data 
(shown with the bin centered values at the orange points within the defined bins shown by the grid) in terms of $Q^2$ vs.~$W$.}
\label{kin_cover}
\end{figure}

In this section we present our final inclusive electron scattering cross sections. Our starting point is a comparison of the CLAS12
results with the previous CLAS results measured for $W < 2.0$~GeV and $Q^2 < 3.6$~GeV$^2$~\cite{osipenko} shown in
Fig.~\ref{fig:XSEC_res_3bins}. Figure~\ref{kin_cover} compares the $(Q^2,W)$ kinematic coverage of the CLAS data (shown by the green 
lines) to our CLAS12 data (shown by the orange points). The CLAS results were interpolated into the $(W,Q^2)$ grid of our experiment 
accounting for the difference in electron beam energy. This difference has an impact on both the virtual photon flux $\Gamma_v$ and the 
virtual photon polarization parameter $\epsilon$ defined by Eqs.(10,11) in Ref.~\cite{skorodumina18}. The unpolarized cross sections 
$\sigma_U$ were computed from the CLAS results on the $F_2$ structure function according to Eq.(4) in Ref.~\cite{blin21} with parameter 
$\epsilon$ for the CLAS measurements and the ratio $R_{LT}=\sigma_L/\sigma_T$ taken from Ref.~\cite{ricco99}. The unpolarized cross 
sections $\sigma_U$ from the CLAS measurement have been converted into the form used here for the CLAS12 data according to Eq.(5) in 
Ref.~\cite{blin21} with the $\sigma_T$ and $\sigma_L$ values determined from the CLAS measurements and with $\epsilon$ parameters for an 
electron beam energy of 10.6~GeV. The CLAS data shown in Fig.~\ref{fig:XSEC_res_3bins} were computed as the product of the virtual photon 
flux and the unpolarized $\sigma_U$ cross section for the electron beam energy of 10.6~GeV.
 
All details of the interpolation/extrapolation of the inclusive structure functions can be found in Ref.~\cite{golubenko}. The website 
of Ref.~\cite{sfcs-web} provides for an evaluation of the interpolated $p(e,e')X$ observables onto a $(W,Q^2)$ kinematic grid for an
incoming electron beam energy defined by the user for $W < 2.0$~GeV and $Q^2 < 7.0$~GeV$^2$. Our new CLAS12 measurement is consistent
within uncertainties with the previous CLAS measurement.

Measurements of the inclusive $p(e,e')X$ cross sections in the resonance region were also carried out in Hall~C at JLab with a 
small-acceptance spectrometer at high luminosity achieved with an electron beam current up to 100~$\mu$A~\cite{malace09}. Because of 
the small acceptance of the Hall~C spectrometer, those inclusive cross sections are only available within bins for highly correlated 
values of $W$ and $Q^2$ shown by the blue data points in Fig.~\ref{kin_cover} compared to the kinematic coverage of our CLAS12 data. 
The Hall~C data below $W$=2.5~GeV are only available within the limited $Q^2$ range from 3.5--7.5~GeV$^2$ in comparison with the CLAS12 
RG-A results obtained for $Q^2$ from 2.55--10.4~GeV$^2$. Furthermore, the large electron scattering angle acceptance of CLAS12 allows us 
to obtain the $p(e,e')X$ cross sections for each given $Q^2$ bin with a broad coverage over $W$ from the pion threshold up to 2.5~GeV. 
This feature is important for the exploration of the $Q^2$ evolution of the partonic structure of the nucleon ground states. On the 
other hand, the high luminosity achievable in the Hall~C measurements allows for exploration of both the $W$ and $Q^2$ evolution of the 
$p(e,e')X$ cross sections in more detail, collecting data for finer $(W,Q^2)$ bins than achievable in our CLAS12 measurements. Therefore, 
a combined analysis of the $p(e,e')X$ cross sections from CLAS12 and Hall~C would be beneficial for gaining insight into the structure of 
the nucleon ground states at large values of the fractional parton momenta $x$ in the resonance region. In addition, the $p(e,e')X$ data 
obtained at different beam energies in the measurements with CLAS12 and in Hall~C for a given $(W,Q^2)$ bin offer the opportunity for 
separation between the longitudinal and transverse contributions to the unpolarized virtual photon+proton cross sections.

\begin{figure}[t]
\centering
\includegraphics[width=0.98\columnwidth]{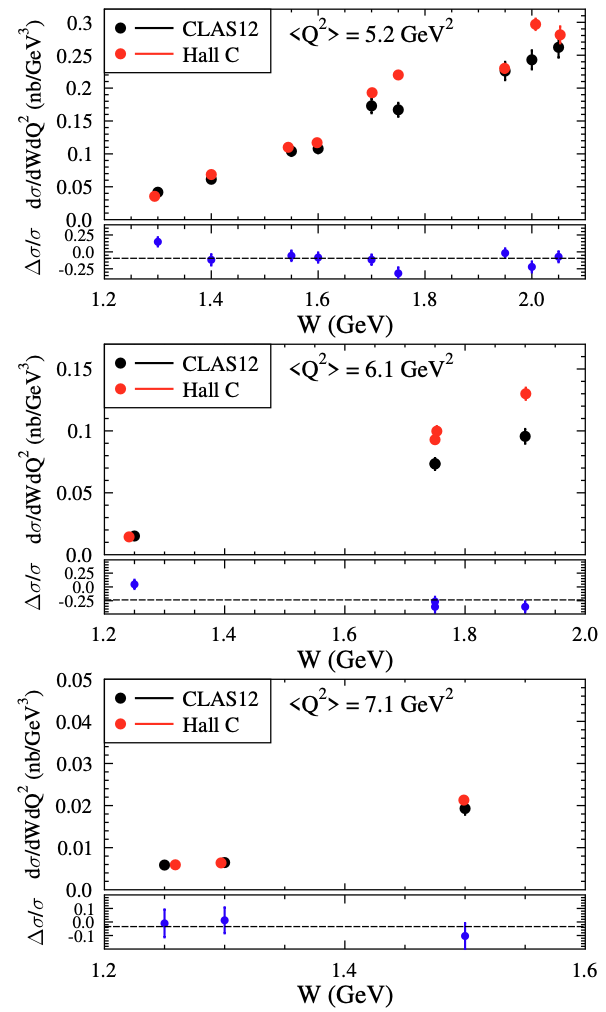}

\caption{Comparison of the CLAS12 inclusive electron scattering cross sections from this work to those from Hall~C~\cite{malace09}.
The Hall~C data have been converted from $d\sigma/d\Omega dE'$ to $d\sigma/dW dQ^2$ as described in the text accounting for the different 
Hall~C beam energy (5.5~GeV) compared to this CLAS12 dataset (10.604~GeV). Only selected data points from the Hall~C measurement were 
compared to the CLAS12 cross sections satisfying $\Delta Q^2 < 0.1$~GeV$^2$ and $\Delta W < 0.01$~GeV. The data can be compared for 3
average $Q^2$ bins of 5.2~GeV$^2$ (top), 6.1~GeV$^2$ (middle), and 7.1~GeV$^2$ (bottom). The bottom sub-plot with each $Q^2$ bin shows 
the relative cross section difference, $(\sigma_{\rm CLAS12}-\sigma_{\rm Hall~C})/\sigma_{\rm CLAS12}$, between the two measurements with 
the dashed line representing the average difference for each $Q^2$ bin.}
\label{data-comp}
\end{figure}

A comparison between the CLAS12 RG-A and Hall~C $p(e,e')X$ cross sections was carried out. The $\sigma_U$ cross sections were derived 
from the Hall~C $d\sigma/d\Omega dE'$ cross section data~\cite{malace09} by multiplying the inverse virtual photon flux values 
$1/\Gamma_v$ for the kinematics in the Hall~C measurements. The $\sigma_T$ and $\sigma_L$ contributions were deduced from $\sigma_U$ 
in a similar way as described above for the CLAS results. After that, the $\sigma_U$ cross sections were re-evaluated from the $\sigma_T$ 
and $\sigma_L$ components with $\epsilon$ computed using the beam energy $E_b$ and $(W,Q^2)$ from the CLAS12 data. Finally, the
appropriate Jacobian was introduced. Appropriately selected data points from the Hall~C measurement were compared to the CLAS12 cross 
sections for those bins satisfying the criteria $\Delta Q^2 < 0.1$~GeV$^2$ and $\Delta W < 0.01$~GeV. In total there are 16 data points 
for which a comparison between the two datasets can be made. Figure~\ref{data-comp} shows a direct comparison of the cross sections from 
the two measurements for these selected kinematic points. The two datasets are in agreement within uncertainties for most data points.
However, there is a bit of tension in the comparison for the $Q^2 = 6.1$~GeV$^2$ bin where the scale difference is $\sim$25\%. We have
not attempted to assign an uncertainty due to the mis-match of the $(W,Q^2)$ bin centers between the CLAS12 and Hall~C data or to the
parameterized value of $R_{LT}$ in the transformation of the Hall~C cross sections. In reality, these considerations introduce 
additional sources of systematic uncertainty that must be considered and accounted for before a fully meaningful direct comparison 
can be performed. The data points from Hall~C within the $(W,Q^2)$ bins of the RG-A CLAS12 measurements offer complementary information 
on the cross section evolution within these bins that can be used in the future for improvement in the evaluation of the averaged cross
sections.

\begin{figure*}[htp]
\centering
\includegraphics[width=1.0\textwidth]{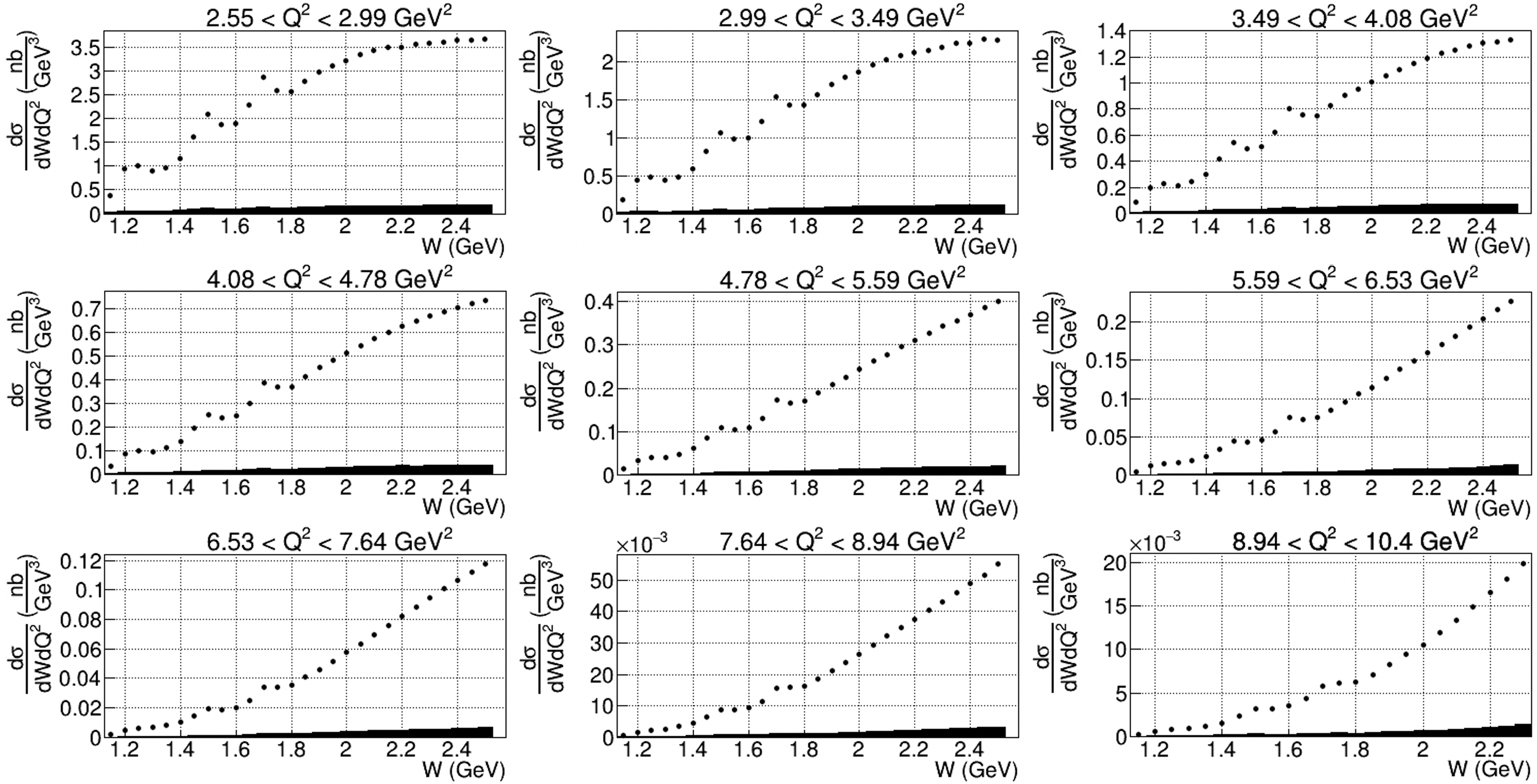}
\caption{Inclusive electron scattering cross sections determined from CLAS12 RG-A data. The statistical uncertainties on the CLAS12 
data are shown but they are smaller than the data point size for the majority of the data points. The bin-by-bin systematic uncertainty 
is shown by the filled area at the bottom of each plot.}
\label{fig:XSEC_res}
\end{figure*}

Our final results for the extracted cross sections from the CLAS12 RG-A dataset are shown in Fig.~\ref{fig:XSEC_res}. These data are 
also included into the CLAS Physics Database~\cite{physicsdb}. The cross sections $d\sigma/dWdQ^2$ are shown as a function of $W$ over
the range from 1.125 to 2.525~GeV in 50-MeV wide bins for our 9 bins of $Q^2$ from 2.55 to 10.4~GeV$^2$. Our analysis includes a bin
centering correction to evolve the measured cross sections to the geometric centers of the given $W$ and $Q^2$ bins. The statistical
uncertainties on the CLAS12 data are shown but they are smaller than the data point size for the majority of the data points. The 
bin-by-bin systematic uncertainty is shown by the filled area at the bottom of each plot. 

\begin{figure}[t]
\centering
\includegraphics[width=0.98\columnwidth]{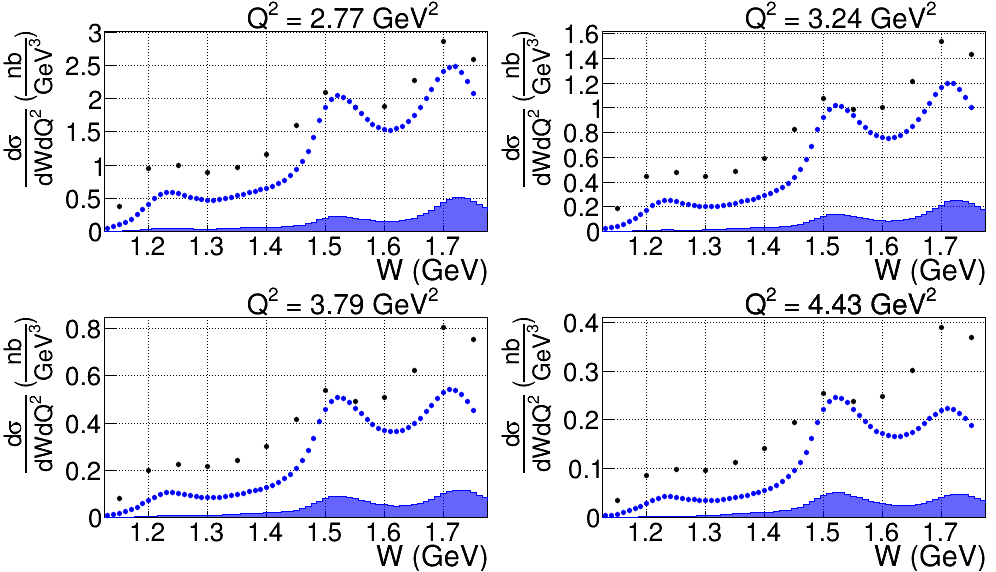}
\caption{Inclusive electron scattering cross sections from CLAS12 data at a beam energy of 10.6~GeV as a function of $W$ for selected
bins in $Q^2$ as shown. The blue points represent the computed resonant contributions from the experimental results on the resonance
electrocouplings from the studies of $\pi N$, $\eta N$, and $\pi^+\pi^- p$ electroproduction off protons with CLAS~\cite{blin19,blin21}.
The shaded areas at the bottom of each plot show the systematic uncertainties for the evaluation of the resonant contributions.}
\label{incl-cs}
\end{figure}

The CLAS12 measurements of the inclusive electron scattering cross sections as a function of $W$ for selected bins in $Q^2$ are shown 
in Fig.~\ref{incl-cs}. The values of $Q^2$ selected for display range from 2.77 to 4.43~GeV$^2$. The error bars shown are
statistical only. The blue points represent the resonance contributions evaluated from the CLAS results on the resonance electrocouplings
available from the studies of $\pi N$, $\eta N$, and $\pi^+\pi^- p$ electroproduction off protons within the framework of the approach
developed in Refs.~\cite{blin19,blin21}. The resonant amplitudes were computed within the Breit-Wigner ansatz with the $\gamma_vpN^*$
electrocouplings available from the studies of meson electroproduction data in the resonance region~\cite{Car20}. The $N^*$ total decay
widths were taken from the experiments with hadronic probes reported in the PDG~\cite{PDG23}. So far this is the only available 
evaluation of the resonant contributions from the experimental results on the $N^*$ electrocouplings and their hadronic decay widths. 
Currently, as the resonance electrocouplings from the available CLAS data span the mass range of $W < 1.75$~GeV, the blue points do not
extend beyond this value. The resonant contributions to the inclusive $p(e,e')X$ cross sections remain significant over the entire range 
of $Q^2 < 5$~GeV$^2$ where the experimental results on the $N^*$ electrocouplings are available.

The unique capability of the CLAS12 detector in providing the inclusive electron scattering observables in the full range of $W$ from 
the pion threshold to 2.5~GeV at any given $Q^2$ opens up promising opportunities for the exploration of the ground state nucleon PDFs 
at large $x$ in the resonance region with the resonant contributions estimated from experimental data. The structures seen in the
$p(e,e')X$ data in Fig.~\ref{incl-cs} in the first, second, and third resonance regions for $Q^2 < 5$~GeV$^2$ are related to the 
resonant contributions. The three resonance region peaks demonstrate pronounced differences in their evolution with $Q^2$ because of
differences in the $Q^2$ evolution of the electrocouplings of the contributing $N^*$ and $\Delta^*$ states. This observation 
emphasizes the importance of obtaining information on the electrocouplings of all prominent resonances to enable a credible evaluation 
of the resonant contributions into the $p(e,e')X$ cross sections. Notably, the resonance-like structures in Fig.~\ref{fig:XSEC_res} are
clearly seen over the entire range of $Q^2$ covered in the CLAS12 data, suggesting opportunities for extraction of the $N^*$ 
electrocouplings for $Q^2$ up to 10~GeV$^2$ from the studies of the exclusive electroproduction channels.

\begin{figure*}[ht]
\centering
\includegraphics[width=0.75\textwidth]{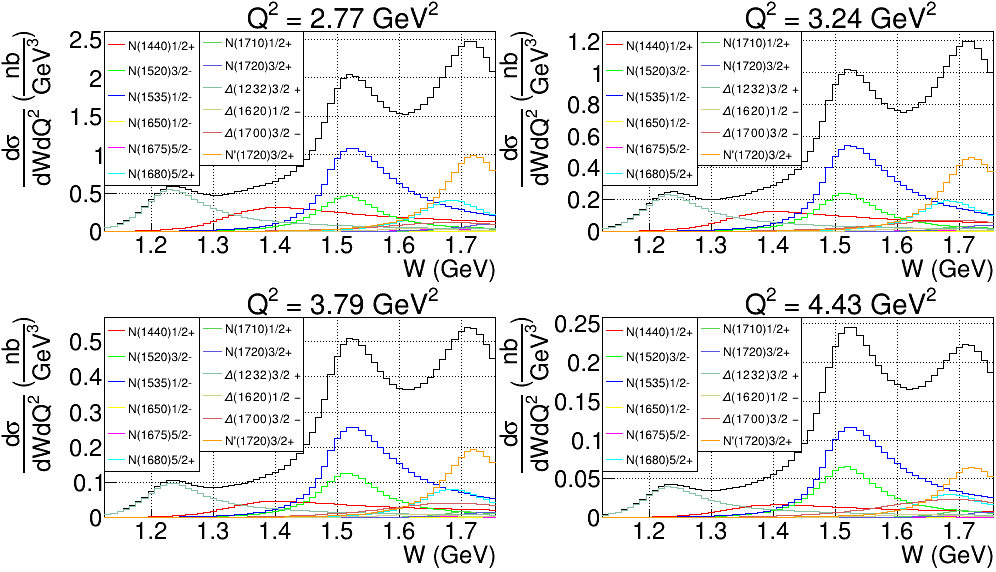}
\caption{Decomposition of the total resonant contribution (black curves) to the inclusive electron scattering cross section in the 
resonance region at different $Q^2$ for a beam energy of 10.6~GeV, showing the separate contributions of each resonance in the mass 
range up to 1.75~GeV (see legend on the plots).}
\label{breakdown}
\end{figure*}

Figure~\ref{breakdown} shows the decomposition of the resonant contributions to the total cross section for the same selected $Q^2$ 
bins in terms of the contributions from the individual excited nucleon states~\cite{blin19}. Interference between different $N^*$ and 
$\Delta^*$ states was taken into account using the approach described in Ref. \cite{blin21}. Each of the three resonance region 
maxima seen in our data arise from several excited states of the nucleon. Even the first resonance maximum, which is dominated by 
the $\Delta(1232)3/2^+$, has significant $N(1440)1/2^+$ contributions. The second resonance region maximum is created by the
$N(1440)1/2^+$, $N(1520)3/2^-$, and $N(1535)1/2^-$. As $Q^2$ increases the contribution from the $N(1535)1/2^-$ becomes the largest 
since the $A_{1/2}$ electrocoupling of the $N(1535)1/2^-$ decreases more slowly with $Q^2$. The third resonance region maximum 
comprises several nucleon excited states as shown in Fig.~\ref{breakdown}, with the biggest contribution from the new $N'(1720)3/2^+$
baryon state that was discovered from the combined studies of $\pi^+\pi^-p$ photo- and electroproduction data measured in the 
experiments of 6-GeV era with CLAS~\cite{np1720}. The contributions from the $\Delta(1700)3/2^-$ becomes increasingly important 
as $Q^2$ increases with impact on the resonant contributions not only in the third, but also in the second resonance region. The 
resonant cross sections show a pronounced evolution with $Q^2$ in the first, second, and third resonance regions, although they show a
stronger $Q^2$ fall-off in the first and third resonance regions compared to the second. This suggests that the different excited 
nucleon states display distinctively different structural features in the $Q^2$ evolution of their electrocouplings. This underscores 
the necessary efforts on the extraction of the $N^*$ electrocouplings of all prominent resonances for $Q^2 > 4$~GeV$^2$ from the 
upcoming data from CLAS12, bridging the efforts between analyses in the $N^*$ and deep inelastic physics regimes
\cite{accardi22,Brod20}.

Our data on the inclusive $p(e,e')X$ cross section allow for extraction of the inclusive $F_2$ structure function in any given bin of 
$Q^2$ within the coverage over $W$ from the pion threshold up to 2.5~GeV. Consequently, the truncated moments of the $F_2$ structure
function can be obtained by direct integration of the data at each given $Q^2$. The information on the evolution of the truncated 
$F_2$ structure function moments within the resonance region over the broad range of $Q^2$ covered in our measurements will allow us 
for the first time to explore the evolution of the partonic structure of the ground state of the nucleon for large values of $x$ 
within the resonance excitation region in the range of distances where the transition from strongly coupled to perturbative QCD 
regimes is expected~\cite{Car23,accardi22,ding23}.

\section{Summary and Conclusions}
\label{sec:summary}

In this paper we present results from the first absolute cross section measurements with the new large-acceptance CLAS12 spectrometer
in Hall~B at Jefferson Lab. Inclusive electron scattering cross sections are provided over a broad range of invariant mass of the final 
state hadrons $W$ from 1.125 to 2.525~GeV and four-momentum transfer squared (or photon virtuality) $Q^2$ from 2.55 to 10.4~GeV$^2$ 
collected at a beam energy of 10.6~GeV. The measured cross sections agree well with the world data available at $W$ from the $\pi N$ 
threshold to 2.5~GeV for $Q^2$ from 2.5--7.0~GeV$^2$~\cite{osipenko,malace09}. The large electron scattering angle acceptance of the 
CLAS12 detector makes it possible in each bin of $Q^2$ to cover the entire range of $W$ from the $\pi N$ threshold to 2.5~GeV. The wide 
$W$-coverage at any given $Q^2$ is of particular importance for extending insight into the ground state nucleon partonic structure at 
large values of fractional parton momenta $x$ within the resonance excitation region. The broad coverage over $Q^2$ allows for 
exploration of the evolution of the partonic structure of the nucleon within the range of distances where the transition from the
strongly coupled to perturbative QCD regimes is expected~\cite{Brod20,Cr22}.

The CLAS results on the electroexcitation amplitudes of most excited nucleon states in the mass range up to 1.8~GeV for 
$Q^2 < 5$~GeV$^2$ allow for the evaluation of the resonance contribution into the inclusive electron scattering observables
\cite{blin19,blin21,blin23}. The realistic evaluation of the resonance contributions into the inclusive $p(e,e')X$ cross sections with
the $N^*$ electrocouplings and hadronic decay widths available from the experimental data extends the capabilities for understanding
the ground state nucleon partonic structure within the resonance region and its evolution with distance. The structures in the first,
second, and third resonance regions observed in the $W$-dependence of the inclusive $p(e,e')X$ cross sections measured with CLAS12 over
the range of $Q^2 < 10$~GeV$^2$ suggest the opportunity to extend the information on the $Q^2$ evolution of the nucleon resonance
electrocouplings for $Q^2$ over this full range from the measurements of exclusive meson electroproduction in the resonance region that
are currently in progress for data collected with CLAS12. Furthermore, these results offer a new opportunity to extend the scope for 
the exploration of quark-hadron duality~\cite{meln05}.

\vskip 0.3cm

The authors would like to acknowledge the outstanding efforts of the JLab staff that made this experiment possible. This work was
supported in part by the Chilean Comisi\'on Nacional de Investigaci\'on Cient\'ifica y Tecnol\'ogica (CONICYT), the Italian Istituto
Nazionale di Fisica Nucleare, the French Centre National de la Recherche Scientifique, the French Commissariat \`{a} l'Energie 
Atomique, the Scottish Universities Physics Alliance (SUPA), the United Kingdom's Science and Technology Facilities Council, the 
National Research Foundation of Korea, and the Skobeltsyn Nuclear Physics Institute and Physics Department at the Lomonosov Moscow 
State University.. This work was supported in part by the the U.S. Department of Energy, Office of Science, Office of Nuclear Physics 
under contract DE-AC05-06OR23177 and the National Science Foundation (NSF) under grants PHY 10011349 and PHY 2209421.


\begin{thebibliography}{0}%
\makeatletter
\providecommand \@ifxundefined [1]{%
 \@ifx{#1\undefined}
}%
\providecommand \@ifnum [1]{%
 \ifnum #1\expandafter \@firstoftwo
 \else \expandafter \@secondoftwo
 \fi
}%
\providecommand \@ifx [1]{%
 \ifx #1\expandafter \@firstoftwo
 \else \expandafter \@secondoftwo
 \fi
}%
\providecommand \natexlab [1]{#1}%
\providecommand \enquote  [1]{``#1''}%
\providecommand \bibnamefont  [1]{#1}%
\providecommand \bibfnamefont [1]{#1}%
\providecommand \citenamefont [1]{#1}%
\providecommand \href@noop [0]{\@secondoftwo}%
\providecommand \href [0]{\begingroup \@sanitize@url \@href}%
\providecommand \@href[1]{\@@startlink{#1}\@@href}%
\providecommand \@@href[1]{\endgroup#1\@@endlink}%
\providecommand \@sanitize@url [0]{\catcode `\\12\catcode `\$12\catcode `\&12\catcode `\#12\catcode `\^12\catcode `\_12\catcode `\%12\relax}%
\providecommand \@@startlink[1]{}%
\providecommand \@@endlink[0]{}%
\providecommand \url  [0]{\begingroup\@sanitize@url \@url }%
\providecommand \@url [1]{\endgroup\@href {#1}{\urlprefix }}%
\providecommand \urlprefix  [0]{URL }%
\providecommand \Eprint [0]{\href }%
\providecommand \doibase [0]{http://dx.doi.org/}%
\providecommand \selectlanguage [0]{\@gobble}%
\providecommand \bibinfo  [0]{\@secondoftwo}%
\providecommand \bibfield  [0]{\@secondoftwo}%
\providecommand \translation [1]{[#1]}%
\providecommand \BibitemOpen [0]{}%
\providecommand \bibitemStop [0]{}%
\providecommand \bibitemNoStop [0]{.\EOS\space}%
\providecommand \EOS [0]{\spacefactor3000\relax}%
\providecommand \BibitemShut  [1]{\csname bibitem#1\endcsname}%
\let\auto@bib@innerbib\@empty
\end{thebibliography}%


\begin{thebibliography}{99}

\bibitem{accardi}
A. Accardi {\it et al.}, Phys. Rev. D {\bf 93}, 114017 (2016).

\bibitem{alekh}
S. Alekhin {\it et al.}, Phys. Rev. D {\bf 96}, 041011 (2017).

\bibitem{ball}
R.D. Ball {\it et al.} {\it (NNPDF Collaboration)}, Eur. Phys. J. C {\bf 77}, 663 (2017).

\bibitem{revpdf}
J. Gao, L. Harland-Lang, and J. Rojo, Phys. Rep.  {\bf 742}, 1 (2018).

\bibitem{tmc1}
H. Georgi and H.D. Politzer, Phys. Rev. D {\bf 14}, 1829 (1976).

\bibitem{tmc2}
I. Schienbein {\it et al.}, J. Phys. G {\bf 35}, 053101 (2008).

\bibitem{bodek}
A. Bodek {\it et al.}, Phys. Rev. D {\bf 20}, 1471 (1979).

\bibitem{osipenko}
M. Osipenko {\it et al.} {\it (CLAS Collaboration)}, Phys. Rev. D {\bf 67}, 092001 (2003).

\bibitem{malace09}
S.P. Malace {\it et al.}, Phys. Rev. C {\bf 80}, 035207 (2009).

\bibitem{tvaskis}
V. Tvaskis {\it et al.}, Phys. Rev. C {\bf 97}, 045204 (2018).

\bibitem{liang}
Y. Liang et al. (Jefferson Lab E94-110 Collaboration), Phys. Rev. C {\bf 105}, 065205 (2022).

\bibitem{mecking}
B.A. Mecking {\it et al.}, Nucl. Inst. and Meth. A {\bf 503}, 513 (2003).

\bibitem{christy10}
M.E. Christy and P.E. Bosted, Phys. Rev. C {\bf 81}, 055213 (2010).

\bibitem{meln05}
W. Melnitchouk, R. Ent, and C. Keppel, Phys. Rep. {\bf 406}, 127 (2005).

\bibitem{bh1}
E.D. Bloom and F.J. Gilman, Phys. Rev. Lett. {\bf 25}, 1140 (1970).

\bibitem{bh2}
E.D. Bloom and F.J. Gilman, Phys. Rev. D {\bf 4}, 2901 (1971).

\bibitem{Bu12}
I.G. Aznauryan and V.D. Burkert, Prog. Part. Nucl. Phys. {\bf 67}, 1 (2012).

\bibitem{Mokeev:2018zxt}
V.I. Mokeev {\it et al.} {\it (CLAS Collaboration)}, Few Body Syst. {\bf 59}, 134 (2018).

\bibitem{Mo19}
V.I. Mokeev {\it et al.} {\it (CLAS Collaboration)}, Eur. Phys. J. Web Conf. {\bf 241}, 03003 (2020).

\bibitem{Car20}
D.S. Carman, K. Joo, and V.I. Mokeev, Few Body Syst. {\bf 61}, 29 (2020).

\bibitem{mokeev23}
V.I. Mokeev {\it et al.}, Phys. Rev. C {\bf 108}, 025204 (2023).

\bibitem{Car24}
V.I. Mokeev and D.S. Carman, Nuovo Cimento C {\bf 47}, 216 (2024).

\bibitem{Mo12} 
V.I. Mokeev {\it et al.}  {\it (CLAS Collaboration)}, Phys. Rev. C {\bf 86}, 035203 (2012).

\bibitem{Mo16}
V.I. Mokeev {\it et al.}, Phys. Rev. C {\bf 93}, 025206 (2016).

\bibitem{Car23}
D.S. Carman, R.W. Gothe, V.I. Mokeev, and C.D. Roberts, Particles 6, 416 (2023).

\bibitem{blin19}
A.N. Hiller Blin {\it et al.}, Phys. Rev. C {\bf 100}, 035201 (2019).

\bibitem{blin21}
A.N. Hiller Blin {\it et al.}, Phys. Rev. C {\bf 104}, 025201 (2021).

\bibitem{blin23}
A.N. Hiller Blin {\it et al.}, Phys. Rev. C {\bf 107}, 035202 (2023).

\bibitem{clas12nim}
V.D. Burkert {\it et al.} {\it (CLAS Collaboration)}, Nucl. Inst. and Meth A {\bf 959}, 163419 (2020).

\bibitem{Qiu2019}
Y.-Q. Ma and J. Qiu, Phys. Rev. Lett. {\bf 120}, 022003 (2018).

\bibitem{Radyushkin2019}
A.V. Radyushkin, Int. J. Mod. Phys. A {\bf 35}, 2030002 (2020).

\bibitem{Lin21}
H-W. Lin, PoS LHCP2021, 352 (2021).

\bibitem{Ale23}
C. Alexandrou, J. Phys. Conf. Ser., {\bf 2586}, 012001 (2023).

\bibitem{Cr22}
Y. Lu {\it et al.}, Phys. Lett. B {\bf 830}, 137130 (2022).

\bibitem{Cr23}
P.-L. Yin {\it et al.}, Chin. Phys. Lett. {\bf 40}, 091201 (2023).

\bibitem{croberts1}
J. Segovia {\it et al.}, Few Body Syst. {\bf 59}, 26 (2018).

\bibitem{croberts2}
C. Mezrag {\it et al.}, Phys. Lett. B {\bf 783}, 263 (2018).

\bibitem{Bu18}
V.D. Burkert, Ann. Rev. Nucl. Part. Sci {\bf 68}, 405 (2018).

\bibitem{Bu20}
V.D. Burkert, EPJ Web Conf. {\bf 241}, 01004 (2020).

\bibitem{dc-nim}
M.D. Mestayer {\it et al.}, Nucl. Inst. and Meth. A {\bf 959}, 163518 (2020).

\bibitem{htcc-nim}
Y.G. Sharabian {\it et al.}, Nucl. Inst. and Meth. A {\bf 968}, 163824 (2020).

\bibitem{ecal-nim}
G. Asryan {\it et al.}, Nucl. Inst. and Meth. A {\bf 959}, 163425 (2020).

\bibitem{ftof-nim}
D.S. Carman {\it et al.}, Nucl. Inst. and Meth. A {\bf 960}, 163629 (2020).

\bibitem{daq-nim}
S. Boyarinov {\it et al.}, Nucl. Inst. and Meth. A {\bf 966}, 163698 (2020).

\bibitem{trigger-nim}
B. Raydo {\it et al.}, Nucl. Inst. and Meth. A {\bf 960}, 163529 (2020).

\bibitem{eb-nim}
V. Ziegler {\it et al.}, Nucl. Inst. and Meth. A 959, 163472 (2020).

\bibitem{cn-2023-001}
R. Capobianco {\it et al.}, CLAS12-Note 2023-001, (2023),
\url{https://misportal.jlab.org/mis/physics/clas12/viewFile.cfm/2023-001.pdf?documentId=83}

\bibitem{simulation-nim}
M. Ungaro {\it et al.}, Nucl. Inst. and Meth. A {\bf 959}, 163422 (2020)

\bibitem{geant4}
S.S. Agostinelli {\it et al.}, Nucl. Inst. and Meth. A {\bf 506}, 250 (2003).

\bibitem{misakEG} 
M. Sargsyan, CLAS-NOTE 90-007 (1990), \url{https://www.jlab.org/Hall-B/notes/clas_notes90/note90-007.pdf}.

\bibitem{moTsaiRC} 
L.W. Mo and Y.S. Tsai, Rev. Mod. Phys. {\bf 41}, 205 (1969).

\bibitem{cn-2020-005}
S. Stepanyan {\it et al.}, CLAS12-Note 2020-005, (2020), 
\url{https://misportal.jlab.org/mis/physics/clas12/viewFile.cfm/2020-005.pdf?documentId=70}

\bibitem{aao_gen} 
AAO\_GEN event generator
\url{https://github.com/JeffersonLab/aao_gen}

\bibitem{RooUnfoldArticle} 
L. Brenner {\it et al.}, Int. J. of Mod. Phys. A {\bf 35}, 2050145 (2020).

\bibitem{bayes_article} 
G. D'Agostini, Nucl. Inst. and Meth. A {\bf 362}, 487 (1995).

\bibitem{RooUnfoldCode} 
RooUnfold github repository, \url{https://gitlab.cern.ch/RooUnfold/RooUnfold}.

\bibitem{clasdis}
H. Avakian and P. Bosted, CLASDIS event generator, \url{https://github.com/JeffersonLab/clasdis/blob/master/README.md}.

\bibitem{lepto}
G. Ingelman, A. Edin, and J. Rathsman, Comp. Phys. Comm. {\bf 101}, 108 (1997).

\bibitem{pythia}
T. Sj\"ostrand, Comp. Phys. Comm. {\bf 246}, 106910 (2020).

\bibitem{bosted}
P. Bosted, CLAS-Note 2004-005, (2004), \url{https://misportal.jlab.org/ul/Physics/Hall-B/clas/viewFile.cfm/2004-005.pdf?documentId=54}

\bibitem{skorodumina18} 
G.V. Fedotov {\it et al.}  {\it (CLAS Collaboration)}, Phys. Rev. C {\bf 98}, 025203 (2018).

\bibitem{ricco99} 
G. Ricco {\it et al.}, Nucl. Phys. B {\bf 555}, 306 (1999).

\bibitem{golubenko}
A.A. Golubenko, V.V. Chesnokov, B.S. Ishkhanov, and V.I. Mokeev, Phys. Part. Nucl. {\bf 50}, 587 (2019).

\bibitem{sfcs-web}
Structure function and cross section interface website, \url{https://clas.sinp.msu.ru/strfun/}

\bibitem{physicsdb}
CLAS Physics Database, \url{http://clasweb.jlab.org/physicsdb}.

\bibitem{PDG23}
R.L. Workman {\it et al.} (Particle Data Group) Prog. Theor. Exp. Phys {\bf 2022}, 083C01 (2022).

\bibitem{np1720}
V.I. Mokeev {\it et al.}, Phys. Lett. B {\bf 805}, 135457 (2020).

\bibitem{accardi22}
A. Accardi {\it et al.}, Eur. Phys. J. A 60, 173 (2024).

\bibitem{Brod20}
S.J. Brodsky {\it et al.}, Int. J. Mod. Phys. E {\bf 29}, 2030006, (2020).

\bibitem{ding23}
M. Ding, C.D. Roberts, and S.M. Schmidt, Particles {\bf 6}, 57 (2023).

\end{thebibliography}
\end{document}